\documentclass[preprint,showpacs,preprintnumbers,amsmath,amssymb,superscriptaddress]{revtex4}
\usepackage{graphicx,color}
\usepackage{amsmath,amssymb}
\usepackage{url}
\usepackage{epstopdf}
\newcommand{\hs}{\hspace*{0.5cm}}

\newcommand{\be}{\begin{equation}}
\newcommand{\ee}{\end{equation}}
\newcommand{\bea}{\begin{eqnarray}}
\newcommand{\eea}{\end{eqnarray}}
\newcommand{\nn}{\nonumber}
\newcommand{\crn}{\nonumber \\}

\newcommand{\al}{\alpha}
\newcommand{\la}{\lambda}

\newcommand{\om}{\omega}

\newcommand{\fr}{\frac}

\newcommand{\bc}{\begin{center}}
\newcommand{\ec}{\end{center}}

\newcommand{\La}{\Lambda}

\newcommand{\ps}{\psi}

\newcommand {\ba}{\begin{array}}
\newcommand {\ea}{\end{array}}
\newcommand{\ben}{\begin{enumerate}}
\newcommand{\een}{\end{enumerate}}

\begin{document}

\title{Low energy phenomena of the lepton sector  in an $A_4$ symmetry model with heavy inverse seesaw neutrinos}


\author{T. Phong Nguyen}\email{thanhphong@ctu.edu.vn }
\affiliation{Department of Physics, Can Tho University, 3/2 Street, Ninh Kieu, Can Tho City 94000, Vietnam}

\author{T.T. Thuc} \email{truongtrongthuck17@gmail.com}
\affiliation{Department of Education and Training of Ca Mau, 70 Phan Dinh Phung, Vietnam}

\author{D.T. Si}\email{dangtrungsi@cantho.edu.vn }
\affiliation{Can Tho Department of Education and Training, Can Tho City 94000, Vietnam }

\author{T.~T.~Hong}\email{tthong@agu.edu.vn}
\affiliation{An Giang University, VNU - HCM, Ung Van Khiem Street,
	Long Xuyen, An Giang 88000, Vietnam }
\author{L. T. Hue}\email{lethohue@vlu.edu.vn}
\affiliation{Subatomic Physics Research Group, Science and Technology Advanced Institute, Van Lang University, Ho Chi Minh City 70000, Vietnam}
\affiliation{Faculty of Technology, Van Lang University, Ho Chi Minh City 70000, Vietnam }

\begin{abstract}%
An extension of the two Higgs doublet model  including  inverse seesaw neutrinos and neutral Higgs  bosons   was constructed based on the $A_4$ symmetry in order to explain the recent neutrino oscillation data. This model can distinguish two well-known normal and inverted order schemes of  neutrino data once both the effective masses  $m_{\beta}$  in  tritium beta decays  and $\langle m\rangle$ in the neutrinoless double beta decay  are observed. The lepton flavor violating decays of the charged leptons $e_b\rightarrow e_a\gamma$, $\mu\rightarrow3e$,  the Standard model-like Higgs boson decays $h\rightarrow e_be_a$, and the $\mu$-e conversions in some nuclei are generated from loop corrections.  The experimental data of  the branching ratio Br$(\mu\rightarrow e\gamma, 3e)$  predict that the upper bounds of  Br$(\tau \rightarrow \mu\gamma,e\gamma)$ and Br$(h\rightarrow e_{a}e_b)$ are much smaller than  the planned experimental sensitivities. In contrast, the $\mu$-e conversions are the promising signals for experiments.
\end{abstract}


\maketitle
\section{Introduction}
\allowdisplaybreaks
Observation of neutrino oscillation
 requires that   models beyond
 the Standard Model (BSM) must be considered  to explain both properties of the tiny active neutrino masses and the structure of the lepton mixing matrix $U_{\mathrm{PMNS}}$ named as  the Pontecorvo-Maki-Nakagawa-Sakata (PMNS) matrix.  The  simple tri-bi maximal (TB) form of $U_{\mathrm{PMNS}}$  was introduced in Refs.~\cite{Harrison:2002er, Harrison:2002kp, Harrison:2002et, Harrison:2003aw}, which can be explained theoretically as the consequence of discrete symmetries such as $A_4$ group~\cite{Ma:2001dn,Babu:2002dz,Altarelli:2005yp, Altarelli:2005yx}. The TB form implies exact zero value of a mixing  angle $\theta_{13}$ defined in the standard form of  $U_{\mathrm{PMNS}}$~\cite{Zyla:2020zbs} which is  inconsistent with non-zero but small $\theta_{13}$ pointed out by experiments so that  this form must be modified, see a recent review in Ref.~\cite{Petcov:2017ggy}. Various  modifications were carried out in order to looking for models as simple as possible~\cite{ Altarelli:2012ss, Ma:2012xp, Ahn:2013mva, Chen:2012st, Karmakar:2015jza,Morisi:2013qna, Karmakar:2014dva, Barry:2010zk, Karmakar:2016cvb, Nguyen:2017ibh, Aoki:2020eqf, Ding:2020vud, Korrapati:2020rao, delaVega:2018cnx, Heinrich:2018nip, Kang:2018txu, Kobayashi:2019mna,   Petcov:2017ggy,  Mukherjee:2015ax, Adhikary:2008au, Pramanick_2016, Mishra:2019oqq, Hernandez:2015tna}. Many of  them are  $A_4$  models generating active neutrino masses based on the well-known  standard seesaw (SS) mechanism~\cite{Mukherjee:2015ax, Pramanick_2016, Karmakar:2015jza, Aoki:2020eqf}, some of them based on the   inverse seesaw (ISS)~\cite{Karmakar:2016cvb}.

 In the $A_4$ models containing heavy neutrinos  to generate active neutrinos through the SS or ISS mechanisms,  the lepton flavor violating (LFV)  couplings  of neutrinos  will give  loop corrections to many  LFV  processes such as the  decays of charged leptons (cLFV) $e_{b}\rightarrow e_a\gamma$, $\mu\rightarrow 3e$,  decays of the Standard Model-like (SM-like) Higgs boson (LFVHD) $h\rightarrow e^+_ae_b^-$, and the $\mu$-e conversions in nuclei. For  the standard SS models, these corrections to the LFVHD are very suppressed~\cite{Arganda:2004bz}, therefore only the  branching ratio (Br) of the decays $\mu\rightarrow e\gamma$ may  reach   experimental sensitivities. In contrast, using  the so-called Casas-Ibarra parametrization~\cite{Casas:2001sr} and the simple diagonal form of  heavy neutrino mass matrix  to determine the mixing parameters and  neutrino masses,   many ISS models  predict large LFV corrections from heavy neutrinos to LFVHD~\cite{Pilaftsis:1992st,Arganda:2014dta,Arganda:2017vdb, Thao:2017qtn}.  Namely,  Br$(h\rightarrow\tau\mu,\tau e)$ can reach the order of $\mathcal{O}(10^{-5})$ under the very small  experimental constraint Br$(\mu\rightarrow e\gamma)< \mathcal{O}(10^{-13})$.  Notice that the recent upper bounds of LFVHD at 95\% confidence level are   $\mathrm{Br}(h\rightarrow \mu e)<6.1\times 10^{-5} (\mathrm{ATLAS}),\;  3.5\times 10^{-4} (\mathrm{CMS})$, and $\mathrm{Br}(h\rightarrow \tau e,\tau \mu)<\mathcal{O}(10^{-3})  (\mathrm{ATLAS},\; \mathrm{CMS} )$~\cite{Khachatryan:2016rke,Sirunyan:2017xzt,Aad:2019ugc, CMS:2021rsq}.  The future sensitivities at $e^+e^-$ colliders \cite{Qin:2017aju} and ATLAS at LHC~\cite{Heinemann:2019trx, Davidek:2020gbw} for the Br$(h\rightarrow \mu e)$ and Br$(h\rightarrow \mu \tau, e\tau)$ are hoped to be order of $\mathcal{O}(10^{-5})$ and $\mathcal{O}(10^{-4})$, respectively. In general, these sensitivities  are still larger than the upper bounds predicted by the models containing only  loop contributions to the LFVHD.

 The most strict constraints from experiments for cLFV decays  are  Br$(\mu\rightarrow e\gamma)<4.2\times 10^{-13}$ and Br$(\mu^+\rightarrow e^+e^+e^-)< 10^{-12}$~\cite{Bellgardt:1987du},  with the respective planned  sensitivities will be $6\times 10^{-14}$ and  $\mathcal{O}(10^{-16})$~\cite{Blondel:2013ia}. These two constraints  often result in  much smaller values of Br$(\tau \rightarrow \mu\gamma,\;e\gamma)$  than the recent experimental upper bounds, or even the planned sensitivities of $\mathcal{O}(10^{-9})$. These cLFV decays and the $\mu$-e conversions in nuclei were also investigated in the standard SS models~\cite{Ilakovac:1994kj, Alonso:2012ji} and ISS model~\cite{Haba:2016lxc}, including the minimal supersymmetric (MSSM) versions~\cite{Ilakovac:2012sh, Abada:2014kba}. Depending on the specific structures of the Higgs and lepton sectors of different discrete symmetric models, the allowed regions satisfying all cLFV experimental bounds will predict different possibility to observe the LFVHD in the future experiments.

 Based on the SS models containing  two  $SU(2)_L$ Higgs doublets transforming as $A_4$ singlets~\cite{Adhikary:2008au, Nguyen:2017ibh}, in this work  we will  introduce a non-supersymmetric $A_4\times Z_3\times Z_{11}$ model with the ISS mechanism ($A_4$ISS) to generate active neutrino masses and $U_{\mathrm{PMNS}}$ enough to explain the recent oscillation data and allow low mass scale of heavy neutrinos. In addition, the LFV signals will be discussed as other promoting channels to constrain the parameter space. Our work also pay attention to the LFVHD, which is often ignored in discrete symmetric  models because  the SM-like Higgs boson is difficult to realize in complicated Higgs potentials of both SUSY and non-SUSY versions. To keep the Yukawa term unchanged in the original SUSY versions, the discrete symmetries $Z_3$ and  $Z_{11}$ are introduced to exclude  unwanted terms appearing in the non-SUSY version.  The  model also consists of additional  $SU(2)_L$ neutral Higgs singlets (flavon) enough to generate the active neutrino mass matrix corresponding to $U_{\mathrm{PMNS}}$ close to the TB form in a special degenerate limit of the two independent parameters in the heavy neutrino mass matrix.  Using a deviate parameter $\tilde{\epsilon}$  to relaxing this limit will result in the  real form of  $U_{\mathrm{PMNS}}$.  The observable  parameters defined by the standard form of  $U_{\mathrm{PMNS}}$ will be determined based on the recent work~\cite{Petcov:2017ggy}.  Using this in the numerical investigation, we collected allowed regions of the parameter space to study low energy observable quantities such  as effective neutrino masses related with the neutrinoless  double beta  ($0\nu\beta\beta$) and Tritium beta decay, the cLFV decays $e_b\rightarrow e_a\gamma$, $\mu\rightarrow 3e$, and $h\rightarrow e_ae_b$.  We note that the decay $h\to e_ae_b$ were rarely discussed in previously in discrete symmetric models, including the SUSY versions. The numerical results of the allowed regions  will be  compared with  previous works as well as the current and future experimental constraints.  In contrast with original SS  models~\cite{Adhikary:2008au, Nguyen:2017ibh}, in this work all allowed values of the  Dirac phase $\delta$ are considered and the  parameter $\tilde{\epsilon}$ is solved exactly in this $A_4$ISS model. Hence the allowed regions of parameters will be determined more exhausted,  leading  different  predictions for the low energy observable quantities. In addition,  different from the  ISS models with the  Casas-Ibarra parameterisation~\cite{Casas:2001sr},  the masses and the particular form of the total lepton mixing matrix of all neutrinos  originated from the $A_4\times Z_3\times Z_{11}$ symmetric breaking  will lead to new predictions for the  LFV signals.

The  $A_4$ISS  contains two $SU(2)_L$ Higgs doublets and only flavons,  therefore it  inherits many properties of the well-known two Higgs doublet models (2HDM) type I and II, see  a review in Ref.~\cite{Branco:2011iw}. Based on many discussions  on  the 2HDMs, we will constrain many important parameters affecting strongly on the signals of LFV processes predicted by the $A_4$ISS model.  Namely, the most important parameters  are the ratio of the two vacuum expectation values (vev) of the two neutral components in the Higgs doublets $t_{\beta}$, the charged Higgs mass,  and the parameter $s_{\delta}$ defining the deviation of the SM-like Higgs boson couplings between the  $A_4$ISS and the  SM.  Based on the experimental results from LHC searches and precision electroweak test,  the recent constraints on the parameter spaces of the 2HDM related with  the $A_4$ISS model were discussed in detailed in Refs.~\cite{Chen:2018shg, Chen:2019pkq,Kling:2020hmi}. In the future project of energy collision of 100 TeV, promising signals of  heavy higgs bosons with masses at $\mathcal{O}(10)$ TeV  were mentioned~\cite{Kling:2018xud}.  In direct signal of heavy Higgs bosons predicted by the 2HDM may also appear in the future $e^+e^-$ colliders \cite{Azevedo:2018llq}, where large allowed $t_{\beta}$ corresponds  to the alignment limit $s_{\delta}\rightarrow 0$.    The predictions on cLFV signals  may    depend strongly on $t_{\beta}$,  leading to another channel to determine which $SU(2)_L$ Higgs doublet generates quark masses, i.e. the information to distinguish the type I and II of the 2HDM.

This work is organized as follows. In section \ref{A4ISS}, we introduce the $A_4$ISS mechanism, constructing the analytic formulas of  active neutrino masses and mixing parameters  as functions of free  parameters. We also give out the allowed regions of parameter space satisfying the recent neutrino oscillation data. The predictions of the effective neutrino masses corresponding to  the two decays, neutrinoless double beta  and Tritium beta are also discussed.  In section \ref{sec_Higgs},  important properties of the SM-like Higgs and charged Higgs bosons are summarized.  Analytic formulas and numerical results  relating to the LFV processes are  separated into two sections \ref{sec_LFV} and \ref{sec_Numerical}.  Finally, the summary of our  new results  is given in section \ref{sec_con}. There are four appendices presenting more details on the product rules of the $A_4$ symmetry, the full Higgs potential, the one-loop formulas contributing to the LFV decay amplitudes.
\section{\label{A4ISS}The $A_4$ISS model}
\subsection{The particle content and lepton masses }
The non-Abelian $A_4$ is a group of even permutations of 4 objects and has $4!/2=12$ elements.  The model has three one-dimension ($\underline{1}$, $\underline{1}'$, $\underline{1}''$) and one three-dimensional ($\underline{3}$) irreducible representations.  The important properties of this group and its representations  needed for model construction were reviewed in the appendix \ref{A4rules}. The transformations for leptons and scalars under the total symmetry $SU(2)_L\times U(1)_Y\times A_4\times Z_3\times Z_{11} \times U(1)_L$ as well as their VEVs of the $A_4$ISS model is shown in Table \ref{particle content}.
\begin{table}[ht]
\caption{\label{particle content} List of fermion and scalar fields, where $\psi^a=(\nu_{aL},\;e_{aL})^T$ ($a=1,2,3$) and $\omega= e^{2i\pi/3}$, and $\omega_{11}= e^{2i\pi/11}$.}
\begin{tabular}{lccccccc}\hline\hline
   Lepton & $SU(2)_L$   &   $U(1)_Y$ &  $A_4$  & $Z_3$ & $Z_{11}$ & $L$\\ \hline
   $\overline{\psi^l}=(\overline{\psi^1}, \overline{\psi^2}, \overline{\psi^3})$ &$ 2^*$ & 1&$\underline{3^*}$  & 1 &  $\omega_{11}^{-5}$& -1\\
   $e_R$ &    1    &  -2  &\underline{1 }  &  1 &  $\omega_{11}^{-2}$ &1\\
   $\mu_R$ &    1    & -2&    $\underline{1}'$   &1  &  $\omega_{11}^{-2}$ & 1\\
   $\tau_R$ &    1    & -2&    $\underline{1}''$   & 1 &  $\omega_{11}^{-2}$ & 1\\
   $N_{R}$ &    1    & 0&  $\underline{3}$    &   $\omega$ & $\omega_{11}^{3}$   & 1\\
   $X_{R}$ &    1    & 0&   $\underline{3}$    &   $\omega$  &  $\omega_{11}$ & -1\\
  \hline
   Scalar   &         &   &   &  &  &  &VEV \\
   \hline
   $h_u$    &    2    &-1 &   \underline{ 1}          &$\om^2$    &  $\omega_{11}^{2}$ & 0 & $\langle h_u\rangle=v_u$                           \\
   $h_d$     &    2    & 1 &    \underline{1}          &    1         &  1  &  0& $\langle h_d\rangle=v_d$                           \\
   $\phi_S$  &    1    & 0 &    \underline{3}          &  $\om$ &  $\omega_{11}^{-4}$ & 0 & $\langle\phi_S\rangle=(v_S,v_S,v_S)$ \\
   $\phi_T$  &    1    & 0 &    \underline{3}          & 1            &   $\omega_{11}^{-4}$  & 0 & $\langle\phi_T\rangle=(v_T,0,0)$                   \\
   $\xi'$     &    1   &0 &    $\underline{1}'$  & $\om$  & $\omega_{11}^{-4}$  &  0& $\langle\xi'\rangle=u'$                             \\
   $\xi''$    &    1   &0 &    $\underline{1}''$ & $\om$  & $\omega_{11}^{-4}$   &  0 &$\langle\xi^{''}\rangle=u''$
 \\
 \hline \hline
\end{tabular}
\end{table}
Here $L$ is the normal lepton number, two Abelian discrete symmetries are added in order to get the minimal Lagrangian generating lepton masses and mixing parameters consistent with experiments.   The two $SU(2)_L$ Higgs doublets are expanded around their VEVs as
\begin{equation}
h_u=\left(\begin{array}{c}
v_u+\frac{S_u+iA_u}{\sqrt{2}}\\
H^-_u \\
\end{array}
\right), \hs   h_d=\left(
\begin{array}{c}
H^+_d \\
v_d+\frac{S_d+iA_d}{\sqrt{2}}\\
\end{array} \right),
\label{eq_Higgsexpand}
\end{equation}
where the electric charge operator is well-known as  $Q= T^3+Y/2$.

Considering  the effective operators up to five dimension (dim.) needed to generate masses of lepton,  the Yukawa Lagrangian respecting the total symmetry consists of  two parts. In particular, the first part is renormalizabe  as follows 
\begin{equation} 
\label{lagrangian}
-{\cal L}^r = f \bar{\psi}_L^lN_Rh_u  +x_A'\xi'(\bar{N}_L^cX_R)''+x_A''\xi''(\bar{N}_L^cX_R)'+x_B(\phi_S\bar{N}_L^cX_R) +\mathrm{H.c.}.
\end{equation}
While the second part consisting of all effective operators of  five dim., including all terms breaking the lepton number $L$ relevant with neutrino masses,  is 
\begin{align}
\label{eq_L5dim}
-{\cal L}^{5d}= \left[y_e (\phi_T\bar{\psi}_L^l)e_R + y_\mu (\phi_T\bar{\psi}_L^l)''\mu_R + y_\tau(\phi_T\bar{\psi}_L^l)'\tau_R \right] \frac{h_d}{\La}
+  \frac{\la_X(\tilde{h}_u h^*_d)}{ 2\Lambda} (\bar{X}_L^cX_R) +\mathrm{H.c.}.	
\end{align}
Here $\Lambda$ is the cut-off scale of the model under consideration, $\tilde{h}_u=i\sigma_2 h^*_u$, the  charge conjugation of the neutral leptons $N_R$ and $X_R$ are  $N_L^c\equiv P_L N^c=(N_R)^c=C (N_R)^T$ and $X_L^c=(X_R)^c$, respectively.  After spontaneous symmetry breaking, the charged lepton mass matrix comes out diagonal with
$m_e=\frac{y_e v_T v_d}{\La}$, $m_\mu=\frac{y_\mu v_T v_d}{\La}$, and $m_\tau=\frac{y_\tau v_T v_d}{\La}$. Correspondingly, the Higgs doublet $h_d$ plays a similar role to the SM Higgs doublet in generating charged lepton masses.  Only the last term in Eq. \eqref{eq_L5dim} breaks the lepton number $L$ with two units,  giving  neutrino mass term $\mu'_X\equiv \left(\lambda_X v_uv_d/\Lambda\right)$, which can be small so that the ISS mechanism can work. The non-renormalizable terms generating charged lepton masses  can be seen as the tree level mass originated the exchange of the heavy vector-like leptons transforming as $E_{L,R}=(E_1,E_2,E_3)^T_{L,R}\sim (1,-2)_{(\underline{3}, 1,1,1)}$ with the total symmetry $(SU(2)_L,\;U(1)_Y)_{(A_4,Z_3,Z_{11},L)}$. 

\begin{figure}[ht]
		\centering
		\includegraphics[width=7cm]{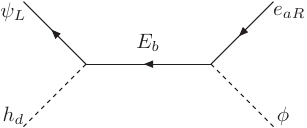}
		\caption{Tree level diagrams generating effective terms  given in Lagrangian \eqref{eq_L5dim}.} \label{fig_effterms}
	\end{figure}

The renormalizable Lagrangian relating with $E_{L,R}$ that respect the total symmetry is
	\begin{align}
		\label{eq_ELR}	
		-\mathcal{L}^Y_{E}= &y^E_{\psi} \left( \overline{\psi^l_L}E_R\right)h_d +  y^E_{e} \left(\overline{E_L}\phi_T\right)e_{R} + y^E_{\mu} \left(\overline{E_L}\phi_T\right)''\mu_{R} + y^E_{\tau} \left(\overline{E_L}\phi_T\right)'\tau _{R}
		\crn&+\Lambda \overline{E_L}E_R +h.c.,
	\end{align}
	where we assume that $\Lambda\gg v_S,v_T,u,u''$ so that all of the above Higgs bosons do not contribute significantly to very heavy vector-like lepton masses $E_{1,2,3}$. Therefore, $m_{Ea}\equiv \Lambda$ for all $a=1,2,3$. The non-renormalizable terms are reduced form the diagrams given in Fig.\ref{fig_effterms}.
	For example in the regions where the vector-like lepton  momenta are  much smaller  than their masses  we the effective term  $ i\frac{y_e}{\La}(\phi_T\bar{\psi^l}_L)e_Rh_d\sim  \left(i y^E_{\psi}\right)\overline{\psi^l_L}h_d\times \frac{i}{\Lambda}\times  \left(iy^E_{e}\right)\phi_Te_{R} $.

 The Lagrangian  for   neutrino mass is
\begin{equation}\label{Lnumass}
-\mathcal{L}^{\nu}=  \frac{1}{2} \begin{pmatrix}
\overline{\nu_L}&  \overline{N^c_L}& \overline{X^c_L}
\end{pmatrix} \mathcal{M}^{\nu\dagger} \begin{pmatrix}
(\nu_L)^c\\
N_R\\
X_R
\end{pmatrix} +\mathrm{h.c.}, \quad \mathcal{M}^{\nu}\equiv  \begin{pmatrix}
0& m^T_D & 0 \\
m_D&0  & M^T_R  \\
0& M_R &\mu_X
\end{pmatrix},
\end{equation}
 where $m_D$, $M_R$, and $\mu_X$ are $3\times3$ Dirac and Majorana neutrino mass matrices have the  following forms
\begin{align}
m_D&= v_u f \times I_3 \equiv v_u Y_{\nu},\label{eq_mD}\\
M_R &= M_0\left( \begin{array}{*{20}{c}}
1&\tilde{\kappa}  - \frac{1}{2}&\tilde{\rho} - \frac{1}{2}\\
\tilde{\kappa} - \frac{1}{2}&\tilde{\rho} + 1&- \frac{1}{2}\\
\tilde{\rho} - \frac{1}{2}&- \frac{1}{2}&\tilde{\kappa} + 1
\end{array} \right) \equiv M_0 M^r_R, \quad
\mu_X  =  \mu'^*_X {\left(\begin{array}{ccc}
	1  & 0  &  0\\
	0  &   0  &  1 \\
	0  &   1  & 0 \end{array}\right)},\label{Majoranamass1}
\end{align}
and
\begin{align}
\label{eq_newnotation}
M_0&\equiv \frac{2x^*_Bv_S}{3}, \quad\, \tilde{\kappa}\equiv \frac{x''^*_Au''}{M_0}=\kappa\,e^{i\phi_1} , \quad\, \tilde{\rho}=\frac{x'^*_Au'}{M_0}\equiv \tilde{\kappa}\left(1 +\tilde{\epsilon}\right) .
\end{align}
The tilde notation in  $\tilde{x}$ implies a complex parameter, and $x\equiv|\tilde{x}|>0$.
Without loss of generality, the  real and positive $M_0$  is assumed in Eq.~\eqref{eq_newnotation} and  $0\leq\phi_1<2\pi$. The  effective neutrino mass matrix $m_{\nu}$ is then obtained by the ISS relations~\cite{GonzalezGarcia:1988rw}, which is a specific of the general SS framework~\cite{Minkowski:1977sc, Mohapatra:1979ia}:
\begin{align}
m_{\nu}&=  m^T_D M_R^{-1}\mu_X\left(M^T_R\right)^{-1}m_D,
\label{R1}
\end{align}
where  \begin{equation}\label{eq_mnu}
m_{\nu}\equiv U^*_{3\nu}\hat{m}_{\nu}U^{\dagger}_{3\nu},\quad  \hat{m}_{\nu}\equiv \mathrm{diag}\left(m_{n_1},m_{n_2},m_{n_3}\right)
\end{equation}
relating with the mixing matrix $U_{3\nu}$ and the masses $m_{n_a}$ ($a=1,2,3$) of the three active neutrinos, leading to the following form of the $U_{\mathrm{PMNS}}$,
\begin{equation}\label{eq_upmns}
U_{\mathrm{PMNS}}=U^{\dagger}_{L,e} U_{3\nu},
\end{equation}
where $U_{L,e}$  is defined by the relation  $U^{\dagger}_{L,e}m_DU_{R,e}=\mathrm{diag}(m_e,m_\mu,m_{\tau})$. The standard form of the $U_{\mathrm{PMNS}}$ is the unitary matrix defined as follows~\cite{Zyla:2020zbs}
\begin{align}
U^{\mathrm{PDG}}_{\mathrm{PMNS}}
&= \begin{pmatrix}
1	& 0 &0  \\
0	&c_{23}  &s_{23}  \\
0&  	-s_{23}& c_{23}
\end{pmatrix}\,\begin{pmatrix}
c_{13}	& 0 &s_{13}e^{-i\delta}  \\
0	&1  &0  \\
-s_{13}e^{i\delta}&  0& c_{13}
\end{pmatrix}\,\begin{pmatrix}
c_{12}	& s_{12} &0  \\
-s_{12}	&c_{12}  &0  \\
0& 0 	&1
\end{pmatrix} \mathrm{diag}\left(1, e^{i\frac{\alpha_{21}}{2}},\,e^{i\frac{\alpha_{31}}{2}}\right) \label{eq_UnuPDG}
\crn&=U^0_{\mathrm{PMNS}} \;\mathrm{diag}\left(1, e^{i\frac{\alpha_{21}}{2}},\,e^{i\frac{\alpha_{31}}{2}}\right),
%
\end{align}
where $s_{ij}\equiv\sin\theta_{ij}$, $c_{ij}\equiv\cos\theta_{ij}$, $i,j=1,2,3$ ($i<j$), $0<\theta_{ij}<180\; [\mathrm{Deg.}]$ and $0<\delta\le 720\;[\mathrm{Deg.}]$.   In this work, $U_{L,e}=U_{R,e}=I_3$, hence
\begin{equation}\label{eq_UPMNS2}
U_{\mathrm{PMNS}}= U_{3\nu}= \mathrm{diag}\left(e^{i\psi_1}, e^{i\psi_2},\,e^{i\psi_3}\right) U^{\mathrm{PDG}}_{\mathrm{PMNS}},
\end{equation}
where the phases $\psi_{1,2,3}$ are absorbed into the charged lepton states. The standard form  $U^{\mathrm{PDG}}_{\mathrm{PMNS}}$ provides the experimental quantities $s_{ij}$ and $\delta$, $\alpha_{21}$ and $\alpha_{31}$.

Let us remind that the TB  framework of the  lepton mixing matrix was predicted in an $A_4$ model with the standard SS mechanism. It also happens in this model in the degenerate condition that $\tilde{\rho}=\tilde{\kappa}$, corresponding to   $\tilde{\epsilon}=0$.  As a result, the neutrino mixing matrix $U_{3\nu}$ given by Eq.~\eqref{R1}  have the TB form:
\begin{align}
U_{\mathrm{TB}}=\left(
\begin{array}{ccc}
\sqrt{\frac{2}{3}} & \frac{1}{\sqrt{3}} & 0 \\
-\frac{1}{\sqrt{6}} & \frac{1}{\sqrt{3}} & -\frac{1}{\sqrt{2}} \\
-\frac{1}{\sqrt{6}} & \frac{1}{\sqrt{3}} & \frac{1}{\sqrt{2}} \\
\end{array}
\right) , \label{eq_UTB}
\end{align}
 corresponding to $s_{13}=0$. This is  in contrast with the
 experimental data of neutrino, which is divided into two cases of normal (NO) and inverted (IO) schemes. The best-fit and  $3\sigma$ values for the NO  are~\cite{Zyla:2020zbs}
\begin{align}
	\label{eq_d2mijNO}
	&s^2_{12}=0.31,\;  0.275\le s^2_{12}\le 0.350;
	\crn&s^2_{23}= 0.558,\;  0.427\le s^2_{23}\le 0.609;
	\crn&s^2_{13}= 0.02241 ,\;  0.02046\le s^2_{13}\le 0.02440;
	\crn&\delta= 222 \;[\mathrm{Deg}] ,\;141\;[\mathrm{Deg}] \le \delta\le 370\;[\mathrm{Deg}] ;
	\crn &\Delta m^2_{21}=7.39\times 10^{-5} [\mathrm{eV}^2], \quad   6.79\times 10^{-5} [\mathrm{eV}^2]\leq\Delta m^2_{21}\leq8.01\times 10^{-5} [\mathrm{eV}^2] ;
	 \crn& \Delta m^2_{32}=2.449\times 10^{-3} [\mathrm{eV}^2], \quad 2.358\times 10^{-3} [\mathrm{eV}^2]\leq\Delta m^2_{32}\leq2.544\times 10^{-3} [\mathrm{eV}^2] .
\end{align}
For the IO case, the values of $s_{12}$ and $\Delta m^2_{21}$ are the same and
\begin{align}
	\label{eq_d2mijIO}
&s^2_{23}= 0.563,\;  0.430\le s^2_{23}\le 0.612;
\crn&s^2_{13}= 0.02261 ,\;  0.02066\le s^2_{13}\le 0.02461;
\crn&\delta= 285 \;[\mathrm{Deg}] ,\;205\;[\mathrm{Deg}] \le \delta\le 354\;[\mathrm{Deg}] ;
\crn& \Delta m^2_{32}=-2.509\times 10^{-3} [\mathrm{eV}^2], \quad -2.603\times 10^{-3} [\mathrm{eV}^2]\leq\Delta m^2_{32}\leq-2.416\times 10^{-3} [\mathrm{eV}^2] .
\end{align}
We consider  the real case, where all $\theta_{ij}$ is in the  $3\sigma$ ranges of the experimental data.  We assume a solution that the  deviation from the TB data arises from
only the condition that  $\tilde{\rho} \ne  \tilde{\kappa}$ in the matrix $M_R$ given in Eq.~\eqref{Majoranamass1}, equivalently $\tilde{\epsilon}\ne 0$.
Then, the mixing matrix $U_{3\nu}$ in Eq.~\eqref{eq_mnu}  is calculated by writing it as
\begin{equation}\label{eq_U3nu1}
U_{3\nu}\equiv\,U_{\mathrm{TB}}U_1U_P,\quad U_P=\mathrm{diag}(e^{-\frac{i\varphi_1}{2}},\; e^{-\frac{i\varphi_2}{2}},\;  e^{-\frac{i\varphi_3}{2}})
\end{equation}
which can be identified with the well-known form given in  Eq.~\eqref{eq_UnuPDG}.
We then have,
\begin{align}
\hat{m}_{\nu}&=U^T_{3\nu} m_{\nu} U_{3\nu}=U^T_{1} m'_{\nu} U_{1}=\mathrm{diag}(m_{n_1},m_{n_2},m_{n_3}), \label{eq_mnu1}\\
&(U_{\mathrm{TB}}U_{1})^T m_{\nu} (U_{\mathrm{TB}}U_{1})=\mathrm{diag}(m_{n_1}e^{i\varphi_1},\; m_{n_2}e^{i\varphi_2},\; m_{n_3}e^{i\varphi_3}),\crn
m'_{\nu}&\equiv U^T_{\mathrm{TB}} m_{\nu} U_{\mathrm{TB}} ,\crn
m'_{\nu}&= m_0e^{-2i\phi_1} \left(
\begin{array}{ccc}
	\frac{16 \left(2 \tilde{\epsilon }^2-2 (3 \tilde{\kappa}^{-1}+2) \tilde{\epsilon }-(3 \tilde{\kappa}^{-1}+2)^2\right)}{\kappa^2 \left(4 \tilde{\epsilon }^2+4 \tilde{\epsilon
		}-9 \tilde{\kappa}^{-2}+4\right)^2} & 0 & \frac{32 \sqrt{3} \tilde{\epsilon }  (\tilde{\epsilon }+2)}{\kappa^2 \left(4 \tilde{\epsilon }^2+4 \tilde{\epsilon }-9
		\tilde{\kappa}^{-2}+4\right)^2} \\
	0 & -\frac{4}{ \kappa^2 (\tilde{\epsilon }+2)^2} & 0 \\
	\frac{32 \sqrt{3} \tilde{\epsilon } (\tilde{\epsilon }+2)}{ \kappa^2 \left(4 \tilde{\epsilon}^2+4 \tilde{\epsilon }-9 \tilde{\kappa}^{-2}+4\right)^2} & 0 & -\frac{16
		\left(2 \tilde{\epsilon }^2+(6 \tilde{\kappa}^{-1}-4) \tilde{\epsilon }-(2-3 \tilde{\kappa}^{-1})^2\right)}{ \kappa^2 \left(4 \tilde{\epsilon }^2+4 \tilde{\epsilon }-9
		\tilde{\kappa}^{-2}+4\right)^2} \\
\end{array}
\right), \label{eq_mnup1}
\end{align}
where $U_{\mathrm{TB}}$ is given in Eq.~\eqref{eq_UTB} and
\begin{align}
m_0&=\frac{f^2 \mu _X v_u^2}{4  M_0^2}>0. \label{eq_fm0}
\end{align}

\begin{figure}[ht]
	\centering
	\minipage{0.485\textwidth}
	\includegraphics[width=7.5cm]{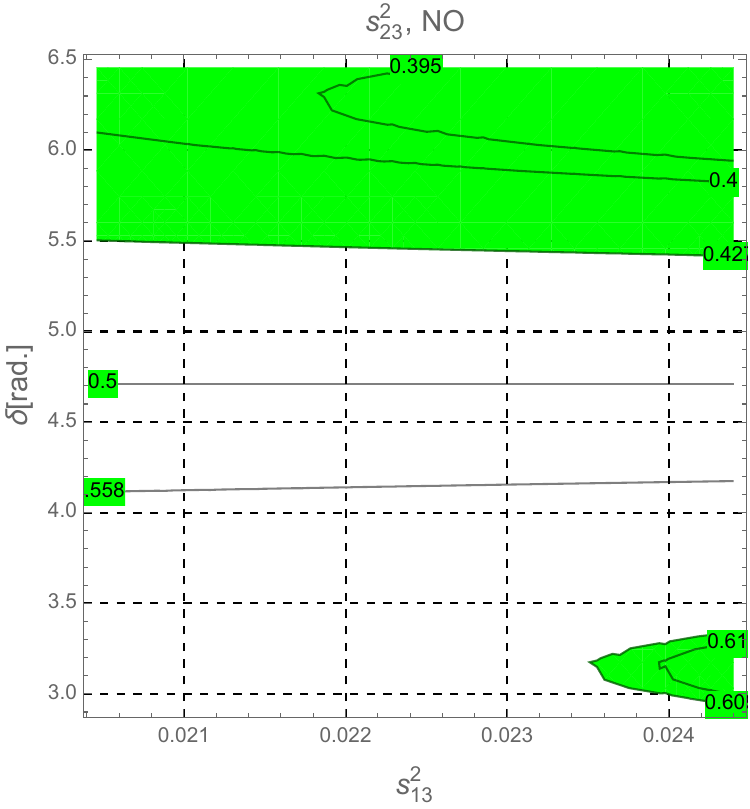}
	\endminipage
	\hfill
	\quad
	\minipage{0.485\textwidth}
	\includegraphics[width=7.5cm]{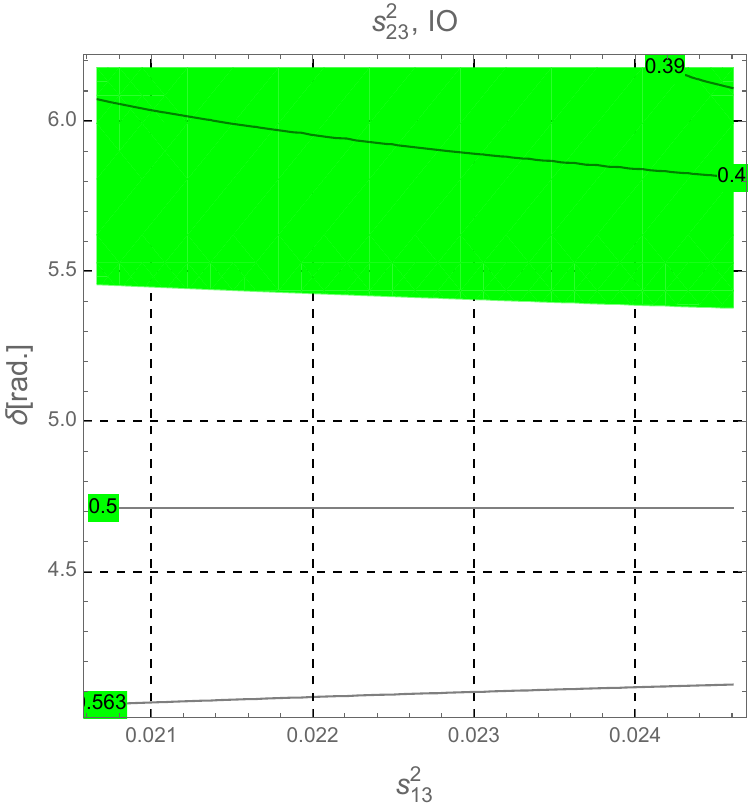}
	\endminipage
	\hfill
	\caption{ Contour plots for $s^2_{23}$ as a function of $s^2_{13}$ and $\delta$ for the NO and IO  schemes. The green regions are excluded by $3\sigma$  data  of  $s^2_{23}$. The black curves show the constant values of $s^2_{23}$. }\label{fig_fs232}
\end{figure}

The form of $m'_{\nu}$ results in the form of $U_1$ as follows:
\begin{align}\label{eq_U1}
U_1=\begin{pmatrix}
c_{\theta}& 0 & s_{\theta} e^{-i\phi_0}\\
0&1  &0  \\
-s_{\theta}e^{i\phi_0}& 0 & c_{\theta}
\end{pmatrix}, \quad U_{\mathrm{TB}}U_1= \left(
\begin{array}{ccc}
\sqrt{\frac{2}{3}} c_{\theta } & \frac{1}{\sqrt{3}} & \sqrt{\frac{2}{3}} s_{\theta } e^{-i \phi _0} \\
\frac{s_{\theta } e^{i \phi _0}}{\sqrt{2}}-\frac{c_{\theta }}{\sqrt{6}} & \frac{1}{\sqrt{3}} & -\frac{c_{\theta }}{\sqrt{2}}-\frac{s_{\theta } e^{-i \phi _0}}{\sqrt{6}} \\
-\frac{c_{\theta }}{\sqrt{6}}-\frac{s_{\theta } e^{i \phi _0}}{\sqrt{2}} & \frac{1}{\sqrt{3}} & \frac{c_{\theta }}{\sqrt{2}}-\frac{s_{\theta } e^{-i \phi _0}}{\sqrt{6}} \\
\end{array}
\right).
\end{align}
where $c_{\theta}\equiv \cos\theta,\; s_{\theta}\equiv \sin\theta$,  and $\phi_0$ are real.  The matrix  $U_1$ is found by diagonalizing the matrix $m'^{\dagger}_{\nu}m'_{\nu}$, leading to the total
the neutrino mixing matrix given in Eq.~\eqref{eq_U1}. It can be identified with the standard form given in Eq.~\eqref{eq_UnuPDG} by the following relation~\cite{Kitabayashi:2015jdj, Petcov:2017ggy}
\begin{align}
\label{eq_thetaPhi0}
s_{\theta}^2&=\frac{3 s_{13}^2}{2},\;
s_{12}^2=\frac{1}{3 c_{13}^2}, \;
%
 \cos\phi_0 =- \frac{\left(1- s_{13}^2\right) \cos2\theta_{23}}{s_{13} \sqrt{2-3 s_{13}^2}},\;
 \cos\delta=\frac{\cos2\theta_{13} \cos2\theta_{23}}{s_{13} \sqrt{2-3 s_{13}^2} \sin2\theta_{23}}.
\end{align}
They were found by the requirement that $|\left(U^0_{\mathrm{PMNS}}\right)_{ij}|= |\left(U_{\mathrm{TB}}U_1\right)_{ij}|$ for all $i,j=1,2,3$.
These equations show that apart from $\theta,\phi_0$, other parameters $s_{12}$ and the Dirac phase can be written in terms of $s_{13}$ and $s_{23}$. The formulas of $\cos\phi_0$ and $\cos\theta$  in \eqref{eq_thetaPhi0} result in the following  well-known relations~\cite{Petcov:2017ggy}: $\sin\phi_0= -\sin2\theta_{23}\sin\delta$,
and
\begin{equation}\label{eq_tp0}
\tan\phi_0= (1-\tan^2\theta_{13}) \tan\delta.
\end{equation}
 Therefore, using $\tan^2\phi_0+1=1/\cos^2\phi_0$ and $\cos\phi_0/\cos\delta<0$ to formulate $\cos\phi_0$  as follows
\begin{equation}\label{eq_cphi0}
\cos\phi_0= -\frac{\cos\delta}{\sqrt{1+ \sin^2\delta  \tan^2\theta_{13} \left( \tan^2\theta_{13}-2\right)}}.
\end{equation}
  Based on Eq.~\eqref{eq_thetaPhi0}, $s^2_{23}$ is written in term of  the following function of $\delta$ and $s_{13}$,
\begin{equation}\label{eq_fs232}
s^2_{23}=\frac{1}{2}-\frac{\cos\delta\; s_{13} \sqrt{2-3 s_{13}^2}}{2 \sqrt{\left(3 \sin^2\delta+1\right) s_{13}^4-2 \left(\sin^2\delta+1\right) s_{13}^2+1}}.
\end{equation}
From now on,  $s_{23}$ and $\phi_0$ are investigated as the above  functions of $s_{13}$ and $\delta$. The requirements $|\cos\phi_0|,|\sin\phi_0|\le 1$ always satisfy under the recent $3\sigma$ data allowing only  very small $s^2_{13}$. In contrast,  $3\sigma$ allowed range of  $s^2_{23}$  gives an upper bound on $\delta$,  more strict than that from the $3\sigma$ experimental data as given in Fig.~\ref{fig_fs232}.
The new upper bound of $\delta$ for both NO and IO schemes  is
\begin{equation}\label{eq_maxDelta}
\delta<5.5\times 180/\pi\simeq 315\; [\mathrm{Deg}.].
\end{equation}
Apart from relations in Eq.~\eqref{eq_thetaPhi0}, the  matrix $U_{\mathrm{TB}}U_1$  casts into the well-known standard form:
\begin{equation}\label{eq_gammai}
U_{\mathrm{TB}}U_1= \mathrm{diag}\left(e^{i\psi_1}, e^{i\psi_2},e^{i\psi_3}\right) U^0_{\mathrm{PMNS}}\mathrm{diag}\left(1, e^{i\gamma_2},e^{i\gamma_3}\right),
\end{equation}
where the unphysical phases $\psi_i$ are  absorbed into the charged lepton states. In contrast, the two phases $\gamma_{2,3}$ will contribute to the Majorana phases $\alpha_{ij}$. By identifying $\mathrm{Arg}[U^0_{\mathrm{PMNS}}]_{ij}\equiv \delta_{ij}=\mathrm{Arg}[\mathrm{diag}\left(e^{-i\psi_1}, e^{-i\psi_2},e^{-i\psi_3}\right)U_{\mathrm{TB}}U_1 \mathrm{diag}\left(1, e^{-i\gamma_2},e^{-i\gamma_3}\right)]_{ij}$,  $\psi_i$ and $\gamma_{2,3}$ are found as follows: $\psi_1=0=\gamma_2=\delta -\phi_0-\gamma_3$, $\psi_{2} =-\delta_{21} +\phi_{21}=\delta_{22}=-\gamma_3 +\phi_{23}$, and $\psi_{3} =-\delta_{31} +\phi_{31}=\delta_{32}=-\gamma_3 +\phi_{33}$,
where $\phi_{ij}\equiv\mathrm{Arg}[U_{\mathrm{TB}}U_1]_{ij}$. These equations are obtained by identifying that $e^{ix}=e^{iy} $ is equivalent with $x=y$ without the unnecessary  repeated term $k2\pi$, $k\in \mathtt{Z}$. Hence, the non-zero contribution to the Majorana phase is
\begin{equation}\label{eq_gamma3}
\gamma_3=\delta -\phi_0,
\end{equation}
 consistent with the result mentioned in Ref.~\cite{Petcov:2017ggy}.

From now on,  three parameters  $\theta_{12}$, $\theta_{23}$, and $\phi_0$ will be written as functions of $s_{13}$ and $\delta$  based on the relations given in Eqs.~\eqref{eq_thetaPhi0},  \eqref{eq_tp0}, \eqref{eq_cphi0}, and \eqref{eq_fs232}.
Taking these functions for  diagonalisation $m_{\nu}$ by  requiring that $[U^T_1 m'_{\nu}U_1]_{13}=0$ will lead to $  s_{2\theta} \left[(m'_{\nu})_{33}\, e^{i\phi_0} -(m'_{\nu})_{11}\,e^{-i\phi_0}\right]=2(m'_{\nu})_{13}\,  c_{2\theta},$
 where $(m'_{\nu})_{11}$ , $(m'_{\nu})_{13}$, and $ (m'_{\nu})_{33}$ are given in Eq.~\eqref{eq_mnup1}. The result is
 \begin{align}
 \label{eq_t2t}
 t_{2\theta} &\equiv \tan(2\theta) =\frac{2(m'_{\nu})_{13}}{(m'_{\nu})_{33}\, e^{i\phi_0} -(m'_{\nu})_{11}\,e^{-i\phi_0} },
 %
 \end{align}
 leading to the following exact solution of $\tilde{\epsilon}$,
  \begin{align}
 \label{eq_eps0}
 \tilde{\epsilon}=&\frac{\sqrt{E}-6 i t_{2\theta } \tilde{\kappa}^{-1} \sin (\phi _0)+4 t_{2\theta } \cos (\phi _0)-4 \sqrt{3}}{4 \left(t_{2\theta } \cos (\phi _0)+\sqrt{3}\right)}, \crn
 E=&8 t_{2\theta } \left(t_{2\theta } \cos (\phi _0)+\sqrt{3}\right) \left(\left(9 \tilde{\kappa}^{-2}+4\right) \cos (\phi _0)-12 i \tilde{\kappa}^{-1} \sin (\phi _0)\right)
 \crn &+\left(6 i t_{2\theta } \tilde{\kappa}^{-1} \sin(\phi _0)-4 t_{2\theta } \cos (\phi _0)+4 \sqrt{3}\right)^2.
 \end{align}
Here, we choose $\tilde{\epsilon}$ guarantees that  $\lim_{t_{2\theta}\rightarrow0}\tilde{\epsilon}=0$,  consistent with our assumption that $\theta=0$  results in the TB form of $U_{\mathrm{PMNS}}$. We note that formula of $\tilde{\epsilon}$ is more general than that given in the original SS models~\cite{Adhikary:2008au, Nguyen:2017ibh}, where $\tilde{\epsilon}$ was assumed to be small for the approximate solutions. The Majorana phases defined in Eq.~\eqref{eq_U3nu1} are also formulated based on the phases of $U^T_1 m'_{\nu}U_1$.  Then, the more precise form of  $U_{\mathrm{PMNS}}$ defined in Eq.~\eqref{eq_UnuPDG} is
\begin{align}
U_{\mathrm{PMNS}}&\equiv U^0_{\mathrm{PMNS}}\;\mathrm{diag}\left(1,\;e^{\frac{i\alpha_{21}}{2}}, e^{\frac{i\alpha_{31}}{2}}\right), \label{eq_UPMNS}
\\ \alpha_{21}&= \mathrm{Arg}[ \left(m'_{\nu}\right)_{11} c^2_{\theta} - \left(m'_{\nu}\right)_{13} s_{2\theta}e^{ i \phi _0} + \left(m'_{\nu}\right)_{33} s^2_{\theta}e^{2 i \phi _0}]- \mathrm{Arg}[\left(m'_{\nu}\right)_{22}] \;[\mathrm{Deg.}], \crn
\alpha_{31}&=\mathrm{Arg}[ \left(m'_{\nu}\right)_{11} c^2_{\theta} - \left(m'_{\nu}\right)_{13} s_{2\theta}e^{ i \phi _0} + \left(m'_{\nu}\right)_{33} s^2_{\theta}e^{2 i \phi _0}]
\crn& -  \mathrm{Arg}[\left(m'_{\nu}\right)_{33} c^2_{\theta} + \left(m'_{\nu}\right)_{13} s_{2\theta}e^{ -i \phi _0} + \left(m'_{\nu}\right)_{11} s^2_{\theta}e^{-2 i \phi _0}] + \gamma_3\; [\mathrm{Deg.}], \label{eq_MajoranaPhase}
\end{align}
where $\gamma_3$ is given in Eq.~\eqref{eq_gamma3} contributes to the phase $\alpha_{31}$, apart from those come from the phases of the neutrino masses in  the diagonal matrix $U^T_1 m'_{\nu}U_1$.

At this step,  all of the active neutrino masses and mixing parameters
can be formulated in terms of the five independent parameters  $s_{13}$, $\delta$, $m_0$  $\phi_1$, and $\kappa$. Until now, we do not now the orders of $m_0$ and $\kappa$, hence it is necessary to take some numerical estimation to figure out these orders.

 The definition  $\Delta m^2_{ij}\equiv m^2_{n_i}-m^2_{n_j}$ gives
 \begin{align} \label{eq_d2m21pm0}
 \frac{\Delta m^2_{21}}{m_0^2}&= \frac{16}{\left|\kappa^2 (2 +\tilde{\epsilon})^2\right|^2} -\left| \left(m'_{\nu}\right)_{11} c^2_{\theta} - \left(m'_{\nu}\right)_{13} s_{2\theta}e^{ i \phi _0} + \left(m'_{\nu}\right)_{11} s^2_{\theta}e^{2 i \phi _0}\right|^2,
 \crn \frac{\Delta m^2_{32}}{m_0^2}&= \left| \left(m'_{\nu}\right)_{33} c^2_{\theta} + \left(m'_{\nu}\right)_{13} s_{2\theta}e^{ -i \phi _0} + \left(m'_{\nu}\right)_{11} s^2_{\theta}e^{-2 i \phi _0}\right|- \frac{16}{\left|\kappa^2 (2 +\tilde{\epsilon})^2\right|^2}.
  \end{align}
 As a result, $\Delta m^2_{32}$  is written  as a function of $\Delta m^2_{21}$ and $\tilde{\kappa}=\kappa e^{i\phi_1}$ for both NO  and IO schemes. The  experimental data of $s_{ij}$,  $\Delta m^2_{21}$, $\Delta m^2_{32}$, and $\delta$  will give  constraints on   $\kappa$ and $\phi_1$.
In  the TB limit,  the right-hand sides of Eqs. in~\eqref{eq_d2m21pm0} depend only on  $\kappa$ and $\phi_1$.  The condition $\Delta m^2_{21}>0$ requires  $\kappa<1.6$ being useful for estimating the deviation of $(\tilde{\rho}-\tilde{\kappa})$ to looking for a real mixing angle $s_{13}$.

\begin{figure}[ht]
	\centering
	\begin{tabular}{ccc}
		\includegraphics[width=4.8cm]{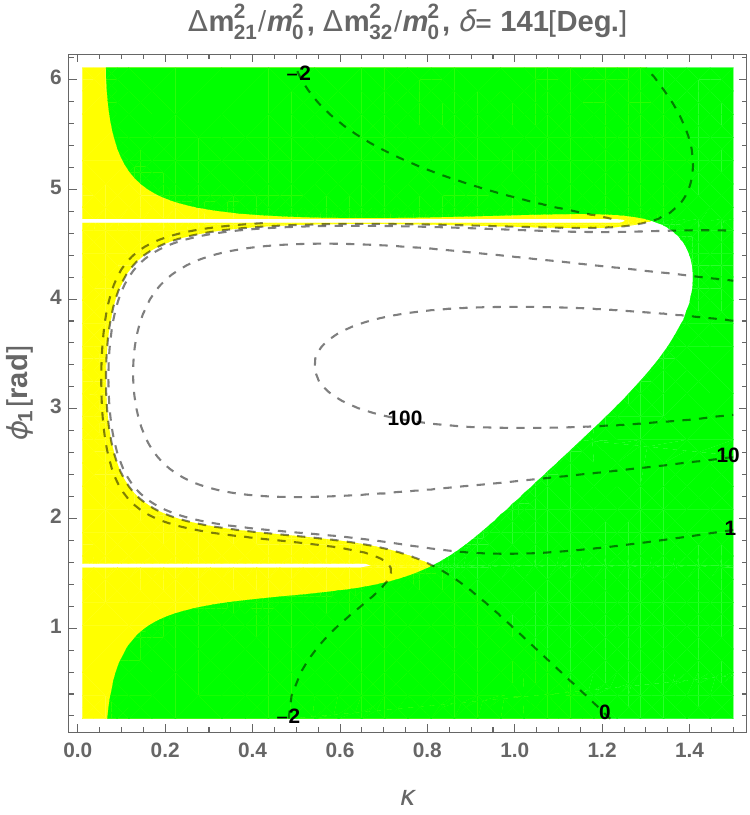}	&	 \includegraphics[width=4.8cm]{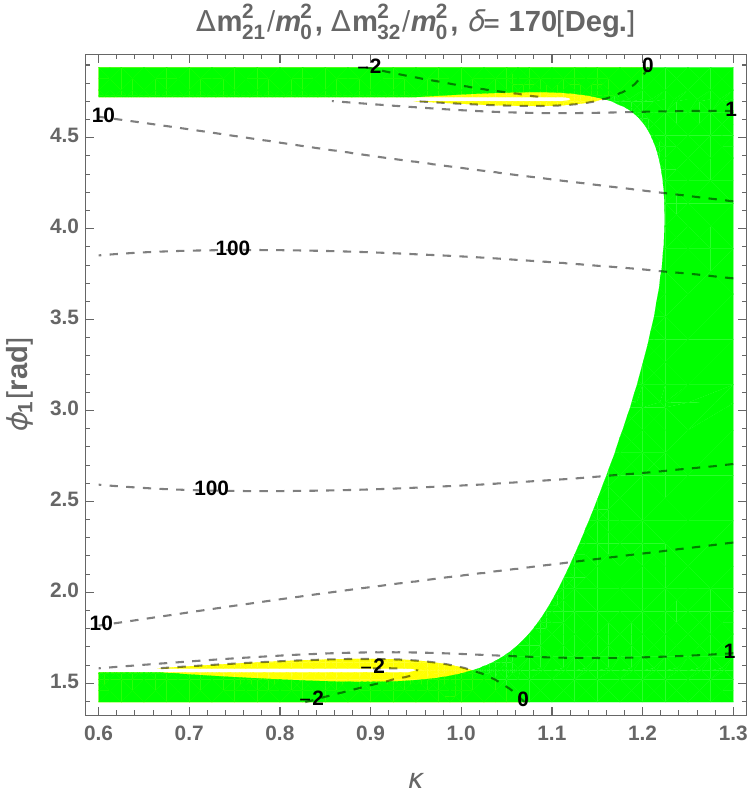}& \includegraphics[width=4.8cm]{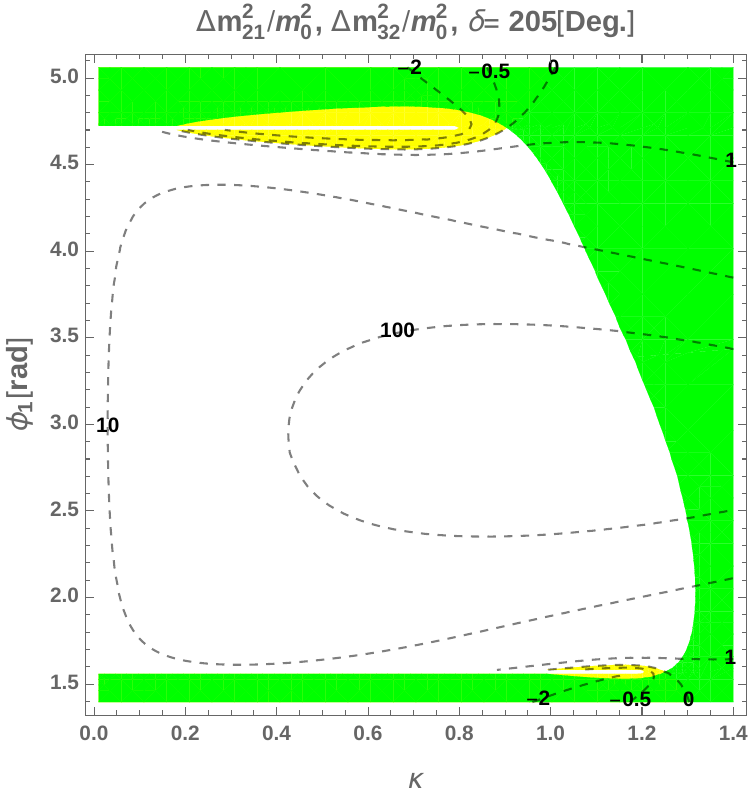}\\
		\includegraphics[width=4.8cm]{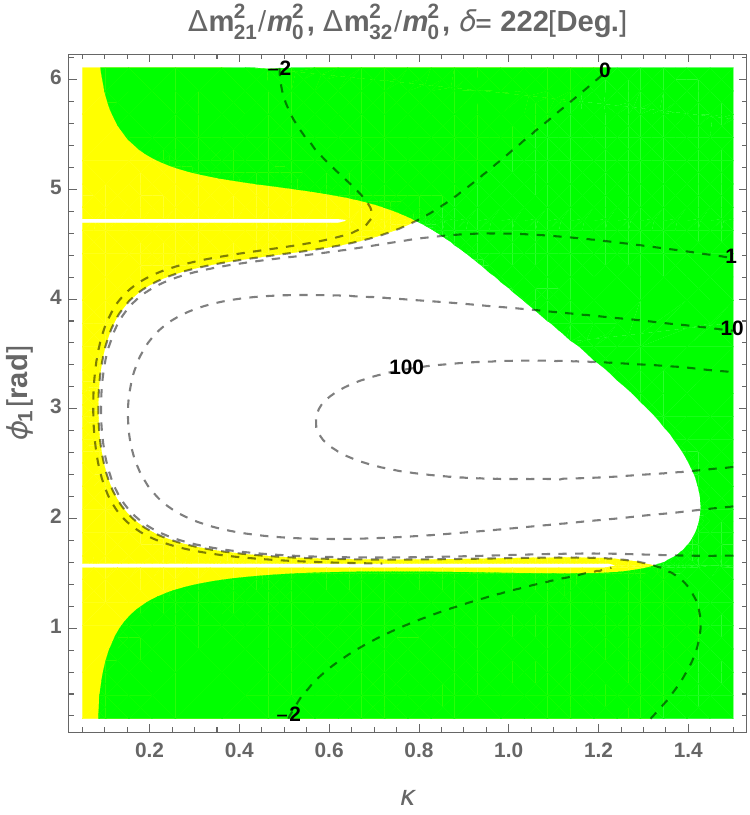} & 	 \includegraphics[width=4.8cm]{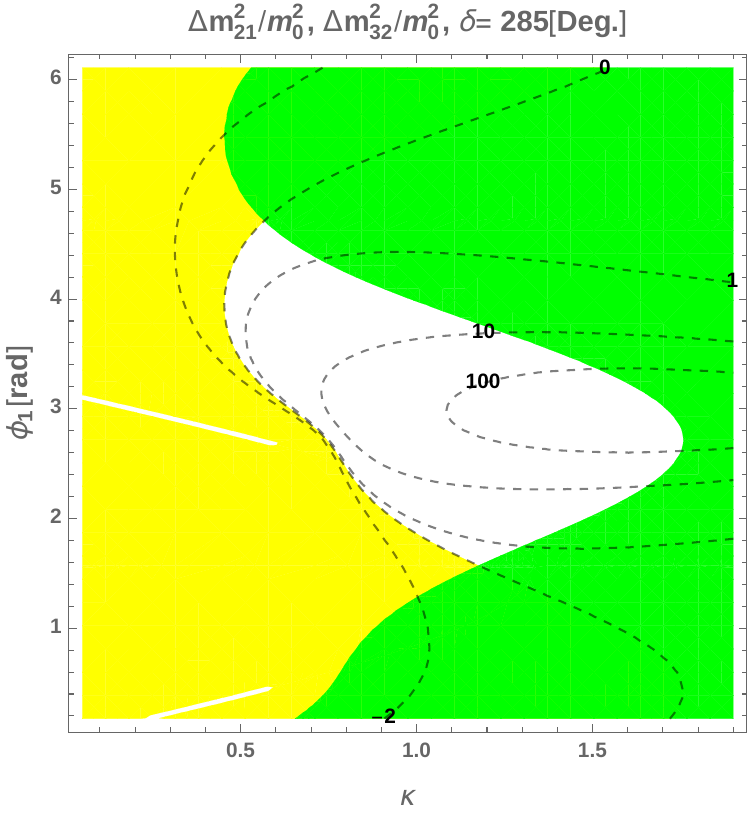}& \includegraphics[width=4.8cm]{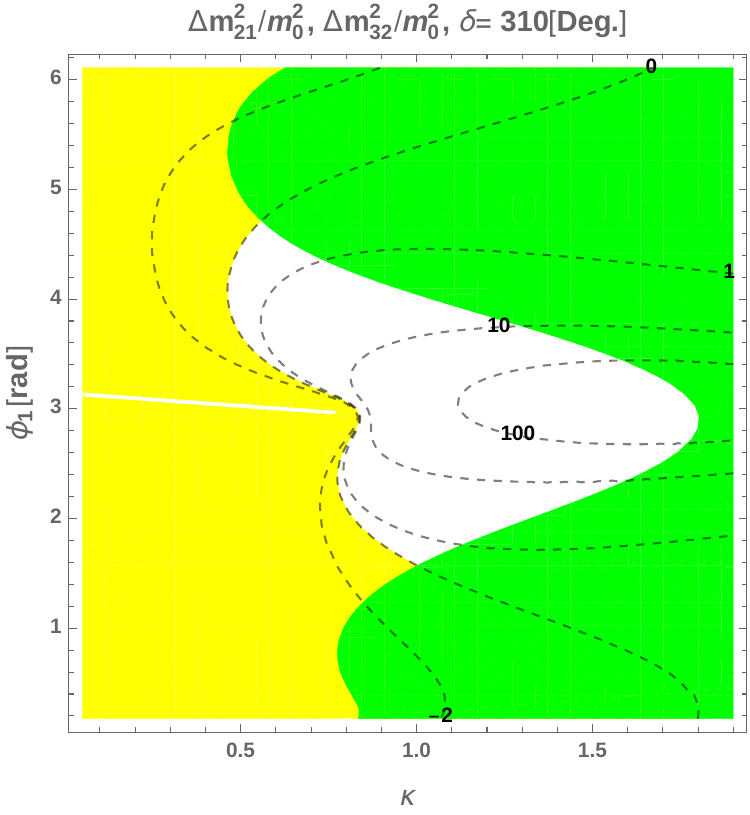}\\
	\end{tabular}
	\caption{ $\Delta m^2_{21}/m_0^2$  and $\Delta m^2_{32}/m_0^2$ as functions of $\kappa$  and $\phi_1$ with different $\delta$ and $s^2_{13}$ at the best-fit point.b The green regions are excluded by the requirement $\Delta m^2_{21}/m_0^2>0$. The yellow regions correspond to $\Delta m^2_{32}/m_0^2\le0$. The dashed black curves show the constant values of $\Delta m^2_{32}/m_0^2$.}\label{fig_dm2ij}
\end{figure}

For the first estimation of $\kappa$ and $\phi_1$ with $\tilde{\epsilon}\neq  0$, we find numerically that $\Delta m^2_{21}/m_0^2$ depend weakly on $s^2_{13}$. Therefore,  it is fixed  at the best-fit point  when $\Delta m^2_{21}/m_0^2$  is plotted as  a function of $\kappa$ and  $\phi_1$ in the range $0\leq \phi_1\leq 360$ [Deg.], while $\delta$ is chosen at some typical values in the allowed range $141 \;[\mathrm{Deg.}]<\delta<340$ [Deg.] for both NO and IO schemes.
Adding a requirement that the formula of $\Delta m^2_{21}/m_0^2 $ given in Eq.~\eqref{eq_d2m21pm0} must be positive,  values of $\kappa>1.9$ are excluded,  see the numerical illustrations  in Fig.~\ref{fig_dm2ij}.
 The allowed regions of $\{ \kappa,\; \phi_1\}$ depend rather complicatedly  on $\delta$.  In Fig.~\ref{fig_dm2ij}, the contour plot of $\Delta m^2_{32}/m_0^2$  are also shown, where the two yellow  and non-color regions supporting the respective IO and NO schemes  are  separated by the constant curve   $\Delta m^2_{32}/m_0^2=0$.   It can be estimated that, the allowed regions for the NO scheme always require  that $1.4\times 180/\pi<\phi_1<4.8\times 180/\pi$ [Deg.].

\begin{figure}[ht]
	\centering
	\begin{tabular}{ccc}
		\includegraphics[width=4.8cm]{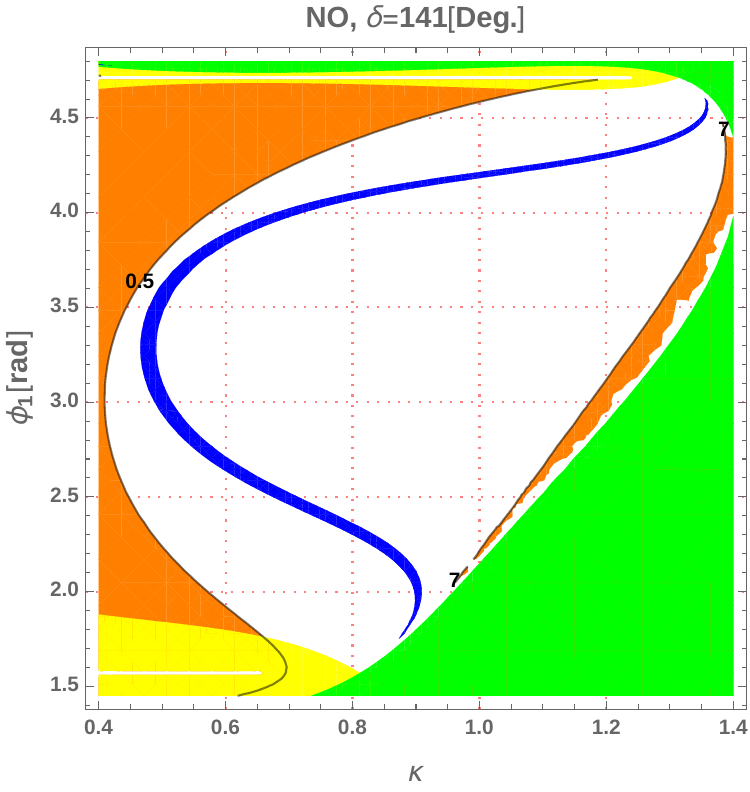}	&
		\includegraphics[width=4.8cm]{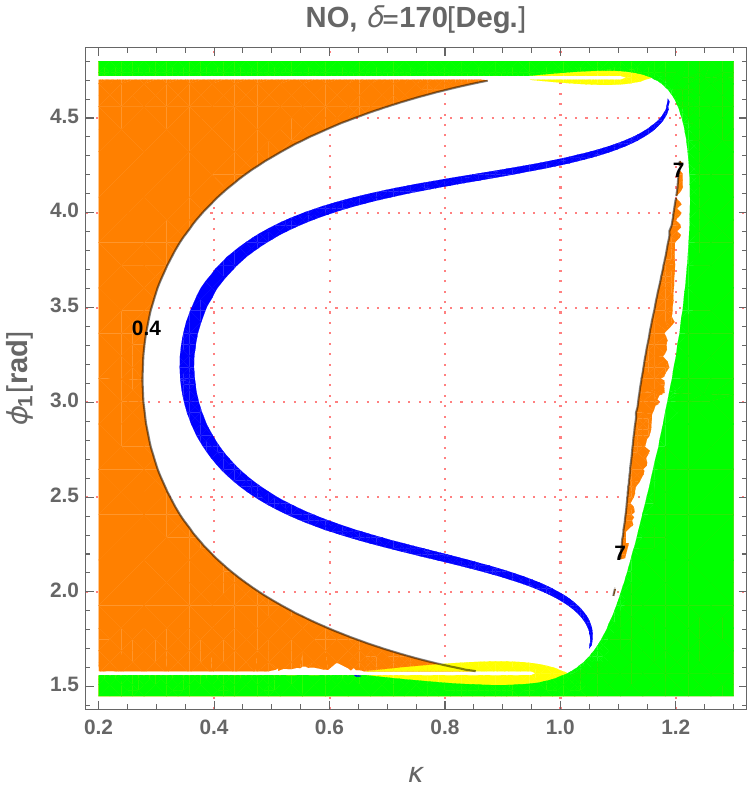}&
		\includegraphics[width=4.8cm]{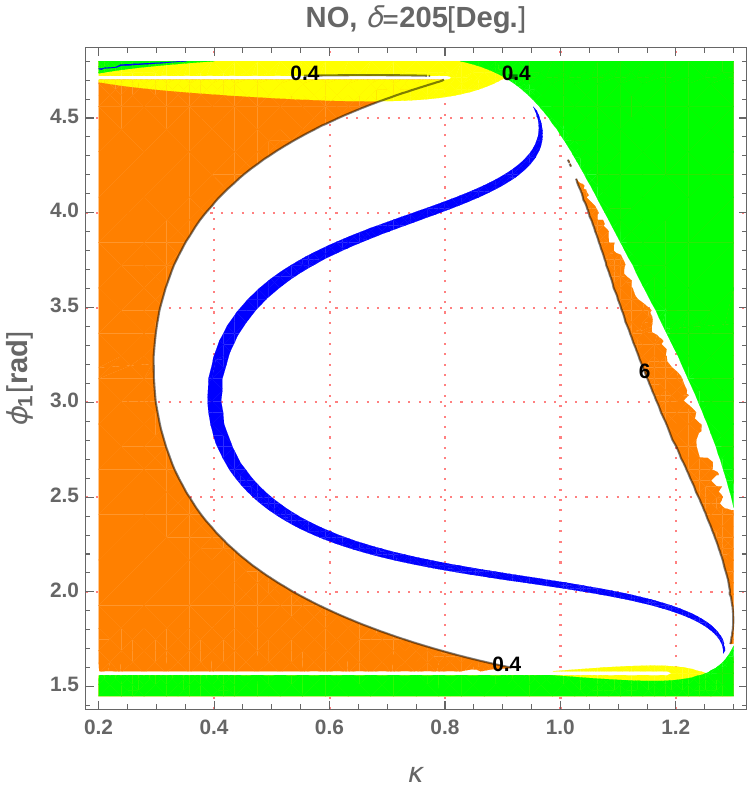}\\
		\includegraphics[width=4.8cm]{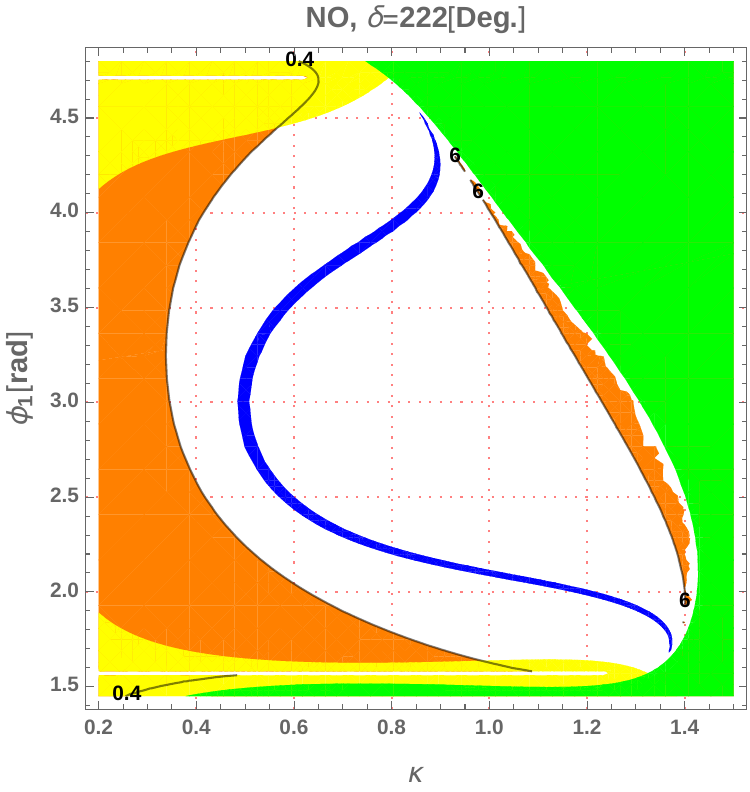} & 	\includegraphics[width=4.8cm]{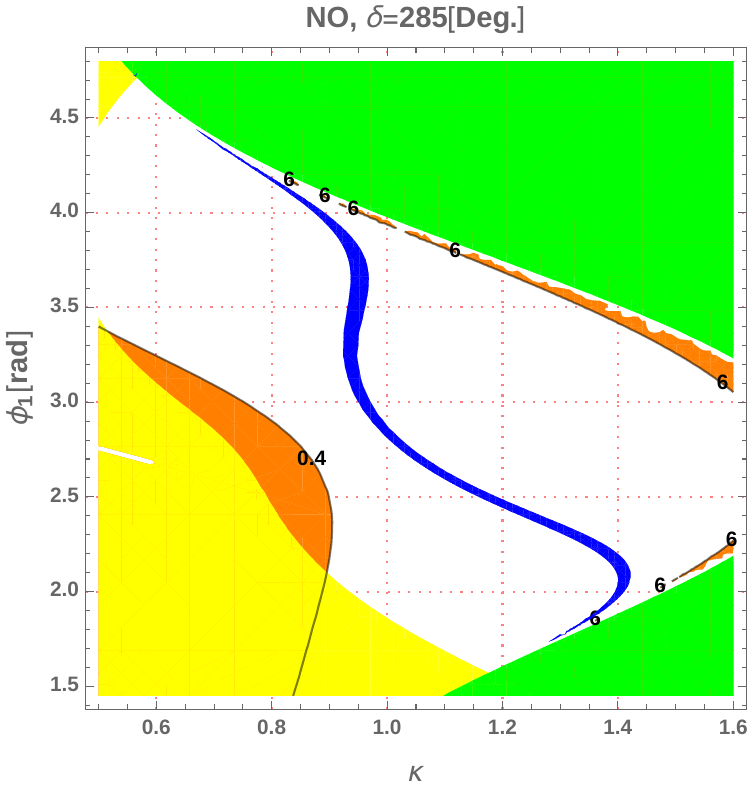}& \includegraphics[width=4.8cm]{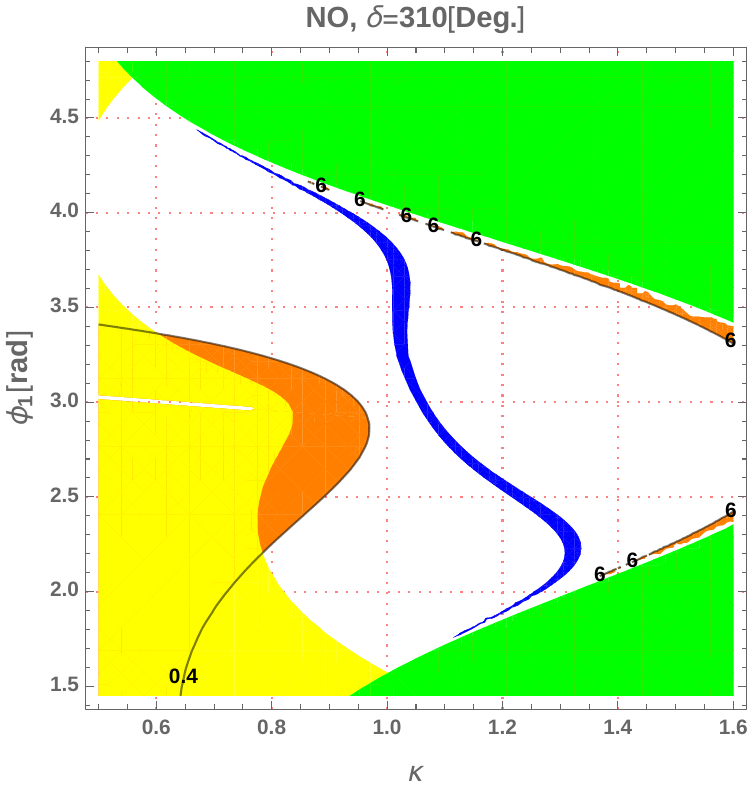}\\
	\end{tabular}
	\caption{ Contour plots of  $ R_{32,21}$ and $m_0$ as functions of $\kappa$ and $\phi_1$ in the NO case with fixed $s^2_{13}$ at best-fit point and different $\delta$.  The blue regions satisfy the condition in Eq.~\eqref{eq_Rconstraint}. The yellow regions and green regions are the same as those in Fig.~\ref{fig_dm2ij}.  The two black curves  show constant values $x_{1,2}$ of $m_{0}\times 10^{11}[\mathrm{GeV}]$ and separate the non color region having $m_0$ in the range $x_1<m_0<x_2$ with the orange regions out side this range. }\label{fig_contourfRNO}
\end{figure}

More details for the NO scheme, we require that $ R_{32,21}\equiv \Delta m^2_{32}/\Delta m^2_{21}$ derived from Eq.~\eqref{eq_d2m21pm0} must in the allowed experimental range
\begin{equation}\label{eq_Rconstraint}
 R_{32,21}\in \left[ \frac{\mathrm{ min^{exp}}(\Delta m^2_{32})}{\mathrm{max^{exp}}(\Delta m^2_{21})}\simeq 29.4 ,\;\frac{\mathrm{max^{exp}}(\Delta m^2_{32})}{\mathrm{min^{exp}}(\Delta m^2_{21})}\simeq 37.5 \right].
\end{equation}
Now, from numerical illustrations shown in Fig.~\ref{fig_contourfRNO},  the allowed regions (blue) are more constrained.
We also add here the contour plots of $m_0$ as a function of $\phi_1$ and $\kappa$ given in Eq.~\eqref{eq_d2m21pm0} and $\Delta m^2_{21}$ is fixed at the best-fit point. The numerical investigation above shows that for the NO scheme,  $\kappa$ must satisfies $0.4<\kappa<1.6$. And the order of $m_0$ is around $10^{-10}$ eV.
In addition,  every allowed $\phi_1$ corresponds to a very narrow allowed range of $\kappa$.
Regarding the IO scheme, the allowed regions  are very narrow  for fixed $\delta$ values, and  require that  $\phi_1$ must be in the two ranges $[0,\; 90]$ and $[270,\; 360]$ [Deg.].  In addition, the allowed ranges of $\kappa$ and $m_0$ are similar to those from the NO scheme, hence  will be used to  scan for determining  more general allowed regions in the following numerical investigation.

\subsection{\label{sec_alloweddata}Low energy observables}

\begin{figure}[h]
	\minipage{0.485\textwidth}
	\centering
	\includegraphics[width=7.5cm]{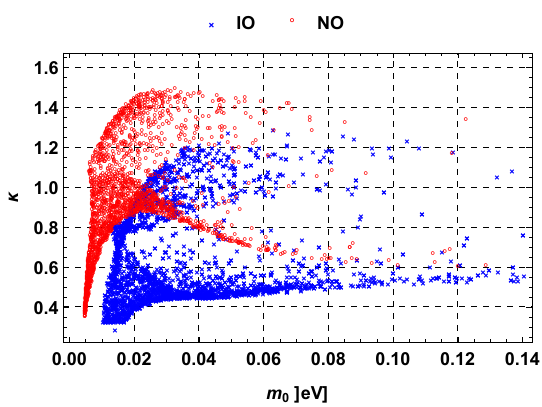}
	\endminipage
	\hfill
	\quad
	\minipage{0.485\textwidth}
	\centering
	\includegraphics[width=7.5cm]{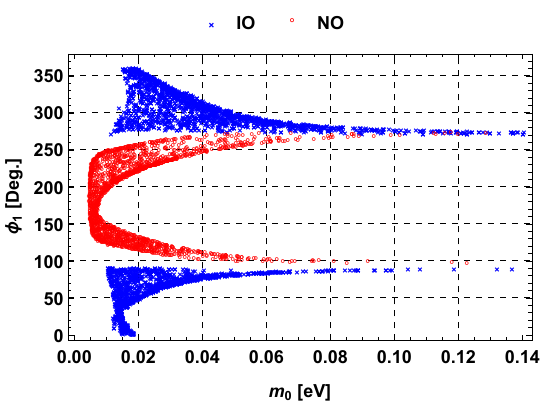}
	\endminipage
	\hfill
	\minipage{0.485\textwidth}
	\centering
	\includegraphics[width=7.5cm]{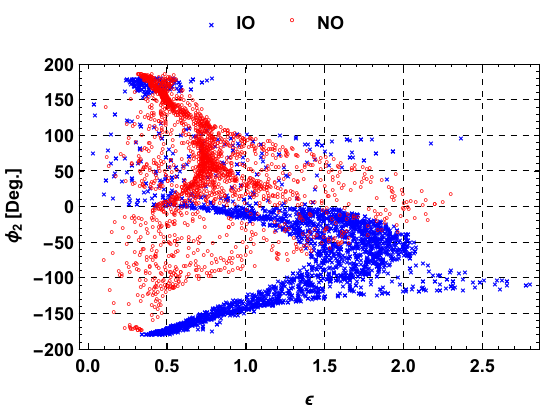}
	\endminipage
	\hfill
	\quad
	\minipage{0.485\textwidth}
	\centering
	\includegraphics[width=7.5cm]{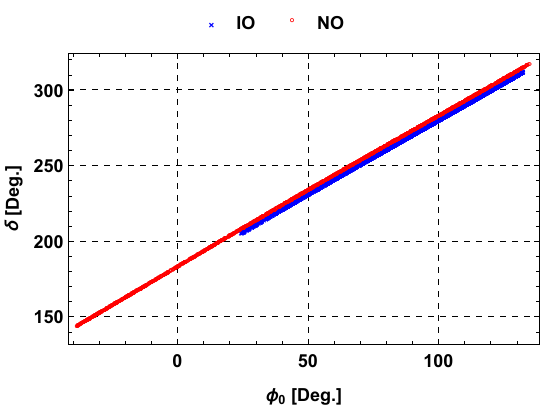}
	\endminipage
	\hfill
	\caption{The allowed ranges  of the  parameters appearing in the model under consideration.
}\label{fig_modelparameters1}
\end{figure}
As we discussed above, the neutrino mass matrix in Eq.~\eqref{eq_mnup1} depends on
the set of five free real parameters $(s_{13}, \delta, \phi_1, \kappa, m_0)$ used to scan numerically. By fixing $s_{13}$, $\delta$, and $\Delta m^2_{21}$, we have estimated the reasonable ranges of all parameters  $m_0$, $\kappa$, and $\phi_1$.   They will be used to scan  all  five independent parameters $(s_{13}, \delta, m_0,  \kappa, \phi_1)$ to collect all allowed points satisfying  the $3\sigma$ experimental data in the general case.  For both NO  and IO schemes,  the unknown parameters  get  random values in the following ranges:
\begin{align}
	\label{eq_Input1}
	0.001 \;\mathrm{eV}\leq m_0\leq 0.18\;\mathrm{eV},
	\quad  0.05\leq \kappa \leq 1.65,
	\quad 0\leq \phi_{1} < 2\pi \; \mathrm{ [rad.]},
\end{align}
and the two parameters $s_{13}$ and $\delta$ runs over $3\sigma$ allowed range of the experimental data.  
We stress that wider ranges  of $m_0$ and $\kappa$ were investigated before  choosing  the best ranges for  the following numerical illustration.  The parameter spaces ($\kappa, \epsilon, \phi_0, \phi_1,  \phi_2 =\arg[\tilde{\epsilon}]$) and their correlations are respectively plotted in  Fig.~\ref{fig_modelparameters1}, where the red and the blue patterns represent the  allowed regions predicted by the NO and the IO schemes, respectively. Hereafter, we continue using these conventions if there is no more explanation. We note that the upper bound values of $m_0$ for the typical NO (IO) scheme was estimated from our numerical investigation. We can find that the allowed regions of the model's parameters are separated completely for the two schemes. In particularly, the ranges $0.35<\kappa<1.5$, $0.1<\epsilon<2.3$, $-40<\phi_0<140$ [Deg.], $90<\phi_1<270$ [Deg.], $-180<\phi_2<180$ [Deg.] support the NO scheme while the IO  implies the allowed values of the parameters as  $0.3<\kappa<1.3$, $0<\epsilon<3$,  $25<\phi_0<135$ [Deg.], $(0<\phi_1<90)\cup(270<\phi_1<360)$ [Deg.], $-180<\phi_2<180$ [Deg.]. The predicted ranges  for the CP phase $\delta$ in the two schemes are more narrow than  the  respective $3\sigma$ experimental data, namely   $141< \delta<320$ [Deg.] for the NO and $205< \delta<310$ [Deg.] for the IO. The lower bounds of $\delta$ are matched with that of experimental bounds for both cases of neutrino mass hierarchy. Whereas, the upper bounds of $\delta$ are lower than about 50 [Deg.] comparing with their experimental values as given in Eqs. (\ref{eq_d2mijNO}) and (\ref{eq_d2mijIO}).

\begin{figure}[ht]
	\minipage{0.485\textwidth}
	\centering
	\includegraphics[width=7.5cm]{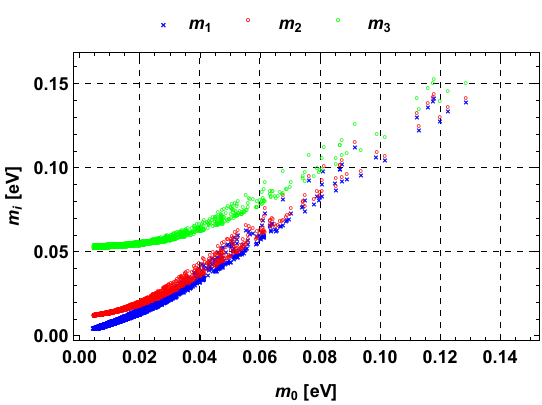}
	\endminipage
	\hfill
	\quad
	\minipage{0.485\textwidth}
	\centering
	\includegraphics[width=7.5cm]{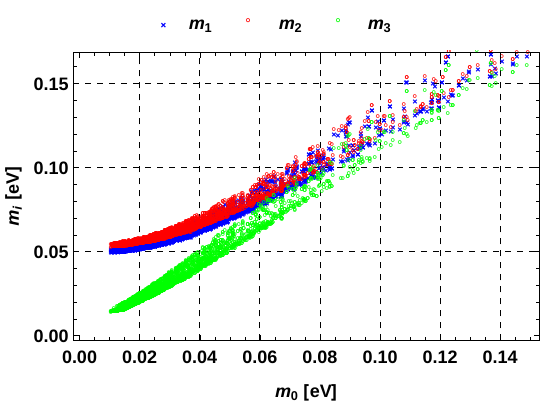}
	\endminipage
	\hfill
	\caption{The active neutrino masses $m_i$ as  functions of  $m_0$ in the left (right) panel predicted by the NO (IO) scheme.}\label{active masses}
\end{figure}

The light neutrino masses predicted by the model are respectively plotted in Fig. \ref{active masses} as function of the light neutrino mass scale $m_0$ for the NO (left panel) and the IO (right panel). There the red, blue, green plots represent for $m_1, m_2, m_3$, respectively.  We can recognize that the neutrino masses are strong hierarchy with small values of $m_0$ and they can be quasi degenerate $m_1 \cong m_2 \cong m_3 \geq 0.12$ eV  for both hierarchies if $m_0$ approaching about 0.12 eV.

\begin{figure}[ht]
	\minipage{0.485\textwidth}
	\centering
	\includegraphics[width=7.5cm]{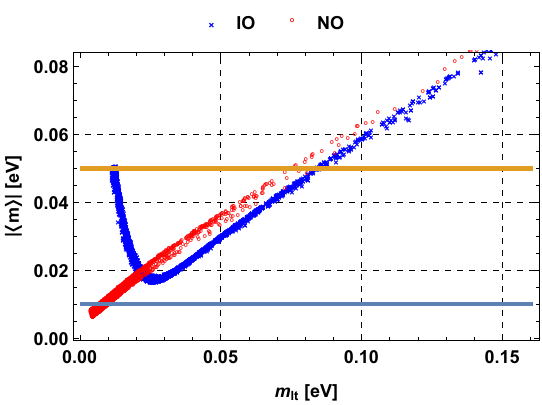}
	\endminipage
	\hfill
	\quad
	\minipage{0.485\textwidth}
	\centering
	\includegraphics[width=7.5cm]{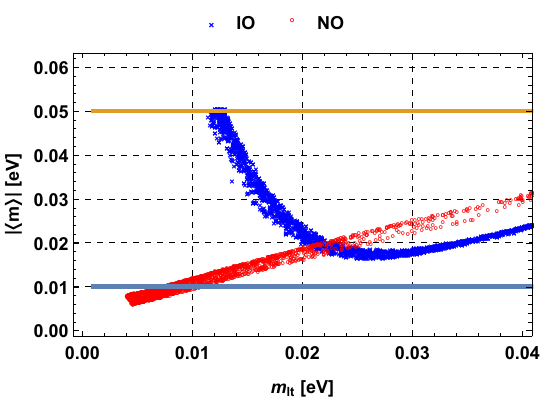}
	\endminipage
	\hfill
	\caption{The relations between $|\langle m\rangle|$  with the lightest mass $m_{lt}$. The solid horizontal lines are the sensitivities of $|\langle m\rangle|$ of the new generation of neutrinoless double beta decay experiment
	\cite{Tanabashi:2018oca}.}\label{fig_mee}
\end{figure}

\begin{figure}[ht]
	\minipage{0.485\textwidth}
	\centering
	\includegraphics[width=7.5cm]{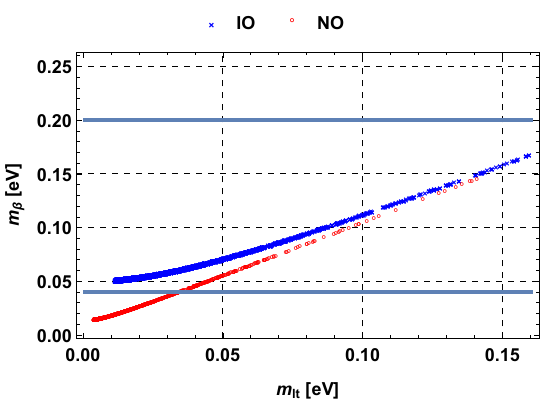}
	\endminipage
	\hfill
	\quad
	\minipage{0.485\textwidth}
	\centering
	\includegraphics[width=7.5cm]{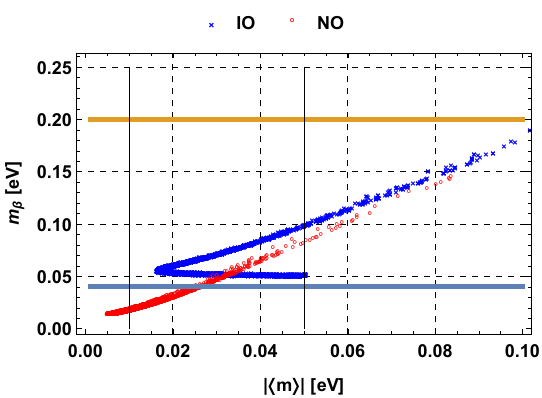}
	\endminipage
	\hfill
	\caption{ The relations between $m_{\beta}$ with  $m_{lt}$ ($|\langle m\rangle|$) in the left (right) panel. The solid horizontal lines are the prospect sensitivities of Katrin  (0.2 eV) \cite{Angrik:2005ep} and Project 8 (0.04 eV) \cite{Asner:2014cwa}. The solid vertical lines are the sensitivities of $|\langle m\rangle|$ of the new generation of neutrinoless double beta decay experiment
	\cite{Tanabashi:2018oca} .}\label{fig_mbeta}
\end{figure}

We all know that neutrino oscillation experiments could not determine the absolute scale of active neutrino masses. Instead, it can be measured by non-oscillation neutrino experiment such as Tritium beta decay \cite{Otten:2008zz},  neutrinoless double beta decay ($0\nu\beta\beta$) \cite{Dolinski:2019nrj}, or by cosmological and astrophysical observations \cite{Lattanzi:2017ubx}. As the model consequences, we would like to study the effective neutrino masses associate with $0\nu\beta\beta$ ($|\langle m\rangle|$) and beta decay ($m_\beta$) which are defined as
\bea
|\langle m\rangle| &=&\left | \sum_{i=1}^3 m_i(U_{\rm PMNS})_{ei}^2 \right|=  \left|\Big(m_1 c^2_{12}+ m_2 s^2_{12}e^{i\alpha_{21}}\Big) c^2_{13} + m_3 s^2_{13}e^{i(\alpha_{31}-2\delta)}\right|,\label{eq_meff}\\
m_\beta &=& \sqrt{ \sum_{i=1}^3 m_i^2|(U_{\rm PMNS})_{ei}|^2} =  \sqrt{m_1^2 c_{13}^2c_{12}^2+ m_2^2c_{13}^2s_{12}^2 + m_3^2s_{13}^2}\;. \label{eq_mbeta}
\eea
In Eq.~\eqref{eq_meff}, we can use  directly $U_{3\nu}$ given in Eq.~\eqref{eq_U3nu1} instead of $U_{\rm PMNS}$ defined in Eq.~\eqref{eq_UPMNS}.
The predictions of the allowed regions presenting the relation between effective masses $|\langle m\rangle|$ and $m_\beta$ with the lightest active neutrino mass, $m_{lt}$, are respectively plotted in Fig.~\ref{fig_mee} and Fig.~\ref{fig_mbeta} (left panel). Whereas, the correlation between $m_\beta$ and $|\langle m\rangle|$ is shown in the right panel of the Fig. ~\ref{fig_mbeta}.

The consequence is that the IO scheme is excluded for $m_{\beta}<0.05$ eV, or $|\langle m \rangle|<0.015$ eV. In contrast, the NO is excluded for  $m_{\beta}<0.01$ eV or $|\langle m \rangle| < 0.005$ eV.  In addition,  the sharped allowed regions in the right panel of Fig.~\ref{fig_mbeta} implies  the very strict relations between $m_{\beta}$ and $|\langle m \rangle|$   for both schemes. On the other word, the model is very predictive. Because once both of these quantities are observed, the reality of the model is immediately confirmed. If it survive, based on  the two separated regions corresponding the NO and IO schemes, the model can  figure out only one of the two schemes survive.

\section{\label{sec_Higgs} The Higgs sector}
The same as  the SM, the covariant derivative for  local $SU(2)_L\otimes U(1)_Y$ symmetry is
 \be D_{\mu} = \partial_{\mu} - igT^aW^a_{\mu}-i\frac{g'}{2}B_{\mu}Y. \label{derivative}\ee
 Here $T^a$ ($a=1,2,3$) are the generators of the $SU(2)_L$ symmetry.
  The kinetic terms of the two Higgs doublets generating gauge boson masses  are:
 \bea  \mathcal{L}^H_{\rm{kin}}&=& \left(D_{\mu}h_u\right)^{\dag}\left(D^{\mu}h_u\right) + \left(D_{\mu}h_d\right)^{\dag}\left(D^{\mu}h_d\right).
  \label{Hkinetic}\eea
The SM charged gauge bosons are $W^{\pm}\equiv (W^1\mp i W^2)/\sqrt{2}$. The same relation known in two Higgs doublet models (2HDMs) for the two vevs,
\begin{equation}
v^2=v_u^2+v_d^2=174^2\mathrm{ GeV}^2, \hs t_W=\frac{s_W}{c_W}\equiv \frac{g'}{g}, \label{SMmatching}
\end{equation}
where $s_W^2=0.231$.  The model also contains the SM $Z$ boson satisfying   $W^{3}_{\mu}= c_W Z_{\mu} +s_W A_{\mu}$  and  $B_{\mu}=-s_W Z_{\mu} +c_W A_{\mu}$.

A detailed analytical calculation on the mass and mixing of the Higgs in the 2HDMs  was done previously, for example in Ref.~\cite{Gunion:2002zf}.   Similarly to the very important parameter $t_{\beta}$ defined by the  ratio of the two neutral Higgs VEVs  in the MSSM and the 2HDM,   $t_\beta$ in the model under consideration  is defined as follows:
\begin{equation}
t_{\beta}\equiv\tan\beta= \frac{v_u}{v_d}.
\label{eq_tbeta}
\end{equation}
 It is noted that from now on we will use the following conventions for any angles $x=\delta,\alpha,\beta, 2\alpha,2\beta$, and  $(\beta-\alpha)$: $t_x \equiv \tan(x), \; s_x\equiv \sin (x), \; c_x\equiv \cos (x)$.
Because the Yukawa couplings of charged leptons with $h_d$ in Eq.~\eqref{eq_L5dim} is fixed,  $t_{\beta}$ defined in this model is consistent with Refs.~\cite{Ilakovac:2012sh,Popov:2013xaa}, but  equivalent to $1/t_{\beta}$ defined in  other MSSM and 2HDMs discussed previously.

For the Higgs spectrum, this model must contain at least one SM-like Higgs boson observed by the LHC. It is one of those included in the squared mass matrix of the CP-even Higgs boson, which is a $10\times 10$ matrix  consisting of a large number of the independent Higgs self-couplings. In this work, we will choose a simple case of the realistic Higgs potential,
 which  is  summarized in the appendix \ref{Hcondition}.   Here we will focus on the identification of SM-like Higgs boson.  In particular,  a simple regime  is chosen that these Higgs doublets decouple to other Higgs singlets,  and a soft term  generate masses for CP-odd neutral Higgs and results in heavy charged Higgs bosons, the same as those appear in the well-known 2HDM~\cite{Branco:2011iw}.

The model contains two singly charged Higgs bosons $\varphi^\pm$ and two massless states $G^\pm_W$ which are Goldstone bosons eaten by gauge bosons $W^\pm$. The masses and relations between the original and physical states of the charged Higg components are:
\begin{align}
m^2_{G_W}&=0, \quad m^2_{\varphi}=\left(\la_4 v^2 +\frac{\mu_{12}^2 }{s_{\beta}c_{\beta}}\right), \crn
   H_u^\pm&=s_\beta G^\pm_W +c_\beta \varphi^\pm, \;
\hs H_d^\pm =-c_\beta G^\pm_W +s_\beta \varphi^\pm.
\label{eq_cHiggs}
\end{align}
Regarding CP-odd neutral Higgs components,  the squared mass matrix  has a zero determinant, which implies  exactly a massless state corresponding to the Goldstone boson of the $Z$ boson in the SM.  One of  the remaining CP-odd neutral Higgs boson  relate to the two Higgs doublets with mass satisfying $ m_A^2=\frac{\mu_{12}^2 }{s_{\beta}c_{\beta}}=m^2_{\varphi}-\lambda_4 v^2,$
 implying that $\mu_{12}^2>0$.

 For the CP-even neutral Higgs corresponding to the simple case we assume above,  the total squared mass matrix will separate into two submatrices, namely a $2\times2$ and a $8\times8$ ones. The $8\times8$ matrix gives eight physical heavy Higgs bosons with masses depending on heavy VEVs $v_S$ and $v_T$, see appendix \ref{Hcondition} for the details. In the original basis $(S_u,\;S_d)^T$ given in Eq.~\eqref{eq_Higgsexpand}, the $2\times2$ matrix containing the SM-like Higgs boson has the following form,
\begin{equation}
\mathcal{M}^2=\left(\begin{array}{cc}
2\la_1v_u^2 +\mu^2_{12}t^{-1}_\beta &  2 \la_3v_uv_d -\mu^2_{12}\\
2 \la_3v_uv_d -\mu^2_{12} &  2\la_2v_d^2 +\mu^2_{12}t_{\beta}\\
\end{array}\right).
\end{equation}
It gives two mass eigenstates, denoted as $H$ and $h_1$ where the lighter $h_1\equiv h$ will be identified with the SM-like Higgs boson. Their masses and relations with the original states  are
\begin{align}
 m^2_{H,h}&= \frac{1}{2}\left[\mathcal{M}^2_{11}  +\mathcal{M}^2_{22} \pm \sqrt{ \left(\mathcal{M}^2_{11} - \mathcal{M}^2_{22}\right)^2 +4 \left(\mathcal{M}^2_{12}\right)^2}\right],\crn
 S_u &= c_{\alpha} h +s_{\alpha} H, \;  S_d = -s_{\alpha} h +c_{\alpha} H.
 \label{NHiggs}
\end{align}
where $\alpha$ is defined based on the following relation
\begin{equation}\label{eq_t2al}
t_{2\alpha}\equiv\tan2\alpha=  -\dfrac{2 t_\beta \left[ m^2_{\varphi} -\left( 2\lambda_3 +\lambda_4\right)v^2\right]}{\left(m^2_{\varphi} -\lambda_4 v^2\right)\left(t^2_{\beta} -1\right)+ 2\left(\la_2 -\la_1 t^2_\beta\right)v^2}.
\end{equation}
Because   the SM-like Higgs boson $h$ were found experimentally at LHC,  we require that the its couplings  with normal fermions and gauge bosons have   small deviations with those predicted by the SM, see  the Feynman rules given in table ~\ref{tabe_SMHiggscoupling}.
\begin{table}[ht]
	\centering
	\caption{ Coulings of the SM-like Higgs with SM particles $H_1\equiv h$}
	\begin{tabular}{|c|c|c|c|}
		\hline
		Vertex & Coupling & Vertex & Coupling \\
		\hline
		$h\overline{e_a}e_a $ &$\dfrac{ig m_{e_a}}{2m_W } \dfrac{s_{\alpha}}{c_{\beta}}$ & $h\overline{q_a}q_a $ & $\dfrac{ig m_{q_a}}{2m_W } \dfrac{s_{\alpha}}{c_{\beta}}$ \\
		\hline
		$hW^+_\mu W^-_\nu$ &$igm_W\sin(\beta-\al)g^{\mu\nu}$  & $hZ_\mu Z_\nu$ &$\frac{igm_W}{c_W^2} \sin(\beta-\al)g^{\mu\nu}$\\
		\hline
	\end{tabular}\label{tabe_SMHiggscoupling}
\end{table}
 Here all quarks multiplets are assumed to be singlets under all $A_4$ and other discrete symmetries listed in table~\ref{particle content}. The Yukawa Lagrangian of quark is
\begin{equation}\label{eq_Lyq}
-\mathcal{L}^Y_{q}= y^{d}_{ab}\overline{Q_{aL}}h_dd_{bR} + y^{u}_{ab}\overline{Q_{aL}}\tilde{h}_du_{bR} +h.c.,
\end{equation}
where $\tilde{h}_d=i\sigma_2h_d^*=(v_d+ \frac{S_d -iA_d}{\sqrt{2}},\;-H^-_d)^T$.  Correspondingly, the top quark mass is $m_t = v_d y_t$, leading to the perturbative limit  $m_t/v_d=y_t<\sqrt{4\pi}$. As a result, we have $c_{\beta}=v_d/v>m_t/(v\sqrt{4\pi})$, equivalently  $t_{\beta}\le 3.4$. The model now treats like the 2HDM type-I, which the recent constraints on the model parameters were given in Ref.~\cite{Chen:2019pkq,Kling:2018xud}. The typical constraints on the free parameters are $ |s_{\delta}|\le 0.05$ and $t_{\beta}\le3.3$.

 There is another assignment of quark sector corresponds to the case of the 2HDM type-II and the  MSSM:  all upper quarks only couple with  $h_u$, leading to the condition $t_{\beta}\ge 0.3$ for the perturbative limit.  The recent constraints of the model parameters were given in Refs.~\cite{Chen:2018shg,Kling:2018xud}, where the typical ranges of free parameters are $|s_{\delta}|<0.008$ and $0.2\le t_{\beta}\le 5$ after casting into the model under consideration. The  Lagrangian contains  Yukawa couplings of quarks with charged Higgs boson is:
 \begin{align}
 	\label{eq_Lq}
 	\mathcal{L}_{ud\varphi}&=- \frac{g}{\sqrt{2}m_W}V^*_{ab} \overline{d_b} \left( t_{\beta}  m_{d_b} P_L +t^{-1}_{\beta}  m_{u_b} P_R \right) u_{a} \varphi^- +\mathrm{H.c.},
 \end{align}
 where $V$ is the Caibibbo-Cobayashi-Maskawa matrix, $m_{q_a}$ with $q=u,d$ and $i=1,2,3$ is the mass of the quark $q_a$.

 We note that $t_{\beta}$ defined in this work   is equivalent with $1/t_{\beta}$ defined in Ref.~\cite{Chen:2018shg,Chen:2019pkq} for both type-I and II of the 2HDM.  Our numerical calculation does not depend on the assignments of quark, except the $\mu-e$ coversion rate, which was discussed for the MSSM, including the contributions from the charged Higgs bosons in the 2HDM type II. But the phenomenology investigated in this work strongly depend on $t_{\beta}$,  leading to another channel to determine which $SU(2)_L$ Higgs doublet generates quark masses.

From the recent experimental data, the coupling of the SM-like boson with normal charged leptons, namely $h\overline{e_a}e_a$ given in Table \ref{tabe_SMHiggscoupling}, must be consistent with the coupling predicted from in the SM, leading to a consequence that $s_{\alpha}\simeq -c_{\beta}$.  A small deviation from the SM couplings correspond to a small parameter $\delta$ satisfying
\begin{equation}\label{eq_delta}
\sin(\beta-\alpha)\equiv\cos\delta\simeq 1,
\end{equation}
 equivalently $\delta\equiv \frac{\pi}{2} +\alpha-\beta\simeq0$, which is known in the 2HDMs. The constraints of $\delta$, $t_{\beta}$, and charged Higgs bosons comes from the electroweak and Higgs precision measurements discussed recently for the 2HDMs~\cite{Azevedo:2018llq, Chen:2018shg, Chen:2019pkq},

  In conclusion, in the numerical investigation, the free parameters in the Higgs sector we will use are $\beta$, $\delta$,   $m_h\equiv m_{h}$, $m_{\varphi}$, $\lambda_2$,  and $\lambda_4$. The SM-like Higgs mass  $m_h\equiv m_{h}\simeq 125.09$ GeV was determined experimentally from LHC. Three   parameters $\alpha$ and $\lambda_{1,3}$ are determined from three  Eqs.~\eqref{eq_delta}, \eqref{NHiggs}, and \eqref{eq_t2al}.  We are interesting in the regions allowing large range of $t_{\beta}$, hence we will fix $\lambda_4=0$, and $m_A=m_{H_2}=m_{\varphi}$, which were shown from previous discussions~\cite{Chen:2019pkq,Kling:2020hmi,Kling:2018xud}.
 Based on the general formulas of the three parameters $\lambda_{1,2,3}$ given in Ref.~\cite{Gunion:2002zf}, they can be written as follows:
\begin{align}
\label{eq_l123}
\lambda_1&= \frac{c^2_{\alpha} m^2_{h} + m^2_{\varphi}\left( s^2_{\alpha} -c^2_{\beta}\right)}{2 v^2s^2_{\beta}},\;
\lambda_2 =  \frac{s^2_{\alpha} m^2_{h} + m^2_{\varphi}\left( c^2_{\alpha} -s^2_{\beta}\right)}{2 v^2s^2_{\beta}},\crn
\lambda_3&= \frac{-s_{\alpha}c_{\alpha}  m^2_h    +m^2_{\varphi}\left( s_{\beta}c_{\beta} +s_{\alpha}c_{\alpha}\right) }{2 v^2c_{\beta}s_{\beta}}.
\end{align}
The analytic formulas of these three Higgs self couplings will be used to calculate the coupling  $h\varphi^+\varphi^-$ that contributes to the LFVHD.

\section{\label{sec_LFV}LFV decays}
\subsection{\label{subsec_neutrino}Neutrino sector: ISS relations vs the exact numerical solution}
In this section we study effects of the allowed regions of parameter space  on the LFV decays.  The neutrino mixing is the only source of the LFV processes.  The original basis of  the nine left-handed  neutral neutrinos are denoted as $\nu'_L=(\nu_L, (N_R)^c,\;(X_R)^c)^T$, where $\psi_L\equiv (\psi_{1L},\; \psi_{2L},\; \psi_{3L})^T$, for all all leptons $\psi_L\equiv \nu_L, (N_R)^c,(X_R)^c$. Also, the original right- handed neutrino basis is  $\nu'_R=(\nu'_L)^c=((\nu_L)^c, N_R,\; X_R)^T$ with $\psi_R=(\nu_{aL})^c,N_{aR},X_{aR}$. A  four-component spinor for a Majorana neutrino is then $\psi =(\psi_L,\;\psi_R)^T$ where $\psi=\nu_a, N_a,X_a$ satisfying $\psi^c=C\overline{\ps}^T=\psi$, where $C$ is the charge conjugation operator.  The  relations between a four-component Majorana  spinor  with the respective left- and right-handed components are $\psi_{L,R}=(\psi_{R,L})^c=P_{L,R}\psi$, where $P_{L,R}=(1\mp\gamma_5)/2$.  The mass term of these nine neutrinos were given in  two Eqs.~\eqref{lagrangian} and \eqref{eq_L5dim}.  The corresponding mass matrix is diagonalized by a total $9\times9$ unitary mixing matrix  $U^{\nu}$  determined as follows:
\be U^{\nu T}\mathcal{M}^{\nu} U^{\nu}=\hat{\mathcal{M}}^{\nu}=\mathrm{diag}(m_{n_1},\;m_{n_2},...,\;m_{n_9})=\mathrm{diag}(m_1,\,m_2,\,m_3,\,M_1,\,M_2,...,\,M_6),   \label{diamnu}\ee
where the first three masses  $m_{n_a}$ ($a=1,2,3$) and respective eigenvectors $n_{a}$ are identified with those of active neutrinos observed by experiments defined in Eq.~(\ref{eq_mnu}).  The remaining ones belong to those of the  six heavy neutrinos $n_{I+3}$ with $I=1,2,...,6$.  Relations between original and mass  neutrino bases are
\begin{equation}\label{eq_nu-relation}
\nu'_{iL}= U^{\nu}_{ij}n_{jL}=U^{\nu}_{ij} P_L n_j,\;  \mathrm{and} \; \nu'_{iR}= U^{\nu*}_{ij} n_{jR}=U^{\nu*}_{ij} P_R n_j,
\end{equation}
where $n=(n_1,\; n_2,\;...,\; n_9)^T$ and $n_i=(n_{iL},\;n_{iR})^T$ ($i=1,2,..,9$). The matrix $U^{\nu}$ is parameterized as follows \cite{Casas:2001sr, Ibarra:2010xw, Dreiner:2008tw},
\begin{align}
 U^{\nu}= \exp\left(
\begin{array}{cc}
\textbf{O} &  R\\
-R^{\dagger}&\textbf{O} \\
\end{array}
\right)\left(
\begin{array}{cc}
U_{3\nu} & \textbf{O}\\
\textbf{O}&V_6  \\
\end{array}
\right),
\label{Unugen}
\end{align}
where $\textbf{O}$ is the $3\times3$ matrix with all elements being zeros, $V_6$ is the unitary matrix, $\exp(x)\equiv \sum_{i=0}^{\infty}x^i/i!$, and $R$ is the $3\times6$ matrix satisfying the ISS condition that  max$|R_{ai}|\ll1$ for all $a=1,2,3$ and $i=1,2,..,9$.  Accordingly, the ISS relations given in Eq.~\eqref{R1} are obtained from the expansion $U^{\nu}$ up to the order $\mathcal{O}(R^2)$. Correspondingly, the sub-matrix $R$ and heavy neutrino masses in this case are
\begin{align}\label{eq_heavyN}
R^*&\simeq  \left(-m^T_DM^{-1},\hs m^T_DM_R^{-1} \right), \crn
V_6^*\hat{M}_NV_6^{\dagger}&=M_N +\frac{1}{2}R^TR^*M_N+M_N\frac{1}{2}R^\dagger R, \quad M_N\equiv \begin{pmatrix}
0& M^T_R \\
M_R&\mu_X
\end{pmatrix},
\end{align}
where $M\equiv  M^T_R\mu_X^{-1}M_R$ and $\hat{M}_N\simeq \mathrm{diag}(M_1,M_2,...,M_6)$ presents six  heavy neutrino masses.

Before determining Feynman rules to calculate Br of LFV decays, we remark some requirements to guarantee the ISS relations given in Eqs.~\eqref{R1} and~\eqref{eq_heavyN}.   In principle, in order to consistent between the approximation solution from the ISS relations and the exact one obtained directly from calculating numerically the total neutrino mass matrix given in Eq.~\eqref{Lnumass}, the condition max$(|R_{ai}|)\ll1$ must satisfy, equivalently max$\left(|m_D(M_R^T)^{-1}|_{ai}\right)\ll1$, leading to a consequence that the two scales $v_uf$ and $M_0$ given in \eqref{Majoranamass1} must satisfies $v_uf/ M_0\ll1$.

Direct numerical calculation to derive masses and mixing of the active neutrino from the total neutrino mass $\mathcal{M}^{\nu}$, it is necessary to know the input as the following set of free parameters  $\{m_d,\, M_0,\, \mu_X, \tilde{\kappa},\,\tilde{\rho} \}$.  We limit our investigation in the allowed regions of the parameter space indicated in section~\ref{sec_alloweddata}.  Namely, two parameters $\tilde{\kappa}$ and $\tilde{\rho}$ reduce to two allowed values of $\kappa$ and $\phi_1$.  The parameter $\mu_X$ is considered as a function of $m_0$ from Eq.~\eqref{eq_fm0}, then allowed values of $m_0$ can be determined from the numerical investigation.  These allowed values will be used as the inputs. The remaining parameters $m_D\equiv  vf$, and $M_0$ do not  appear in the approximate solution of active neutrino data using ISS relation~\eqref{eq_mnu} because it is absorbed in $m_0$. Instead, the constraint of $m_0$ is used to determine $\mu_X$. In contrast, determining  $\mathcal{M}^{\nu}$ requires all  of the three parameters. In numerical investigation,  the exact solution is found by using the software mathematica 11 to find the mixing matrix $U^{\nu}$ relating with the eigenvectors of the matrix $\mathcal{M}^{\nu\dagger}\mathcal{M}^{\nu}$ and the neutrino  masses.

\subsection{\label{subsec_LFVFeynmanrule} Feynman rules for LFV decays}
In the Yukawa Lagrangian parts \eqref{lagrangian} and \eqref{eq_L5dim}, couplings relating to LFV decays are
\begin{equation}
	-\mathcal{L} \rightarrow \frac{m_{e_a}}{\sqrt{2}v_d} S_d \overline{e_a}e_a+ \frac{m_{e_a}}{v_d}\left(\overline{\nu_{La}}e_{aR}H^+_d+ \mathrm{H.c.}\right)
+ \frac{S_u}{\sqrt{2}}f\left(\overline{\nu_{aL}}N_{aR}+\mathrm{H.c.}\right)  +f\left[ H^-_u \overline{e_{aL}}N_{aR} + \mathrm{H.c.} \right],\nn
\end{equation}
where  the third term results in the coupling $h\overline{n_i}n_j$ for the ISS mechanism  discussed in detail in Ref.~\cite{Korner:1992zk, Pilaftsis:1992st}.  We can prove that
\begin{equation}\label{eq_fnuN}
f\overline{\nu_{aL}}N_{Ra}=\frac{1}{v_u}\overline{n_i}\left( D^*_{ij}m_{n_j}P_R\right) n_j, \quad D_{ij}\equiv \sum_{a=1}^3U^{\nu*}_{ai}U^{\nu}_{aj},
\end{equation}
which was introduced firstly in Ref.~\cite{Pilaftsis:1992st}.
Using the definition $ \lambda_{ij}\equiv \left( D_{ij}m_{n_i} +D^*_{ij}m_{n_j}\right)$ needed to write the right  Feynman rules for Majorana neutral lepton  in terms of Dirac spinors~\cite{Dreiner:2008tw,Arganda:2004bz,Arganda:2014dta}, the  coupling $h\overline{n_i}n_j$ is written in the symmetric form as follows:
$$\frac{gc_{\alpha}}{4s_{\beta}m_W}h\sum_{i,j=1}^9\overline{n_i}\left( \lambda_{ij}P_L +\lambda^*_{ij}P_R\right)n_j.$$
In the unitary gauge, the Feynman rules for vertex couplings relating with the decay processes in this work are  given in table~\ref{table_LFVcoup},
\begin{table}[ht]
	\centering
	\caption{ Feynman rules for couplings  contributing to LFV decays, $ \Gamma = s_\beta c_\beta(c_{\alpha} c_{\beta}\lambda_1-s_{\alpha}s_{\beta}\lambda_2)+
		(c_{\alpha}s^3_{\beta}-s_{\alpha}c^3_{\beta})\lambda_3+\sin(\beta-\alpha) \lambda_4$}\label{table_LFVcoup}
	\begin{tabular}{|c|c|c|c|}
		\hline
		Vertex & Coupling & Vertex & Coupling \\
		\hline
		$\overline{e_a}e_a h$ &$-\dfrac{ig m_{e_a}}{2m_W } \left( c_{\delta} -t_{\beta}s_{\delta}\right)$ & $ h\varphi^+\varphi^-$ & $-\dfrac{2im_W}{g} \Gamma$\\
		\hline
		$\overline{n_i}n_jh$ & $-\frac{ig}{2m_W} \left( c_{\delta} +t^{-1}_{\beta} s_{\delta}\right)\left( \lambda_{ij}P_L + \lambda^*_{ij}P_R\right) $ &	$hW^+_\mu W^-_\nu$ &$igm_Wc_{\delta}g^{\mu\nu}$  \\
			\hline
		$\overline{e_a}n_i\varphi^-$ &  $\dfrac{-igU^{\nu}_{ai}}{m_W\sqrt{2}}\left(m_{e_a}t_{\beta} P_L+m_{n_i}t^{-1}_{\beta} P_R \right)$&$\overline{n_i}e_a\varphi^+$ &  $\dfrac{-igU^{\nu*}_{ai}}{m_W\sqrt{2}}\left(m_{e_a}t_{\beta} P_R+m_{n_i}t^{-1}_{\beta} P_L \right)$ \\
		\hline
		 $ h\varphi^+W_\mu^-$&$\dfrac{igs_{\delta}}{2}\;(p_{h}-p_{\varphi^+})^\mu$&  $h\varphi^-W_\mu^+$&$\dfrac{-igs_{\delta}}{2}\;(p_{h}-p_{\varphi^-})^\mu$\\
			\hline
		$\overline{e_a}n_iW^-_\mu$&$\dfrac{ig}{\sqrt{2}}U^{\nu}_{ai}\gamma^\mu P_L$&$\overline{n_i}e_aW^+_\mu$&$\dfrac{ig}{\sqrt{2}} U^{\nu*}_{ai}\gamma^\mu P_L$\\
	\hline
	\end{tabular}
\end{table}
 consistent with those mentioned in the 2HDMs~\cite{Branco:2011iw}.

\subsection{LFV decays $e_b\rightarrow e_{a}\gamma$}
In the limit $m^2_a/m^2_b\ll1$,  $m_{a,b}$ being  masses of charged leptons $e,\mu$ and $\tau$, the Brs of the cLFV decays $e_b\rightarrow e_a\gamma$   is  determined  as follows ~\cite{Lavoura:2003xp,Hue:2017lak},
\begin{equation}
\mathrm{Br}(e_b\rightarrow e_a\gamma)=  \frac{3\alpha_{\mathrm{e}}}{2\pi}\left(\left| D_{(ba)L} \right|^2+ |D_{(ba)R}|^2\right)\times \mathrm{Br}(e_b\rightarrow  e_a\bar{\nu}_{a}\nu_b), \label{brlfvdecay2}
\end{equation}
where $D_{(ba)L,R}$ are scalar factors arising from loop corrections, $\alpha_e\simeq 1/137$  in numerical investigations,  and the experimental values of the Br$(e_b\rightarrow\,e_a\bar{\nu}_{a}\nu_{b})$ are $\mathrm{Br}(\tau\rightarrow\mu\bar{\nu}_{\mu}\nu_{\tau}) \simeq 17.41\%$,  $\mathrm{Br}(\tau\rightarrow e\bar{\nu}_{e}\nu_{\tau}) \simeq 17.83\%$ and  $\mathrm{Br}(\mu\rightarrow e\bar{\nu}_{e}\nu_{\mu}) \simeq 100\%$.
 The analytic expressions   $D_{(ba)L,R}= D^{W}_{(ba)L,R}+D^{\varphi}_{(ba)L,R}$   are determined in Appendix \ref{app_eijkl}.   The one loop contributions were established the same way as those mentioned in the SS case~\cite{Nguyen:2017ibh}, and  consistent with \cite{Lavoura:2003xp,Hue:2017lak}. The two loop contributions mentioned in the 2HDM type-X model given in Ref.~\cite{Vicente:2019ykr} do not appear in our model.
 \subsection{Decays $\mu\rightarrow eee^+\equiv \mu \rightarrow 3e$. }
 The analytic formulas base on the analytical results  for non-supersymmetric contributions to LFV decays in the 2HDM presented in Refs.~\cite{Alonso:2012ji, Ilakovac:2012sh, Abada:2014kba},  and results  Refs.~\cite{delAguila:2008zu,delAguila:2019htj, Hernandez-Tome:2019lkb} which general formulas can be used for the 2HDM.  The $\mu$-e conversion rate for the ISS model were given in Ref.~\cite{Haba:2016lxc}.
  The partial decay width $\mu\rightarrow3e $ is taken from  Ref.~\cite{Ilakovac:2012sh}, namely only the non-suppersymmetric contributions are collected. The decay rate is
  \begin{align}
  \label{eq_brei3j}
  \mathrm{Br}(\mu\rightarrow3e)&=  \left\{ 2\left| \frac{1}{2} F^{\mu eee,LL}_{\mathrm{box}} +F^{\mu e,L}_Z -2s^2_W\left( F^{\mu e,L}_Z -F^{\mu e,L}_\gamma\right)\right|^2 + 4 s^4_W \left| F^{\mu e,L}_Z- F^{\mu e,L}_\gamma\right|^2
  \right. \crn& \left.+ 16 s^2_W \mathrm{Re}\left[ \left( F^{\mu e,L}_Z + \frac{1}{2} F^{\mu eee,LL}_{\mathrm{box}}\right) G^{\mu e,L*}_\gamma \right]  -48 s^4_W \mathrm{Re}\left[ \left( F^{\mu e,L}_Z  -F^{\mu e,L}_\gamma \right) G^{\mu e,L*}_\gamma \right]
  \right. \crn& \left.+  32 s^4_W \left| G^{\mu e,L}_{\gamma}\right|^2 \left[ \ln\frac{m^2_{\mu}}{m^2_e} -\frac{11}{4}\right]   \right\} \times \frac{\pi \alpha^2_{W}}{64 },
  \end{align}
  where $\Gamma_{\mu}=\alpha_W^2 m^5_{\mu}/(384\pi m^4_W)$, $\alpha_W=g^2/(4\pi)$, and the loop functions are listed in the appendix~\ref{app_eijkl}.  In the limit of the ISS model, the Eq.~\eqref{eq_brei3j} is consistent with that given in Refs.~\cite{Ilakovac:1994kj,Alonso:2012ji}. Apart from the gauge boson and Higgs  contributions taken from Refs.~\cite{Ilakovac:1994kj,Alonso:2012ji}, the Higgs contributions were checked with the results given in Refs.~\cite{Arganda:2005ji, Toma:2013zsa}.
\subsection{$\mu-e$ conversion in nuclei}
Based on the results given in Refs.~\cite{Alonso:2012ji,Ilakovac:2012sh,Popov:2013xaa}, we collect  all one-loop non-supersymmetric contributions to the $\mu-e$ conversion rates, see the detailed  formulas listed in appendix~\ref{app_eijkl}.   In the model under consideration  the $\mu-e$ conversion rate $R^J_{\mu\rightarrow e}$ in a nuclei $J$ consisting of $Z$ protons and $N$ neutrons is
\begin{align}
\label{eq_RmueISS2}
R(J)\equiv R^J_{\mu \rightarrow e}=\frac{G_F^2 \alpha_W^2 \alpha^3 m^5_{\mu}}{8\pi^4 \Gamma_{capt}} \times \frac{Z^4_{eff}}{Z} F^2_p  \left( \left|Q^L_W\right|^2 +\left|Q^R_W\right|^2\right),
\end{align}
where $Q^X_W=(2Z+N) V^X_u +(Z+2N) V^X_d$,
 $X=L,R$.     $V^{L}_{u,d}$ is defined as follows
\begin{align}
\label{eq_VLRq}
V^{L}_q&= Q_q s^2_W \left( F^L_{\gamma, \mu e} +G^R_{\gamma,\mu e}\right) + F^L _{Z,\mu e} \left( \frac{1}{2}I^3_q -Q_q s^2_W\right) +\frac{1}{4} F^{\mu eqq}_{\mathrm{Box}}, \; q=u,d,  \crn
V^{R}_q&= Q_q s^2_W \left( F^R_{\gamma, \mu e} +\frac{m_e}{m_{\mu}}G^L_{\gamma,\mu e}\right) + F^R _{Z,\mu e} \left( \frac{1}{2}I^3_q -Q_q s^2_W\right),
\end{align}
where $Q_q$ and $I_3$ are the electric charge and iso spin of the quark $q=u_a,d_a$.
To determine $V^X_q$ with $X=L,R$,  we just consider the limit of the 2HDM type II so that the couplings of the charged Higgs bosons with all quarks in our model are the same as those given in Ref.~\cite{Ilakovac:2012sh}.  Well-known values of $Z_{eff}$, $F_p$, and $\Gamma_{capt}$ corresponding to various nuclei are given in table~\ref{t_mueparameter}~\cite{Alonso:2012ji,Sun:2020puo}.
\begin{table}[ht]
		\caption{Effective atomic charges, nuclear form factors and capture rates, $N=A-Z$.} \label{t_mueparameter}
\begin{tabular}{cccccc}
	\hline
${^A_Z}J$	& $Z_{eff}$& $F_p(-m_{\mu}^2)$& $\Gamma_{capt}(10^6s^{-1})$ [$10^{-18}$ GeV] & $R_{\mu\rightarrow e}^{J}$& $R_{\mu\rightarrow e}^{J}$ \\
	&&&&  (current bound) & (future sensitivity)\\
\hline
$^{17}_{13}$Al& $11.5$ & $0.64$& $0.7054$ [$ 0.4641$] && $10^{-16}-10^{-17}$\\
$^{48}_{22}$Ti& $17.6$& $0.54$& $2.59$ [$ 1.7042$]&$\le4.3\times  10^{-12}$& $10^{-18}$\\
$^{197}_{79}$Au& $33.5$& $0.16$& $13.07$ [$8.5987 $]& $\le7\times  10^{-13}$&  \\
$^{208}_{82}$Pb& $34.0$& $0.15$& $13.45$ [$ 8.8487$] & $\le4.3\times  10^{-11}$&\\
\hline
\end{tabular}
\end{table}
 The form factors $F^X_{\gamma,\mu e}$, $F^X_{Z,\mu e}$, $G^X_{\gamma,\mu e}$, and $F^{\mu e qq}_{\mathrm{box}}$ are collected in appendix~\ref{app_eijkl}.

\subsection{LFV decays of the SM-like Higgs boson $h\rightarrow\,e_ae_b$}
\begin{figure}[h]
  \centering
 \includegraphics[width=15cm]{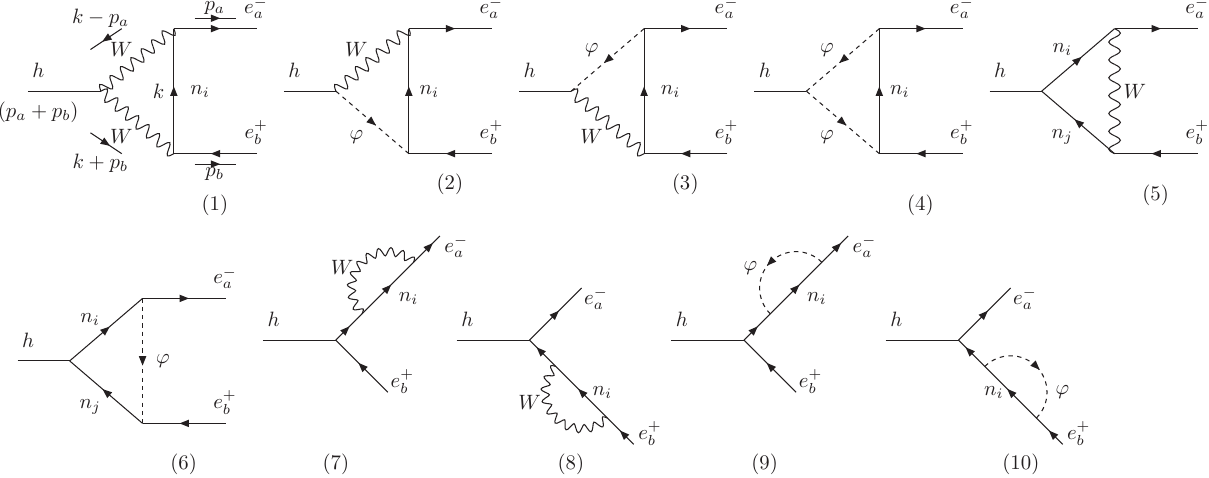}\\
 \caption{One-loop Feynman diagrams for $h\rightarrow e_ae_b$ in the unitary gauge.}\label{hlilj1}
\end{figure}

In the unitary gauge, the Feynman diagrams corresponding to one-loop contributions to the LFV decay amplitudes of the SM-like Higgs boson $h\rightarrow e_ae_b$ are shown in Fig.~\ref{hlilj1}.
The effective Lagrangian of  the decay   is  written as
$ \mathcal{L}^{\mathrm{LFV}h}= h \left(\Delta_L \overline{e_a}P_L e_b +\Delta_R \overline{e_a}P_R e_b\right) + \mathrm{H.c.}$,
  where   $\Delta_{(ba)L,R}$ are scalar factors arising from the loop contributions.
The partial width of  the decay is
\be
\Gamma (h\rightarrow\,e_a e_b)\equiv\Gamma (h\rightarrow e_a^{-} e_b^{+})+\Gamma (h\rightarrow e_a^{+} e_b^{-})
=  \fr{ m_{h} }{8\pi }\left(\vert \Delta_{(ba)L}\vert^2+\vert \Delta_{(ba)R}\vert^2\right), \label{LFVwidth}
\ee
 with the condition  $m^2_{h}\gg m^2_{a,b}$.
 The  LFVHD decay rate  is  Br$(h\rightarrow e_ae_b)= \Gamma (h\rightarrow \,e_ae_b)/\Gamma^{\mathrm{total}}_{h}$ where $\Gamma^{\mathrm{total}}_{h}\simeq 4.1\times 10^{-3}$ GeV is the SM value.  The deviation of $\Gamma^{\mathrm{total}}_{h}$ from SM  is small with   $|s_{\delta}|\le0.05$ defined in Eq.~\eqref{eq_delta}, hence we can ignore this change in our numerical investigation. In notations constructed in \cite{Hue:2015fbb}, the $\Delta_{(ba)L,R}$ can be written as
\begin{equation}
 \Delta_{(ba)L,R} =\sum^{10}_{i=1} \Delta^{(i)}_{(ba)L,R},  \label{deLR}
\end{equation}
where $\Delta^{(i)}_{(ba)L,R}$ is listed in the Appendix \ref{DeltaLR}. Detailed calculations are based on  \cite{Hue:2015fbb}, where  modifications have been made in appendix~\ref{DeltaLR} to make the Passarino-Veltman (PV) functions to be consistent with LoopTools.  In the SM limit given by Eq.~\eqref{eq_delta}, formulas relating contributions from only $W$ boson  are consistent with those shown in~\cite{Thao:2017qtn} calculated in the unitary gauge, and  consistent with previous  analytic formulas performed in the 't Hooft-Feynam gauge~\cite{Arganda:2004bz}.
\section{ \label{sec_Numerical} Numerical investigation of LFV decays}
\subsection{Setup parameters}
In this section we will apply the allowed regions of the parameter space mentioned above to investigate the LFV decays. First, we discuss the independent parameters needed to determine numerically the masses and mixing matrix of all neutrinos from the total mass matrix given in Eq~\eqref{Lnumass}.  The five independent parameters $s_{13}$, $\delta$, $\phi_1$, $\kappa$ and $m_0$ will be constrained from the experimental data, as we have presented precisely.  In exact numerical calculation using the direct neutrino mass matrix~\eqref{Lnumass},  more unknown parameters are  $m_D\equiv v f$, $t_{\beta}$,  and $M_0$.

Apart from the free parameters mentioned above, the are unknown parameter relating with the Higgs sector. In particularly,  contributions of charged Higgs bosons $\varphi^{\pm}$ depends on $m_{\varphi}$, the   mixing angle $\delta$ between the CP-even neutral Higgs bosons, and four Higgs-self couplings $\lambda_{1,2,3,4}$ in coupling factor $h\varphi^+\varphi^-$. But all of them are not independent and we can choose the new independent parameters are $m_{h},\,m_{\varphi}$, and $\delta$ as we discussed previously.  The perturbative constraints for these Higgs self-couplings of the model are~\cite{Gunion:2002zf, Chen:2018shg,Ginzburg:2005dt}:
\begin{align}
\label{eq_laConstraint}
0&<\lambda_{1,2},\; |\lambda_{3}|<4\pi,\; \lambda_3 +\sqrt{\lambda_1\lambda_2}>0,\;  \lambda_3 +\sqrt{\lambda_1\lambda_2}>0.
\end{align}
We fix $m_{h}=125.09$ GeV.  The Dirac mass scale $m_D=vf<174\times\sqrt{4\pi}\simeq 616$ GeV. The heavy neutrino masses are originated from the $A_4$ breaking scale $M_0$  hence they can be very large. This situation is completely different from that discussed previously in Refs.~\cite{Arganda:2014dta}, where heavy neutrino masses are bounded from above because of the  Casas-Ibarra parametrization~\cite{Casas:2001sr} specializing the structure of Dirac mass term  $m_D$,  resulting in  the perturbative limit of the Yukawa coupling. In the numerical investigation, we require $M_0/(m_Ds_{\beta})\ge 10\gg1$,   necessary to obtain the consistent ISS relations given in Eqs.~\eqref{R1} and \eqref{eq_heavyN}.

Discussions on the lower bounds of the charged Higgs boson in  2HDMs were discussed on Ref.~\cite{Arbey:2017gmh}.  Constraints on the $\varphi^\pm\rightarrow \gamma W^{\pm}$ decay and the $STU$ parameters ~\cite{Song:2019aav}.   Recent experimental data of  charged Higgs decays  $\varphi^+\rightarrow\bar{q}q', \bar{\ell}\nu_{\ell}$~\cite{Sirunyan:2018dvm}.  The model has non zero coupling $hW^\pm \varphi^{\mp}$, which predicts a decay $\varphi^{\pm}\rightarrow hW^{\pm}$ having $\Gamma(\varphi^{\pm}\rightarrow hW^{\pm})\sim \sin^2\delta$. In the alignment limit, $\delta\rightarrow0$, this decay channel vanishes. When  $0\ne \delta \ll1$, the recent experimental constraint must be considered.  A global fit on the 2HDM was discussed on Ref.~\cite{Haller:2018nnx}.

For the case of the 2HDM type-II, based on the recent results given in Ref.~\cite{Chen:2018shg}, the allowed regions of parameters for $m_{\varphi}\le 2$ TeV are: $ 0.2\le t_{\beta}\le5,\; |s_{\delta}|\le 0.008,\; \lambda_4 v< 200\; \mathrm{GeV}$.
 On the other hand, in the case of $\lambda_4=0$ and $s_{\delta}\rightarrow 0$, the values of $t_{\beta}$ is relaxed to a large values of $t_{\beta}\sim50$. In the study on the parameter space corresponding  more heavy masses of the Higgs bosons~\cite{Kling:2018xud},  the large $t_{\beta}$ values  are still allowed.
For the case of the 2HDM type-I, based on the recent results given in Ref.~\cite{Chen:2019pkq}, the allowed regions of parameters are: $ t_{\beta}<3,\; |s_{\delta}|\le 0.05,\; \lambda_4 v< 200\; \mathrm{GeV}$.

In the  numerical investigation for studying the LFV phenomenology using the allowed  values of  the set $\{s_{13},\delta,  m_0,\kappa,\phi_1\}$ that obtained from scanning the ranges given in Eq.~\eqref{eq_Input1},  the remaining unknown parameters as follows. For the cLFV processes, related unknown parameters are $m_D$, $M_0$, and $t_{\beta}$, which do not depend on the quark couplings, i.e. independent with which type-I or II of the 2HDM.
While parameter space of the model under consideration are constrained by recent experimental data of  Br$(\mu\rightarrow \gamma)$ and Br$(\mu\rightarrow3e )$, the two other decay rates   Br$(\tau \rightarrow\mu\gamma, e\gamma)$ seem much smaller. The interesting possibility is they  may be large enough to be detected by the future experimental sensitivities of the order $\mathcal{O}(10^{-9})$.  By collecting only points that satisfy all of the  conditions $10^{-14}\le \mathrm{Br}(\mu\rightarrow e\gamma)<4.2\times 10^{-13}$,   $\mathrm{Br}(\mu\rightarrow3e)<10^{-12}$,  and  max$\left[\log[\mathrm{Br}(\tau\rightarrow \mu \gamma)],\; \log[\mathrm{Br}(\tau\rightarrow e \gamma) \right] $ is as large as possible, we find a requirement that $m_D\ge \mathcal{O}(1)$ GeV which will be  paid attention to  in the following discussion.   The unkonwn parameters  are scanned in the following  ranges
\begin{align}
\label{eq_scanmdM0}
&100 \,\mathrm{ GeV}\le m_D \le 600 \mathrm{ GeV},\; 10\le \frac{M_0}{m_Ds_{\beta}}\le 100, \crn &0.02\le t_{\beta}\le 50, \; 700\; \mathrm{GeV}\leq m_{\varphi} \leq 10^4 \;\mathrm{GeV},
\end{align}
 where the second constraint  bases on the ISS condition~\eqref{eq_heavyN}, namely $|R_{ai}|\ll1$ for all $a=1,2,3$, $i=1,2,..,6$.

 To looking for  regions of the parameter space allows large Br$(h^0_1\rightarrow\tau\mu,\tau e)$ and satisfy the constraints Br$(\mu \rightarrow e\gamma)<4.2\times 10^{-13}$ and Br$(\mu \rightarrow 3e) <10^{-12}$, we need to scan  $s_{\delta}$ in the range $|s_{\delta}|\le 0.05\;(0.008)$ for model type-I (II) and adding a requirement given in Eq.~\eqref{eq_laConstraint}. We note that the small ranges of $m_D$ GeV  were scanned but they give  suppressed values of  Br$(h^0_1\rightarrow e_a e_b)$.  In the numerical investigation, we just collect points satisfying max[Br$(h^0_1\rightarrow \mu\tau,e\tau)\ge 10^{-11}$.

To discuss on the $\mu$-e conversion rates predicted by the 2HDM type-II  only the following allowed range of $t_{\beta}$ is considered:  $ 0.3\leq t_{\beta}\leq 50$. The most interesting allowed range $0.2\le t_{\beta}\le 5$ given in Ref.~\cite{Chen:2018shg}  will also be remarked.
\subsection{Numerical results}
\subsubsection{LFVHD and cLFV processes}
\begin{figure}[ht]
	\centering
	\begin{tabular}{cc}
		\includegraphics[width=7cm]{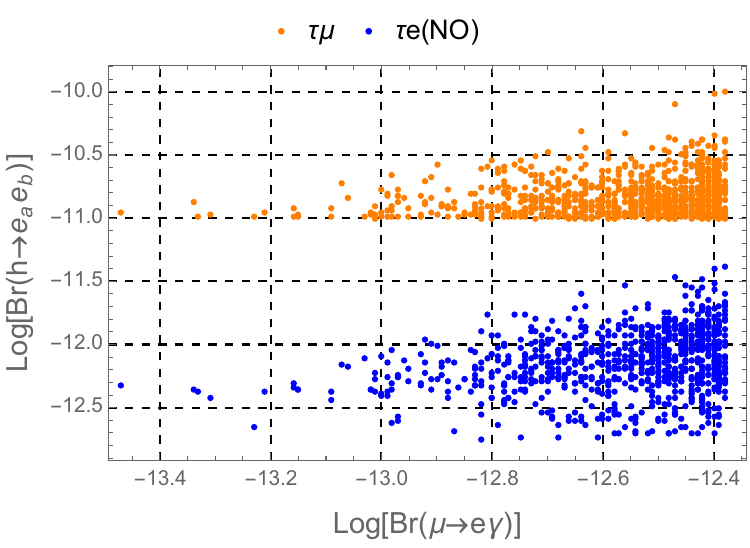}	&	
	\includegraphics[width=7cm]{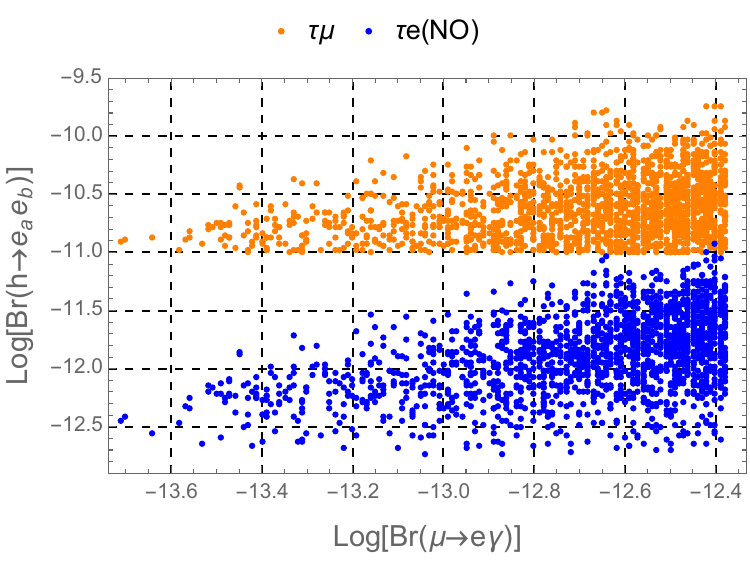}	
		\\
	\end{tabular}
	\caption{The correlation between LFVHD vs. Br$(\mu\rightarrow e\gamma)$ in the NO scheme corresponding two ranges  $0.02\le t_{\beta}\le3.4$ (left) and $0.3\le t_{\beta}\le 50$ (right).}\label{fig_LFVplot2}
\end{figure}

The correlation between Br$(h^0_1\rightarrow e_a e_b)$ and Br$(\mu\rightarrow e\gamma)$ are plotted in the Fig.~\ref{fig_LFVplot2} for the NO scheme.
Only points satisfying  both conditions Br$(\mu\rightarrow e\gamma)<4.2\times 10^{-13}$ and max[Br$(h^0_1\rightarrow e_a e_b)]\ge 10^{-11}$ are collected.  The similar results are found for  the IO scheme, hence we will not show here.
%
Correspondingly, largest values of LFVHD is the Br$(h^0_1\rightarrow \tau\mu)$ satisfying  $10^{-10}\le \mathrm{max}[\mathrm{Br}(h^0_1\rightarrow \tau\mu)]< 10^{-9}$, but  Br$(h^0_1\rightarrow \tau e)\le10^{-11}$.
Both schemes always result in suppressed values of Br$(h^0_1\rightarrow \mu e)$, hence it will not be discussed. The regions corresponding to  the current constraint of Br$(\mu\rightarrow e\gamma)$ give largest values of LFVHD satisfying  $\mathrm{Br}(h^0_1\rightarrow\tau \mu)\le \mathcal{O}(10^{-10})$, and  Br$(h^0_1\rightarrow\tau e)\le \mathcal{O}(10^{-11})$.  These values are much smaller than the values $\mathcal{O}(10^{-7})$ predicted previously by  other models with ISS neutrinos where LFVHD arises from loop corrections. These difference can be explained by the particular parameterization the total neutrino mass matrix. In the model under consideration,  the $A_4$ symmetry results in a very strict structure of the total neutrino mass matrix, which is the origin of the very strict relations between the cLFV and LFVHD decays.

\begin{figure}[ht]
	\centering
	\begin{tabular}{cc}
		\includegraphics[width=7cm]{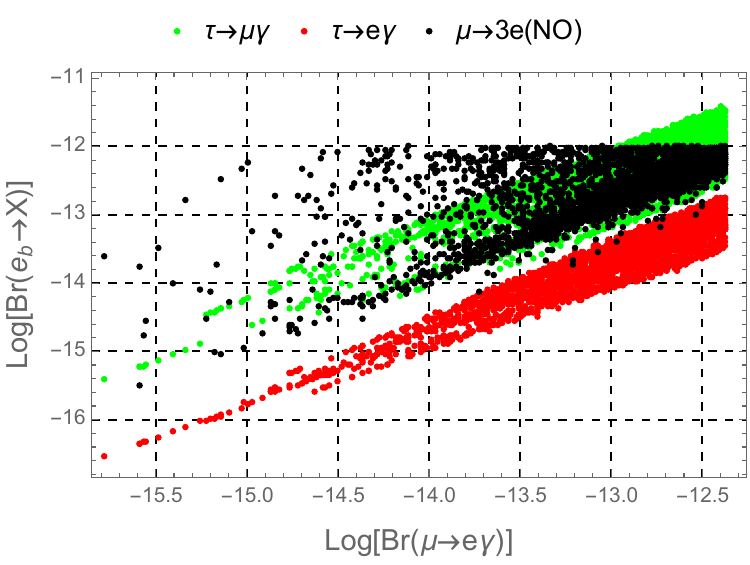}	&	
		\includegraphics[width=7cm]{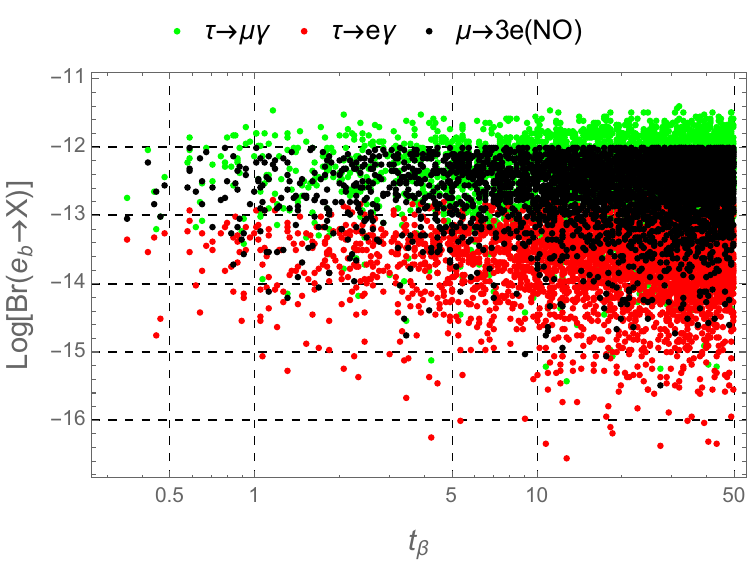}
		\\
	\end{tabular}
	\caption{cLFV rates as functions of Br$(\mu\rightarrow e\gamma)$ ($t_{\beta}$) in the left (right) panel for the NO scheme.}\label{fig_cLFV0}
\end{figure}
\begin{figure}[ht]
	\centering
	\begin{tabular}{cc}
		\includegraphics[width=7cm]{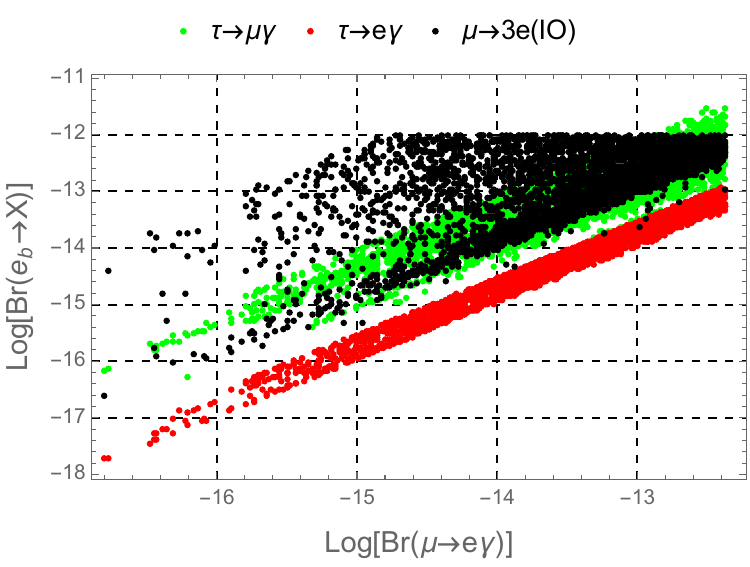}	&	
		\includegraphics[width=7cm]{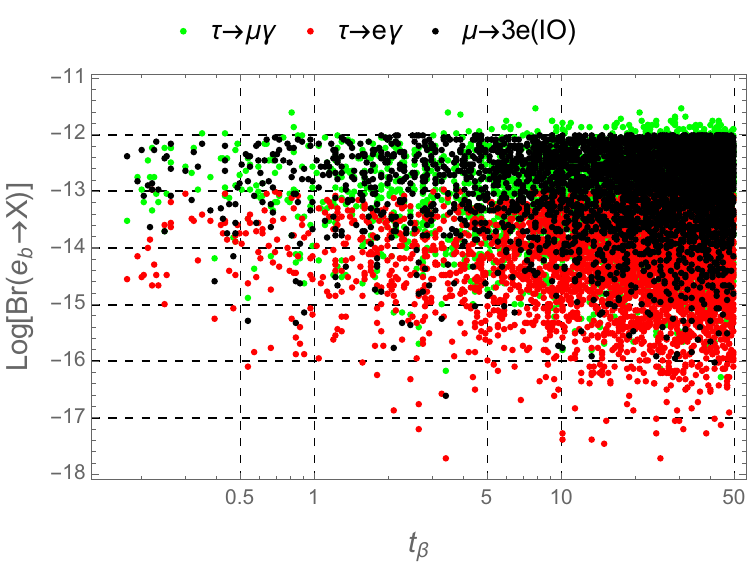}
		\\
	\end{tabular}
	\caption{ cLFV rates as functions of Br$(\mu\rightarrow e\gamma)$ ($t_{\beta}$) in the left (right) panel for IO scheme.}\label{fig_cLFV0I}
\end{figure}

We continue with the numerical results for the cLFV decays Br$(e_b\rightarrow X)$ with  $\{e_b,X\}=\{ \tau, \mu\gamma\},\; \{ \tau, e\gamma\}, \{ \mu,3e\}$. The results are shown in Fig.~\ref{fig_cLFV0} for the  NO  schemes.
 The similar results are found for the IO scheme, see  Fig.~\ref{fig_cLFV0I}.
 The common property for both schemes is that  the constraints  of Br$(\mu\rightarrow e\gamma)$ and Br$(\mu \rightarrow 3e)$ affect strongly on the two decays Br$(\tau\rightarrow \mu\gamma,e\gamma)$, leading to the following upper bounds  Br$(\tau\rightarrow \mu\gamma)<10^{-11}$ and Br$(\tau\rightarrow e\gamma)<10^{-12}$ ($10^{-13}$) for the NO (IO) scheme.

In conclusion for the cLFV decays, we find that Br$(\tau\rightarrow\mu\gamma,e\gamma)$ can reach the order $\mathcal{O}(10^{-11})$, much larger than Br$(\mu\rightarrow e\gamma)$.  But they can not large enough to  be observed  by experiments with  planned sensitivities of $\mathcal{O}(10^{-9})$. Our conclusion for  the LFVHD and cLFV decays  are completely different from the results predicted by the 2HDM type III, where  LFVHD appears at tree level and does not depend on the constraint of Br$(\mu\rightarrow e\gamma)$ ~\cite{Vicente:2019ykr, Hou:2020tgl,Crivellin:2019dun}.

\subsubsection{$\mu$-e conversions in nuclei predicted by  the 2HDM type-II. }
\begin{figure}[ht]
	\centering
	\begin{tabular}{cc}
		\includegraphics[width=7cm]{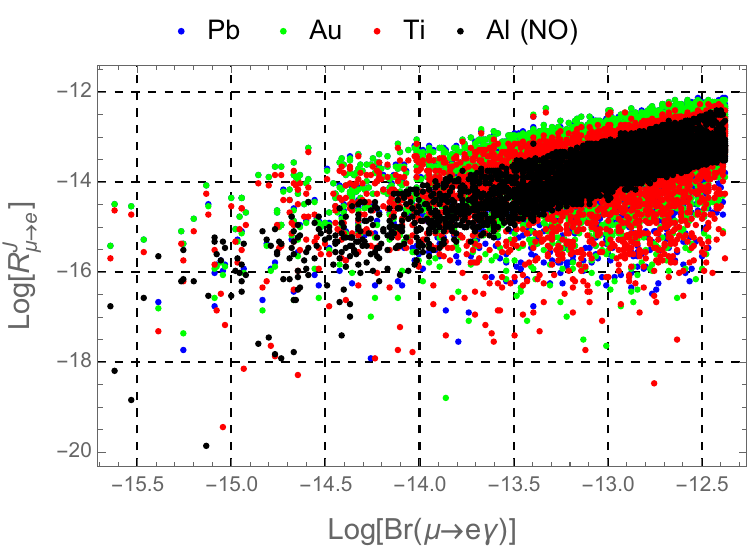}	&	 \includegraphics[width=7cm]{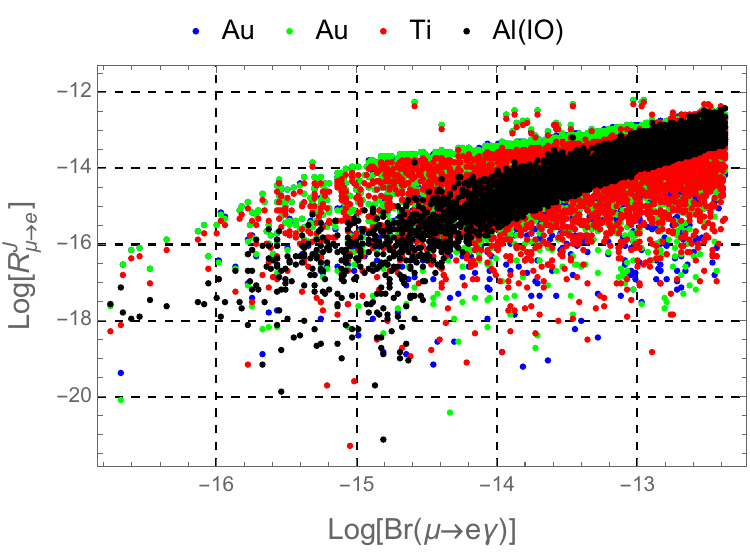}\\
	\end{tabular}
	\caption{Correlations between log[Br($\mu\rightarrow 3e$)] and different $\mu$-e conversion rates for  $0.3\le t_{\beta}\le50$.}\label{fig_mueconversion}
\end{figure}

Regarding to the $\mu-e$ conversion corresponding to the 2HDM type-II, with  $0.3\le t_{\beta}\le 50$.  The correlations  between  Br$(\mu \rightarrow e\gamma)$  and the four $\mu$-e conversions  rates  are  shown precisely  in Fig.~\ref{fig_mueconversion}.
All  $\mu$-e conversion rates are constrained strictly by the data of the decay $\mu\rightarrow \gamma$. They decrease with smaller Br$(\mu\rightarrow \gamma)$. The allowed regions corresponding to the three nuclei Ti, Au, and Pb are nearly the same, while the allowed region for Al is more narrow. The recent constraint of Br$(\mu\rightarrow e\gamma)$ gives an upper bound of $10^{-12}$ for all $\mu$-e conversions rates. The planned sensitivity of $6\times 10^{-14}$ can give the upper bound of $10^{-13}$.

\begin{figure}[ht]
	\centering
	\begin{tabular}{cc}
		\includegraphics[width=7cm]{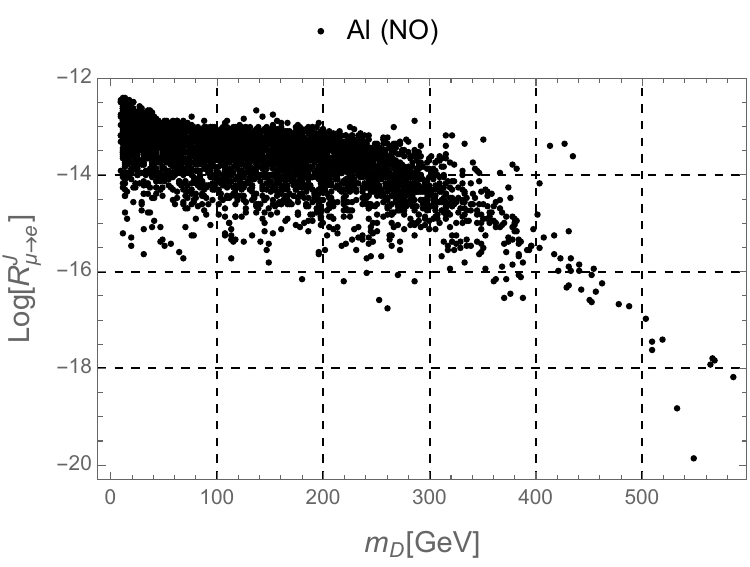}	&	 \includegraphics[width=7cm]{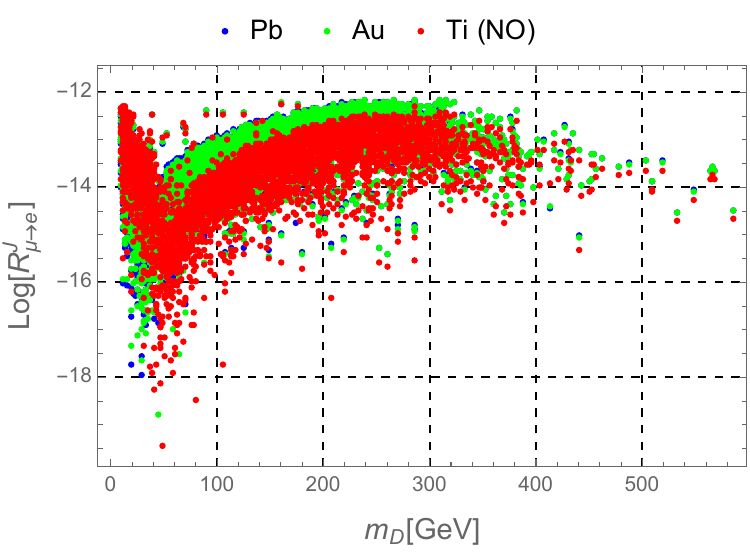}\\
	\end{tabular}
	\caption{Correlations between $m_D$ and different $\mu$-e conversion rates.}\label{fig_muemDNO}
\end{figure}

The results in this case for both NO and IO schemes are nearly the same, hence we just consider the NO scheme in the below discussion. The allowed regions of the parameter space does not changes significantly with $t_{\beta}$ and $m_{\varphi}$. On the other hand, these regions  change for different $\mu$-e conversion rates. Their shapes are the same with $m_D$ and $M_0$. The dependence of $\mu$-e rates on $m_D$ is shown in  Fig.~\ref{fig_muemDNO} for the NO scheme.
The allowed regions  of R$(Al)$ are different from the remaining conversion rates because of  $Z-N$ is positive for Al, in contrast with the three remaining nuclei. Hence the combining the  $\mu$-e results  conversion rates  will give more strict values of $m_D$.

\begin{figure}[ht]
	\centering
	\begin{tabular}{cc}
	 \includegraphics[width=7cm]{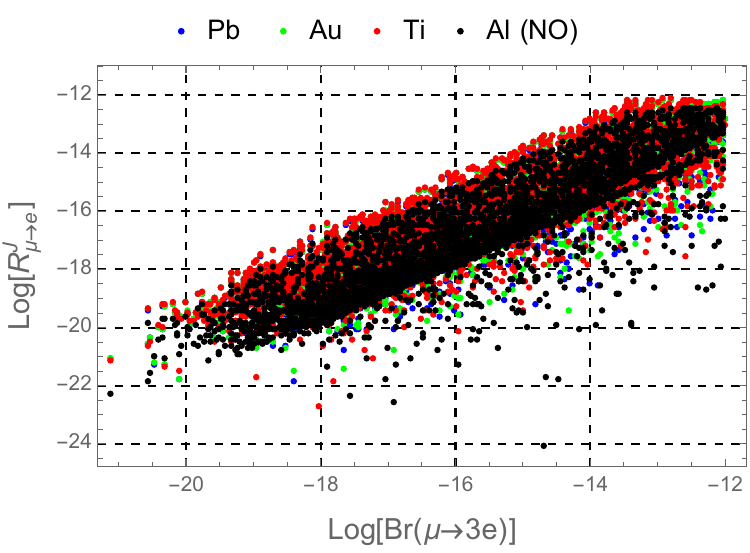}&	\includegraphics[width=7cm]{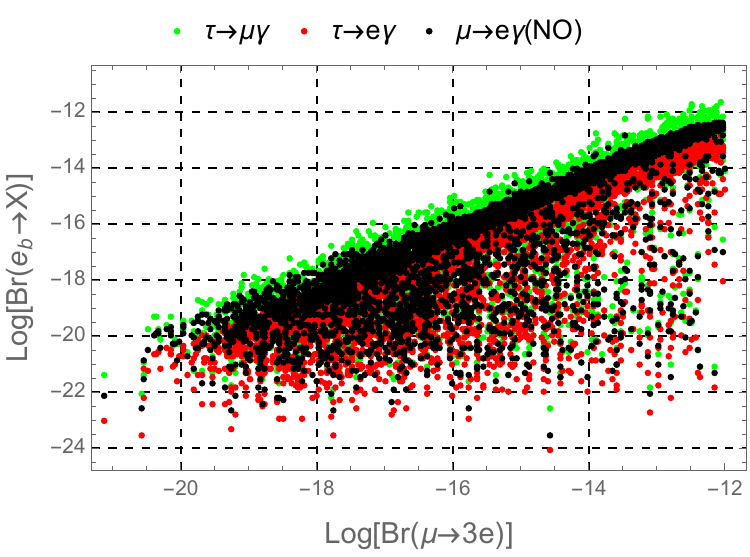}\\
	\end{tabular}
\caption{Correlations between log[Br($\mu\rightarrow 3e$)] and different $\mu$-e conversion rates (cLFV rates) in the left (right) panel.} \label{fig_mueconversion3e}
\end{figure}

Finally,  we remind  that the above allowed regions of parameters satisfy the recent experimental bound  $(\mu\rightarrow 3e)<10^{-12}$.  If this channel is not observed with planned sensitivity of $10^{-16}$, the upper bounds of cLFV decays and  or $\mu$-e conversion rates will be more suppressed, see illustration in  Fig.~\ref{fig_mueconversion3e} for the NO scheme,
where correlations between  Br$(\mu \rightarrow 3e)$  and other cLFV decays the four $\mu$-e conversions  rates  are  shown precisely. The new ranges of $m_D$ and $M_0$ are $10^{-2}\; \mathrm{GeV}\le m_D\le 600 $ GeV and $10\le \frac{M_0}{m_Ds_{\beta}}\le 10^3$. The respective constraints of the cLFV and $\mu$-e conversion rates are: Br$(\tau\rightarrow \mu \gamma),\; R(J)\le 10^{-15}$, and  Br$(\tau,\mu\rightarrow e \gamma)\le  10^{-16}$. In other word, planned experimental upper bound of  Br$(\mu\rightarrow3e)$  predicts more strict upper bounds of Br$(e_b\rightarrow e_a\gamma)$ than those from the respective experimental sensitivities.

\section{\label{sec_con} Conclusion}
In this work, we have introduced the $A_4$ISS model  to explain the recent experimental data of neutrino oscillation. The  total neutrino mass matrix arises from the $A_4$ symmetry,  resulting in the ISS form of the active neutrino mass matrix.  All masses and mixing parameters of the active neutrino corresponding to the oscillation data  were formulated as functions of five independent parameters: $(s_{13}, \delta, \phi_1, \kappa, m_0)$. We have determined all allowed ranges of these parameters satisfying $3\sigma$ experimental neutrino oscillation data. From this,  the model predicts the two following ranges of the $m_{\beta}$ and $\langle m\rangle$ and the following: i) $0.01\;\mathrm{eV}\le m_{\beta}\le 0.15\;\mathrm{eV}$ and $0.005 \;\mathrm{eV}\le |\langle m\rangle| \le 0.09\;\mathrm{eV}$ for the NO scheme, ii)  $0.05\;\mathrm{eV}\le m_{\beta}\le 0.17\;\mathrm{eV}$ and $0.015\;\mathrm{eV}\le |\langle m\rangle| \le 0.1\;\mathrm{eV}$ for IO scheme.  These ranges  can be observed by the forthcoming experiments. More interesting,  the two allowed regions of  these two quantities predicted by the two  NO and IO schemes  shown in the  right panel of Fig.~\ref{fig_mbeta}  are very narrow and nearly distinguishable. As a result, this $A_4$ISS model  will predict which IO or NO schemes is realistic or  the model is ruled out, once both quantities $m_{\beta}$ and $|\langle m\rangle|$ are observed by upcoming experiments.

The active and heavy ISS neutrinos in the $A_4$ISS model result in  cLFV and $\mu$-e conversion nuclei from loop corrections. Numerical results indicated that  Br$(\mu\rightarrow e\gamma,3e)$ give strong upper bounds on other cLFV decays and $\mu$-e conversion rates.  The recent experimental constraints of  Br$(\mu\rightarrow e\gamma,\;3e)$ result in the very suppressed decay rates of Br$(h\rightarrow e_ae_b)<\mathcal{O}(10^{-9})$ and Br$(\tau\rightarrow \mu\gamma,e\gamma)<10^{-11}$, which are much smaller than the planned experimental sensitivities. In the other side, the $\mu$-e conversion rates still reach the orders $\mathcal{O}(10^{-12})$, which is in the observable ranges of experiments. The planned sensitivity of Br$(\mu\rightarrow 3e)\sim \mathcal{O}(10^{-16})$ gives much stronger constraints on  the  cLFV processes that can not be observed, including Br$(\mu\rightarrow e\gamma)<10^{-16}$. The promising signals now are the $\mu$-e conversion rates of  Al and Ti. In conclusion, the $A_4$ISS we introduced here is very predictive. The reality of the model and many interesting predictions on the observable quantities such as $m_{\beta}$, $|\langle m\rangle|$,  cLFV decay rates, and $\mu$-e conversion rates in nuclei will be confirmed or ruled  out by the upcoming experiments.

\section*{Acknowledgments}
This research is funded by Vietnam  National Foundation for Science and Technology Development (NAFOSTED) under grant number  103.01-2018.331. L. T. Hue is thankful to Van Lang University.
\appendix
\section{\label{A4rules}$A_4$ group in the   AF (Altarelli-Feruglio) basis}
The non-Abelian $A_4$ is a group of even permutations of 4 objects and has $4!/2=12$ elements, see a review in~\cite{Altarelli:2010gt, Ishimori:2010au,King:2011zj}. The group is generated by two generators $S$ and $T$ satisfying the relations $S^2 = (ST)^3 = (T^3) = 1$.
The group has  four irreducible representations (rep.), including   three one-dimensional and one three-dimensional ones which are denoted as  $\underline{1}$, $\underline{1'}$, $\underline{1''}$, and $\underline{3}$, respectively.  The multiplication rules for them are as follows
\begin{align} \label{eq_A4product}
\underline{1}\times\underline{R}& = \underline{R} \times \underline{1}= \underline{R}, \; \underline{1'}\times\underline{1''} = \underline{1}, \; \underline{1'}\times\underline{1'} = \underline{1''},\; \underline{1''}\times\underline{1''} = \underline{1'}, \crn  \underline{3}\times\underline{3} &= \underline{3} + \underline{3}_A +\underline{1} + \underline{1'} + \underline{1''},
\end{align}
where $\underline{3}$ and $\underline{3}_A$ imply  the symmetric and anti-symmetric forms of the respective  Clebsch-Gordan coefficients, which particular formulas depend on the  choice of $T$ and $S$. In this work, the three-dimensional unitary representations of $T$ and $S$ are~\cite{Feruglio:2008ht,Feruglio:2009hu}.  Correspondingly,  the Clebsch-Gordan coefficients obtained from tensor products of the two  $A_4$ triplets  $ a = (a_1, \; a_2, \; a_3)$ and $b=(b_1, \; b_2,\; b_3) \sim \underline{3}$ are
\begin{align}
\underline{1} \equiv (ab)_{\underline{1}} &= (a_1b_1+a_2b_3+a_3b_2),\crn
\underline{1'} \equiv (ab)_{\underline{1}'} &= (a_3b_3+a_1b_2+a_2b_1),\crn
\underline{1''} \equiv (ab)_{\underline{1}''}&= (a_2b_2+a_1b_3+a_3b_1), \crn
\underline{3} \equiv (ab)_{\underline{3}} &= \frac{1}{3}(2a_1b_1-a_2b_3-a_3b_2, 2a_3b_3-a_1b_2-a_2b_1, 2a_2b_2-a_1b_3-a_3b_1),\crn
\underline{3}_A \equiv (ab)_{\underline{3}_A} &= \frac{1}{2}(a_2b_3-a_3b_2, a_1b_2-a_2b_1, a_3b_1-a_1b_3).
\end{align}
Only rep. $\underline{3}$  is used for generating the neutrino mass matrix. In the mentioned basis, T is complex and $T^{*}\neq T$ in general so the complex conjugate representation $r^*$ of a representation $r$ ($r=\underline{1'},\; \underline{1''},\; \underline{3}$) is not the same as $r$, although they are all real reps.. It is determined by the following rules \cite{Feruglio:2008ht,Feruglio:2009hu}: $ c \sim \underline{1} \rightarrow c^* \sim \underline{1}$, $ c' \sim \underline{1'} \rightarrow {c'}^* \sim  {\underline{1'}}^*= \underline{1''}$, $ c'' \sim \underline{1'}  \rightarrow {c''}^* \sim  {\underline{1''}}^*= \underline{1'}$, and $a=(a_1,\; a_2,\; a_3)  \sim \underline{3} \rightarrow a^*=(a^*_1,\ a_3^*,\ a_2^*)\sim \underline{3}$.

 \section{\label{Hcondition} The total Higgs potential}
  The Higgs potential respecting all symmetries given in table~\ref{particle content} is
  \begin{align}
  \label{Hpotential}
  V_{Higgs} &= \mu _u^2h_u^\dag {h_u} + \mu _d^2h_d^\dag {h_d}  + \frac{ \lambda_1}{2}{\left( {h_u^\dag {h_u}} \right)^2}+ \frac{\lambda _2}{2}{\left( {h_d^\dag {h_d}} \right)^2}   + {\lambda _3}\left( {h_d^\dag {h_d}} \right)\left( {h_u^\dag {h_u}} \right) \nonumber\\
  & +\lambda _4 \left( {h_d^\dag {h_u}} \right)\left( {h_u^\dag {h_d}} \right) + \left\{ \mu^2_{ud}\left(i\sigma_2h_u\right)^T h_d  +\mathrm{h.c.}\right\} + \sum_{x=u,d} \sum_{H} \lambda ^{xH} \left( {h_x^\dag {h_x}} \right) \left( H^{\dagger} H \right)_1
  \crn
  & +  \mu _{\xi'}^2{{\xi '}^\dag }\xi ' + \mu_{\xi''}^2{{\xi ''}^\dag }\xi '' +
  {\lambda ^{\xi '\xi ''}}\left( {\xi '\xi ''} \right)\left( {{{\xi '}^\dag }{{\xi ''}^\dag }} \right)  + {\lambda ^{\xi '}}{\left( {{{\xi '}^\dag }\xi '} \right)^2} + {\lambda ^{\xi ''}}{\left( {{{\xi ''}^\dag }\xi ''} \right)^2} \nonumber\\
  & + V\left(\phi_T,\phi_S \right)  + V\left( {{\phi _T},\xi '} \right) +V\left( {{\phi _T},\xi ''} \right) +
  V\left( {{\phi _S},\xi '} \right) + V\left( {{\phi _S},\xi ''} \right) \nonumber\\
  & + V\left( {{\phi _T},{\phi _S},\xi '} \right) + V\left( {{\phi _T},{\phi _S},\xi ''} \right)+V\left( {{\phi _T},\xi ',\xi ''} \right) + V\left( {{\phi _S},\xi ',\xi ''} \right),
  \end{align}
where  $H=\xi',\xi'',\phi _S,\phi _T$, and
\begin{align}
V\left( \phi_S, \phi_T \right) &=  
\mu_3^2\left(\phi_S^\dag {\phi_S} \right)_1 + \mu_4^2\left(\phi_T^\dag \phi_T \right)_1  + \sum_{x=S,T} \lambda^{x}\left[ (\phi_x\phi_x^\dag)^2 \right]_1  + \lambda^{ST}\left(\phi_S^\dag \phi_S\phi_T^\dag \phi_T \right)_1,
\crn  V\left(\phi_T,\xi'\right) &=  \lambda^{T\xi'} \left({\xi'}^\dag \xi ' \right) \left(\phi_T^\dag \phi_T \right)_1, 
 \crn
  V\left(\phi_T,\xi'' \right) &=  \lambda^{T\xi ''} \left({\xi''}^\dag \xi''\right) \left( \phi_T^\dag {\phi_T} \right)_1 , \crn
  V\left(\phi_S,\xi' \right) &=  \lambda_2^{S\xi'} {\xi'}^\dag\xi' \left(\phi_S^\dag {\phi_S} \right)_1 +   
  \left\{\lambda_1^{S\xi'}\xi'^2{\left(\phi_S^{\dag2}  \right)}_{1'} +   \lambda_3^{S\xi'}{\xi'}^\dag{\left[ {(\phi^2_S)}_{3s}\phi _S^\dag \right]}_{1'} + h.c. \right\},\crn
  V\left(\phi _S,\xi'' \right) &=  \lambda_2^{S\xi''}{\xi''}^\dag \xi'' \left(\phi_S^\dag \phi_S \right)_1 + 
  \left\{ \lambda_1^{S\xi''} {\xi''}^{\dag2} \left( \phi^2 _S \right)_{1'} +  \lambda_3^{S\xi''}\xi''^{\dagger}\left[(\phi^2_S)_{3s} \phi_S^{\dagger}\right]_{1''} + h.c. \right\}, \nonumber\\
  V\left(\phi_T,\phi_S,\xi' \right) &= 
  \lambda_1^{TS\xi'}{\xi'}^\dag {\left[{\left(\phi_S \phi_T\right)}_{3a} \phi_T^\dag \right]}_{1'}  + \lambda_2^{TS\xi'}{\xi'}^\dag {\left[ {\left(\phi_S \phi_T\right)}_{3s} \phi_T^\dag \right]}_{1'} + h.c.,    \nonumber\\
  V\left(\phi_T,\phi_S,\xi'' \right) &= 
  \lambda_1^{TS\xi''}\xi''^{\dagger} {\left[{\left(\phi_S \phi_T \right)}_{3a} \phi_T^\dag \right]}_{1'} + \lambda_2^{TS\xi''}\xi''^{\dagger} {\left[{\left(\phi_S \phi_T \right)}_{3s}\phi_T^\dag \right]}_{1'} + h.c., \nonumber\\
  V\left(\phi_S,\xi',\xi''\right) &= \lambda^{S\xi'\xi''} \xi'{\xi''}^\dag {\left(\phi_S^\dag {\phi_S} \right)}_{1'} + h.c.,\crn
  V\left( \phi_T,\xi',\xi'' \right)&=  
  	\lambda^{T \xi' \xi''}\xi'{\xi''}^\dag {\left(\phi_T \phi_T^\dag \right)}_{1'} + h.c. ,
\end{align}
where $\left(\phi_S\phi_S\right)_{3a}=(0,0,0)$.

 There are ten neutral Higgs components which will result in ten equations corresponding to the minimal conditions presenting relations between the VEV pattern used in this work with the Higgs self-couplings. Because of the very large number of the Higgs self-couplings, the VEV pattern assumed in this work is easily guaranteed. Therefore, the lengthy and unnecessary minimal conditions will not be presented here.

In general, the squared mass matrix of the CP-even Higgs bosons are $10\times 10$ matrix, where the main contribution to the SM-like Higgs boson arises from the two Higgs doublets $h_u$ and $h_d$. Hence, for simplicity in studying the LFV decay of the SM-lik Higgs boson, we will choose the regime  that these  Higgs doublets decouple to other Higgs singlets, namely
\begin{align}
\la^{u\xi'}&= \la^{d\xi'}= \la^{u\xi''}= \la^{d\xi''}= \la^{Tu}= \la^{Td} = \la^{Su} = \la^{Sd}=0. 
\end{align}
To generate non-zero masses for CP-odd neutral and charged Higgs  bosons we adopt a soft term breaking $Z_3 \times Z_{11}$ in the Higg potential, namely  $\mu^2_{ud}\left(i\sigma_2h_u\right)^T h_d +h.c..$ included in the second line of the Higgs potential~\eqref{Hpotential}. It results in that the masses and eigenstates of all  Higgs bosons arising from two Higgs doublet $h_u$ and $h_u$ are the same as those well-known in the 2HDM.

\section{\label{app_eijkl} Form factors contributing to  the LFV decay rates $e_b\rightarrow e_a\gamma$ ($b>a$),   $e_b\rightarrow 3e_a$ and $\mu-e$ conversion in nuclei}

The  one-loop three-point PV functions~\cite{Passarino:1978jh}, called $C-$ functions, which  specific definitions were given in Ref.~\cite{Nguyen:2017ibh}. For cLFV decay processes $e_b\rightarrow e_a\gamma$,  where are the  masses of  charged leptons  $m_{a,b}$ satisfy   $m^2_{a,b}/m_W^2\ll1$ and $m^2_{a,b}/m_{\varphi}^2\ll1$,  the $C$-functions are
\bea
C_0&=& \frac{t-1-t\ln t}{M_2^2(t-1)^2}, \;
C_1=C_2=- \frac{3t^2-4t+ 1-2t^2\ln t}{4(t-1)^3M_2^2},\crn
C_{11}&=& C_{22}=2C_{12}=\frac{11t^3-18t^2+ 9t -2 -6t^3\ln t}{18M_2^2(t-1)^4},
\label{nCf}\eea
where $t=M_1^2/M_2^2$. The value $t=1$ gives  $C_0=-1/(2M_2^2)$, $C_1=1/(6M_2^2)$,  and $C_{11}=-1/(12M_2^2)$.

Contributions from $W$ and $\varphi^\pm$ bosons to the Br$(e_b\rightarrow e_a\gamma)$ defined as  $D_{(ba)L,R}$  given in Eq.~\eqref{brlfvdecay2} are calculated based on the general form given in Ref.~\cite{Hue:2017lak}, where $C_{(ba)L,R}=\frac{g^2em_b}{32\pi^2 m_W^2}\times D_{(ba)L,R}$  and $C_{(ba)L,R}$ is calculated as follows
\begin{align}
C^{W}_{(ba)L} &= -\frac{ e g^2m_a}{32\pi^2m_W^2}\sum_{i=1}^9U^{\nu*}_{bi}U^{\nu}_{ai}\left[ 2( C_{12} + C_{22} -C_1) m_W^2 + m_b^2(C_{11} + C_{12} + C_1) \right.\crn
& \left. + m_{n_i}^2 (C_0 +C_1 +2 C_2+ C_{12} + C_{22} )\right], \crn
C^{W}_{(ba)R} &= -\frac{ e g^2m_b}{32\pi^2 m_W^2}\sum_{i=1}^9U^{\nu*}_{bi}U^{\nu}_{ai}\left[2( C_{11} + C_{12}- C_2) m_W^2+ m_a^2 (C_{12} + C_{22} + C_2) \right.\crn
& \left. + m_{n_i}^2(C_0 +  2C_1 +C_2+ C_{11} + C_{12} )\right]
\end{align}
with $C_{0,a,ab}=C_{0,a,ab}(m_{n_i},m_W,m_W)$,
and
\begin{align}
C^{\varphi}_{(ba)L} = -\frac{m_a e g^2}{32\pi^2m^2_W}
\sum_{i=1}^9U^{\nu*}_{bi}U^{L}_{ai}&\times \left\{ t_{\beta}^{2}m^2_{b}(C_1 +C_{11}+C_{12})\right.\crn
&+ \left. m^2_{n_i}\left[t_{\beta}^{-2}\left(C_{2}+C_{12} +C_{22}\right) -(C_0 +C_1+C_2) \right]\right\}, \crn
C^{\varphi}_{(ba)R} =-\frac{m_b e g^2}{32\pi^2m^2_W}\sum_{i=1}^9U^{\nu*}_{bi}U^{L}_{ai}&\times \left\{ t_{\beta}^{2}m^2_{a}(C_2 + C_{12}+C_{22})\right.\crn
&+ \left. m^2_{n_i}\left[t_{\beta}^{-2}\left(C_1 +C_{11}+C_{12}\right)  -(C_0 +C_1+C_2 )\right]\right\}
\end{align}
with $C_{0,a,ab}=C_{0,a,ab}(m_{n_i},m_{\varphi},m_{\varphi})$. In the limit $\frac{m_a^2}{m_W^2},\frac{m_b^2}{m_W^2}=0$,  $C^{W}_{(ba)L,R}$ is consistent with that given in \cite{Ibarra:2011xn,He:2002pva,Crivellin:2018qmi}. Also, with  $t_{\varphi,i}=\frac{m^2_{n_i}}{m^2_{\varphi}}$, $C^{\varphi}_{(ba)L,R}$ have consistent forms with Ref.~\cite{Crivellin:2018qmi}. 

Loop functions relating with only gauge bosons are \cite{Ilakovac:1994kj,Alonso:2012ji}:
\begin{align}
\label{eq_ 1loopG}
F_{\gamma}(x)&= \frac{x\left(7 x^2 -x-12\right)}{12(1-x)^3} -\frac{x^2\left(x^2 -10x +12\right)}{6(1-x)^4} \ln x, \crn
G_{\gamma}(x)&= -\frac{x(2x^2 +5x -1)}{4(1-x)^3} -\frac{3x^3}{2(1-x)^4} \ln x,\crn
F_Z(x)&= -\frac{5x}{2(1-x)} -\frac{5x^2}{2(1-x)^2} \ln x, \crn
G_Z(x,y)&= -\frac{1}{2(x-y)}\left[ \frac{x^2(1-y)}{1-x}\ln x - \frac{y^2(1-x)}{1-y}\ln y\right], \crn
H_Z(x,y)&= \frac{\sqrt{xy}}{4(x-y)} \left[ \frac{x^2-4x}{1-x}\ln x -\frac{y^2-4y}{1-y}\ln y\right], \crn
F_{\mathrm{box}}(x,y)&= \frac{1}{x-y}\left\{ \left( 4 + \frac{xy}{4}\right)\left[ \frac{1}{1-x} +\frac{x^2}{(1-x)^2}\ln x  -\frac{1}{1-y} -\frac{y^2}{(1-y)^2}\ln y\right]
\right.\crn&\left. -2xy\left[ \frac{1}{1-x} +\frac{x}{(1-x)^2}\ln x -\frac{1}{1-y} -\frac{y}{(1-y)^2}\ln y\right]\right\}, \crn
F_{X\mathrm{box}}(x,y)&= -\frac{1}{x-y}\left\{ \left(1 +\frac{xy}{4}\right)\left[ \frac{1}{1-x} +\frac{x^2}{(1-x)^2}\ln x  -\frac{1}{1-y} -\frac{y^2}{(1-y)^2}\ln y\right]
\right.\crn&\left. -2xy\left[ \frac{1}{1-x} +\frac{x}{(1-x)^2}\ln x -\frac{1}{1-y} -\frac{y}{(1-y)^2}\ln y\right]\right\}
\crn G_{\mathrm{box}}(x,y)&= \frac{-\sqrt{xy}}{x-y} \left\{ \left(4 +xy\right)\left[ \frac{1}{1-x} +\frac{x\ln x}{(1-x)^2}  -\frac{1}{1-y} -\frac{y\ln y}{(1-y)^2}\right]
\right.\crn&\left. -2\left[ \frac{1}{1-x} +\frac{x^2 \ln x}{(1-x)^2} -\frac{1}{1-y} -\frac{y^2\ln y}{(1-y)^2}\right]\right\}.
\end{align}
The loop functions relating with both  charged gauge  and  Higgs bosons are included in Refs.~\cite{Ilakovac:2012sh,Abada:2014kba}. Here, we use the results given in Ref.~\cite{Ilakovac:2012sh}  and the analytic functions given in Ref.~\cite{Arganda:2005ji} to cast the contributions to the Higgs and gauge bosons into the analytic functions  summarized in the following.

For photon, the off-shell form factors are
\begin{align}
\label{eq_FRga}
F^L_{\gamma}(x)&=F^R_{\gamma}(x)=F_{\gamma}(x),
\crn \overline{F}^L_{\gamma}(x)&=\overline{F}^R_{\gamma}(x)=-\frac{-11 x^3+18 x^2-9 x+2+6 x^3 \ln (x)}{36  (x-1)^4} ,
\crn\left(F^{ba,X}_{\gamma}\right)&=\sum_{i=1}^9 U^{\nu*}_{bi}U^{\nu}_{ai} \left[F^X_{\gamma}(x_{W,i})  +t_{\beta }^{-2}\overline{F}^X_{\gamma}(x_{\varphi,i})\right],
\end{align}
where $X=L,R$, $x_{w,i}\equiv m^2_{n_i}/m_W^2$, and $x_{\varphi,i}\equiv m^2_{n_i}/m^2_{\varphi}$.

The on-shell form factors from photon are:
\begin{align}
\label{eq_GRga}
G^L_{\gamma}(x)&=G^R_{\gamma}(x)=G_{\gamma}(x),
\crn \overline{G}^L_{\gamma}(x)&=\overline{G}^R_{\gamma}(x)=2x\left[f_s(x)  + t^{-2}_{\beta}\tilde{f}_s(x)\right],
\crn G^ {ba,X}_{\gamma}&=\sum_{i=1}^9 U^{\nu*}_{bi}U^{\nu}_{ai} \left[G^X_{\gamma}(x_{W,i})  +\overline{G}^X_{\gamma}(x_{\varphi,i})\right].
\end{align}
For $Z$-boson form factors, non-zero contributions are:
\begin{align}
\label{eq_FRZ}
F^L_{Z}(x)&=F_Z(x), \; G^L_{Z}(x)=G_Z(x), \; H^L_{Z}(x)=H_Z(x),
\crn \overline{G}^L_{Z}(x)&=  -\frac{1}{x-y}\left[ \frac{x\ln(x)}{x-1} -\frac{y\ln(y)}{y-1} \right],
%
%
\crn \overline{G}^R_{Z}(x)&= \frac{1}{x-y}\left[ \frac{x^2\ln(x)}{x-1} -\frac{y^2\ln(y)}{y-1} \right],
\crn F^{ba,L}_{Z}&=\sum_{i,j=1}^9U^{\nu*}_{bj}U^{\nu}_{ai}\left\{\frac{}{} \delta_{ij}F^L_Z(x_{w,i}) \right.
 \crn &\left.+D_{ij} \left[  G_Z(x_{w,i},x_{w,j})  + \frac{m_W^2}{2m^2_{\varphi }t^2_{\beta}} \overline{G}_Z(x_{\varphi,i},x_{\varphi,j})\right] +D^*_{ij} H_Z(x_{w,i},x_{w,j}) \right\},
 \crn F^{ba,R}_{Z}&=\sum_{i,j=1}^9U^{\nu*}_{bj}U^{\nu}_{ai}\left\{D_{ij} \left[   \frac{m_{e_a}m_{e_b} t^2_{\beta}}{4m^2_{W }} \overline{G}^R_Z(x_{\varphi,i},x_{\varphi,j})\right]  \right\}.
\end{align}
 Leptonic Box Formfactors relating with the four body decays into three leptons:
\begin{align}
F^{LL}_{\mathrm{Xbox}}(x,y)&=F_{\mathrm{Xbox}}(x,y), \label{eq_FXbox}
\\ \overline{F}^{LL}_{\mathrm{Xbox}}(x,y,\lambda_{\varphi})&= - \frac{2xy}{t^{2}_{\beta}} \left[\frac{x (x+4)\ln(x)}{4 (x-1) (x-\lambda _{\varphi}) (x-y)} + \frac{y (y+4) \ln(y)}{4 (y-1) (y- \lambda _{\varphi}) (y-x)} \right.
\crn&\left. +\frac{\lambda _{\varphi} (\lambda _{\varphi}+4) \ln(\lambda_{\varphi})}{4 (\lambda _{\varphi}-1) (\lambda _{\varphi}-x) (\lambda _{\varphi}-y)}\right] +  \frac{ xy }{4\lambda_{\varphi} (x-y)t^{4}_{\beta}} \left[ -\frac{x^2 \ln(x)}{(x-1)^2 } + \frac{y^2 \ln(y)}{(y-1)^2}\right],\label{eq_FbXbox}
\\ \overline{F}^{RL}_{\mathrm{Xbox}}(x,y,\lambda_{\varphi})&= \frac{m_{e_a} m_{e_b}t^2_{\beta}}{4 m_W^2} \left[ \frac{x^2 (2 y+1) \ln(x)}{(x-1) (\lambda _{\varphi}-x) (x-y)} + \frac{(2 x+1) y^2 \ln(y)}{(y-1) (\lambda _{\varphi}-y) (y-x)}
 \right.\crn &\left.  -\frac{\lambda _{\varphi} (\lambda _{\varphi}+2 x y) \ln(\lambda_{\varphi})}{(\lambda _{\varphi}-1) (\lambda _{\varphi}-x) (\lambda _{\varphi}-y)}\right], \label{eq_Fb1XboxRL}
\end{align}
where $\lambda_{\varphi} \equiv m^2_{\varphi}/m^2_W$. The total contributions to the $\mu\rightarrow3e$ decay amplitude are
\begin{align}
F^{\mu eee,LL}_{\mathrm{box}}&= \sum_{i,j=1}^9 U^{\nu*}_{bi}U^{\nu}_{ai}U^{\nu*}_{aj}U^{\nu}_{aj}   \left\{G_{\mathrm{box}}(x_{w,i},x_{w,j}) - 2\left[ F_{\mathrm{Xbox}}(x_{w,i},x_{w,j})  + \overline{F}^{LL}_{\mathrm{Xbox}}(x_{\varphi,i},x_{\varphi,j},\lambda_{\varphi}) \right] \right\},  \crn
 F^{\mu eee,RL}_{\mathrm{box}}&=  \sum_{i,j=1}^9U^{\nu*}_{bi}U^{\nu*}_{ai}U^{\nu*}_{aj}U^{\nu}_{aj} \left[ -2\overline{F}^{RL}_{\mathrm{Xbox}}(x_{\varphi,i},x_{\varphi,j},\lambda_{\varphi}) \right].
\end{align}
In the formula Eq.~\eqref{eq_brei3j}, we ignore all terms containing at least one of the following suppressed factors: $m_e^2/m_W^2, \; m_em_{\mu}/m^2_W,\; m_e/m_{\mu}$, for example    $F^{ee,R}_{Z}$, $G^{ee,R}_{Z}$ and $  F^{\mu eee,RL}_{\mathrm{box}}$.

The semi-leptonic box formfactors relating with one-loop contributions to  the $\mu-e$ conversion rate in nuclei are:
\begin{align}
F^{LL}_{\mathrm{box}}(x,y)&=F_{\mathrm{box}}(x,y),\label{eq_Fbox}
\\ \overline{F}^{LL}_{\mathrm{1,box}}(x,y,\lambda_{\varphi})&= -\frac{xy}{2}\left[ \frac{x\ln(x)}{(x-\lambda _{\varphi }) (x-y)} + \frac{y\ln(y)}{( y -\lambda _{\varphi }) (y -x)} +\frac{\lambda _{\varphi }\ln(\lambda _{\varphi })}{(\lambda _{\varphi }-x) (\lambda _{\varphi }-y)}\right]
\crn&+ \frac{\lambda_{\varphi} xy }{4(x-y)t^{4}_{\beta}} \left[ -\frac{x^2 \ln(x)}{(x-1)^2 } + \frac{y^2 \ln(y)}{(y-1)^2}\right], \label{eq_Fbbox}
\end{align}
leading to the following total one-loop contributions
\begin{align}
F^{\mu e uu}_{\mathrm{Box}}&=\sum_{i=1}^9\sum_{d_a=d,s,b}U^{\nu*}_{2i}U^{\nu}_{1i} V_{ud_a}V^*_{ud_a} \left[ F_{\mathrm{box}}(x_{w,i},x_{d_a}) + \overline{F}^{LL}_{1,\mathrm{box}}(x_{\varphi,i},x_{d_a},\lambda_{\varphi})\right], \label{eq_Fmueuu}
\\ F^{\mu e dd}_{\mathrm{Box}}&=\sum_{i=1}^9\sum_{u_a=u,c,t}U^{\nu*}_{2i}U^{\nu}_{1i} V_{du_a}V^*_{du_a}\left[ F_{X\mathrm{box}}(x_{w,i},x_{u_a}) + \overline{F}^{LL}_{\mathrm{Xbox}}(x_{\varphi,i},x_{u_a},\lambda_{\varphi}) \right].  \label{eq_Fmuedd}
\end{align}
In the numerical investigation, numerical values of  the mixing matrix $V$ and quark masses  are collected from Ref.~\cite{Zyla:2020zbs}, namely we use the central values as follows:
\begin{align*}
	V&\simeq
\left(
\begin{array}{ccc}
0.974349 & 0.2265 & 0.00132842\, -0.00336345 i \\
-0.2265 & 0.974349 & 0.0405288 \\
0.00785135\, +0.00336345 i & -0.0405288 & 1. \\
\end{array}
\right),
\end{align*}
and $m_{u,c,t}=2.16\times 10^{-3},\; 1.27,\;172.76$ [GeV],  and $m_{d,s,b}=4.67\times 10^{-3},\; 0.093,\; 4.18$ [GeV].

\section{\label{DeltaLR}$\Delta_{L,R}$ for $h\rightarrow\mu\tau$}
In this part, we will identify our notation to those defined by LoopTools~\cite{Hahn:1998yk}, using the result given in Ref.~\cite{Hue:2015fbb}. For the notations of internal momentum $k$ and external momenta $p_{1,2}$  given in Ref.~\cite{Hue:2015fbb},  the PV functions is redefined as follows
\begin{align}
B^{(i)}_{\mu}&\equiv B^{(i)}_1\times (-1)^i p_{i\mu}, \;
C_{\mu} \equiv \sum_{i=1}^2C_i (-1)^i p_{i\mu},
\end{align}
i.e., the two PV functions $B^{(1)}_1$ and $C_1$ have  opposite signs with those defined in Ref.~\cite{Hue:2015fbb}.  All of the PV-functions in this work  is just the PV-functions denoted by LoopTools: $B^{(i)}_1=B(p_i^2; M_0^2,M_i^2)$, $C_{0,1,2}= C_{0,1,2}(p_1^2, (p_1+p_2)^2,p_2^2; M_0^2, M_1^2,M_2^2)$,  $B^{}_0=B_0(p_i^2; M_0^2,M_i^2)$, and $B^{(12)}_0=B_0((p_1+p_2)^2; M_1^2,M_2^2)$.  We will use these functions for the next calculations. Denoting that $\Delta^{(i)}_{L,R}\equiv \Delta^{(i)}_{(ab)L,R}$ for short, the private contributions of all diagrams in Fig. \ref{hlilj1} to the LFVH decay amplitude are as follows
\begin{align}
\Delta^{(1)}_{L} =& \frac{g^3 m_a}{64\pi^2 m_W^3}\times[\sin(\beta-\alpha)]  \crn
\times&\sum_{i=1}^9U^{\nu}_{ai}U^{\nu*}_{bi}\left\{ m_{n_i}^2\left(B^{(1)}_0+ B^{(2)}_0 + B^{(1)}_1\right) + m_b^2B^{(2)}_1- \left(2m_W^2+m_h^2\right)m_{n_i}^2C_0  \right.\crn
-& \left. \left[m_{n_i}^2 \left( 2m_W^2+m_h^2\right)+2m_W^2 \left( 2 m_W^2+m_a^2-m_b^2\right) \right]C_1\right.\crn
-&\left. \left[ 2 m_W^2\left(m_a^2-m_h^2\right) +m_b^2m_h^2\right]C_2 \frac{}{}\right\}, \label{eq_d1L}\\
\Delta^{(1)}_{R} =& \frac{g^3m_b}{64\pi^2 m_W^3}\times[\sin(\beta-\alpha)]  \crn
\times&\sum_{i=1}^9U^{\nu}_{ai}U^{\nu*}_{bi}\left\{ m_{n_i}^2\left(B^{(1)}_0+ B^{(2)}_0+ B^{(2)}_1\right) +m_a^2B^{(1)}_1- \left(2m_W^2+m_h^2\right)m_{n_i}^2C_0  \right.\crn
-& \left. \left[ 2 m_W^2\left(m_b^2-m_h^2\right) +m_a^2m_h^2\right] C_1\right.\crn
-&\left. \left[m_{n_i}^2 \left( 2m_W^2+m_h^2\right)+2m_W^2 \left( 2 m_W^2-m_a^2+m_b^2\right) \right]C_2 \frac{}{}\right\}, \label{eq_d1R}
\end{align}
where $B^{(i)}_{0,1}=B_{0,1}(p_i^2;m^2_{n_i},m_W^2)$ and $C_{0,1,2}=C_{0,1,2}(p_1^2,m_h^2,p_2^2;m^2_{n_i},m_W^2,m_W^2)$,
\begin{align}
%
\Delta^{(2)}_{L} =& \frac{g^3m_a}{64\pi^2 m_W^3}\times[\cos(\beta-\alpha)]  \crn
\times&\sum_{i=1}^9U^{\nu}_{ai}U^{\nu*}_{bi}\left\{ t_{\beta}^{-1}m_{n_i}^2\left(B^{(1)}_1 +B^{(1)}_0\right) +\left(m_{\varphi}^2+m_W^2-m_h^2\right)\left(t_{\beta}^{-1}m_{n_i}^2C_0-t_{\beta}m_b^2 C_2\right)  \right.\crn
-& \left. \left[t_{\beta}^{-1}m_{n_i}^2 \left( m_W^2+m_h^2-m_{\varphi}^2\right)+2t_{\beta}m_b^2m_W^2 \right]C_1 \frac{}{}\right\}, \label{eq_d2L}\\
\Delta^{(2)}_{R} =& \frac{g^3m_b}{64\pi^2 m_W^3}\times[\cos(\beta-\alpha)]\times (-1) \crn
\times& \sum_{i=1}^9U^{\nu}_{ai}U^{\nu*}_{bi}\left\{ t_{\beta}\left(m_{n_i}^2B^{(1)}_0 + m_a^2B^{(1)}_1 \right) +m_{n_i}^2\left[ t_{\beta} \left(m_{\varphi}^2-m_W^2-m_h^2\right)+2t_{\beta}^{-1}m_W^2 \right]C_0  \right.\crn
-& \left. t_{\beta}\left[m_{a}^2 \left( m_W^2+m_h^2-m_{\varphi}^2\right)+2m_W^2\left(m_b^2-m_h^2\right) \right]C_1 \right.
\crn
+& \left.\left[t_{\beta}^{-1}m_{n_i}^2 \left( m_W^2+m_h^2-m_{\varphi}^2\right)-2t_{\beta}m_W^2m_b^2 \right]C_2 \frac{}{}\right\},\label{eq_d2R}
\end{align}
where $B^{(1)}_{0,1}=B_{0,1}(p_1^2;m^2_{n_i},m_W^2)$ and $C_{0,1,2}=C_{0,1,2}(p_1^2,m_h^2,p_2^2;m^2_{n_i},m_W^2,m_{\varphi}^2)$,
\begin{align}
\Delta^{(3)}_{L} =&  \frac{g^3m_a}{64\pi^2 m_W^3}\times\cos(\beta-\alpha)\times (-1) \crn
\times& \sum_{i=1}^9U^{\nu}_{ai}U^{\nu*}_{bi}\left\{t_{\beta}\left( m_{n_i}^2B^{(2)}_0+ m_b^2B^{(2)}_1\right)+ m_{n_i}^2\left[t_{\beta} \left(m_{\varphi}^2-m_W^2-m_h^2\right)+ 2t_{\beta}^{-1}m_W^2\right]C_0  \right.\crn
+& \left.\left[t_{\beta}^{-1}m_{n_i}^2 \left( m_W^2+m_h^2-m_{\varphi}^2\right)-2t_{\beta}m_W^2m_a^2 \right] C_1 \right.
\crn
-& \left.t_{\beta}\left[m_{b}^2 \left( m_W^2+m_h^2-m_{\varphi}^2\right)+2m_W^2\left(m_a^2-m_h^2\right) \right]C_2 \frac{}{}\right\}, \label{eq_d3L}\\
\Delta^{(3)}_{R} =&\frac{g^3m_b}{64\pi^2 m_W^3}\times\cos(\beta-\alpha) \crn
\times&\sum_{i=1}^9U^{\nu}_{ai}U^{\nu*}_{bi}\left\{t_{\beta}^{-1} m_{n_i}^2\left(B^{(2)}_1+B^{(2)}_0\right)+ \left(m_{\varphi}^2+m_W^2-m_h^2\right)\left(t_{\beta}^{-1}m_{n_i}^2C_0- t_{\beta}m_a^2 C_1\right)  \right.\crn
-& \left. \left[t_{\beta}^{-1}m_{n_i}^2 \left( m_W^2+m_h^2-m_{\varphi}^2\right)+2t_{\beta}m_a^2m_W^2 \right]C_2 \frac{}{}\right\},\label{eq_d3R}
\end{align}
where $B^{(2)}_{0,1}=B_{0,1}(p_2^2;m^2_{n_i},m_W^2)$ and $C_{0,1,2}=C_{0,1,2}(p_1^2,m_h^2,p_2^2;m^2_{n_i},m_{\varphi}^2,m_W^2)$,
\begin{align}
\Delta^{(4)}_{L}=& \frac{g^2m_a}{64\pi^2 m_W^3}\times2\lambda_{h\varphi\varphi}\sum_{i=1}^9U^{\nu}_{ai}U^{\nu*}_{bi} m_W\left[-m_{n_i}^2\left(C_0 -t_{\beta}^{-2}C_1\right) +t_{\beta}^{2}m_b^2 C_2 \right], \label{eq_d4L}\\
\Delta^{(4)}_{R}=& \frac{g^2m_b}{64\pi^2 m_W^3}\times2\lambda_{h\varphi\varphi}\sum_{i=1}^9U^{\nu}_{ai}U^{\nu*}_{bi}m_W\left[-m_{n_i}^2\left(C_0-t_{\beta}^{-2}C_2\right) +t_{\beta}^{2}m_a^2 C_1 \right],\label{eq_d4R}
\end{align}
where $C_{0,1,2}=C_{0,1,2}(p_1^2,m_h^2,p_2^2;m^2_{n_i},m_{\varphi}^2,m_{\varphi}^2)$, and $\lambda_{h\varphi\varphi}=-2m_W\Gamma/g  m$ given in table~\ref{table_LFVcoup}.
\begin{align}
\Delta^{(5)}_{L}=& \frac{g^3m_a}{64\pi^2 m_W^3}\times\frac{c_{\alpha}}{s_{\beta}}\sum_{i,j=1}^9U^{\nu}_{ai}U^{\nu*}_{bj}
\left\{D_{ij}\left[-m_{n_j}^2 B^{(12)}_0 + m_{n_i}^2 B^{(1)}_1 +  m^2_{j}m_{W}^2 C_0\right.\right. \crn
+& \left.  \left(2 m_W^2 (m_{n_i}^2+ m_{n_j}^2)  +2 m_{n_i}^2m_{n_j}^2 -m_a^2 m_{n_j}^2 -m_b^2 m_{n_i}^2\right) C_1 \right]\crn
+&\left. D^*_{ij}m_{n_i}m_{n_j}\left[ -B^{(12)}_0 + B^{(1)}_1 +m_{W}^2 C_0+ \left(4 m_W^2 + m_{n_i}^2+m_{n_j}^2-m_a^2  -m_b^2\right) C_1\right]   \right\}, \crn
%
\Delta^{(5)}_{R}=& \frac{g^3m_b}{64\pi^2 m_W^3}\times\frac{c_{\alpha}}{s_{\beta}}\sum_{i,j=1}^9U^{\nu}_{ai}U^{\nu*}_{bj} \left\{D_{ij}\left[-m_{n_i}^2 B^{(12)}_0 + m_{n_j}^2 B^{(2)}_1 +  m^2_{n_i}m_{W}^2 C_0\right.\right. \crn
+&\left.  \left(2 m_W^2 (m_{n_i}^2+ m_{n_j}^2) +2 m_{n_i}^2m_{n_j}^2 -m_a^2 m_{n_j}^2  -m_b^2 m_{n_i}^2\right) C_2 \right]\crn
+ & \left. D^*_{ij}m_{n_i}m_{n_j}\left[ -B^{(12)}_0  + B^{(2)}_1 + m_{W}^2 C_0 +\left(4 m_W^2 + m_{n_i}^2+m_{n_j}^2-m_a^2 -m_b^2 \right) C_2 \right]\right\},\label{eq_d5R}
\end{align}
where $D_{ij}$ is given in Eq.~\eqref{eq_fnuN},  $B^{(12)}_{0}=B_{0}(m_h^2;m^2_{n_i},m^2_{n_j})$,  $B^{(1)}_{1}=B_{1}(p_1^2;m^2_{W},m^2_{n_i})$, $B^{(2)}_{1}=B_{1}(p_2^2;m^2_{W},m^2_{n_j})$, and $C_{0,1,2}=C_{0,1,2}(p_1^2,m_h^2,p_2^2;m^2_{W},m_{n_i}^2,m_{n_j}^2)$,
\begin{align}
\Delta^{(6)}_{L}=& \frac{g^3m_a}{64\pi^2 m_W^3}\times\frac{c_{\alpha}}{s_{\beta}}\sum_{i,j=1}^9U^{\nu}_{ai}U^{\nu*}_{bj} \left\{D_{ij}\left[m_{n_j}^2 B^{(12)}_0 + \left(t_{\beta}^{2}+1\right) m_{b}^2\left( m_{n_i}^2+m_{n_j}^2 \right) C_2\right.\right. \crn
+& \left. \left[ \left(t_{\beta}^{2}+1\right)m_b^2 m_{n_i}^2+m_{\varphi}^2m_{n_j}^2 +\left(t_{\beta}^{-2}+1\right) m_{n_i}^2m_{n_j}^2\right] C_0\right.\crn
& \left. + \left(m_b^2 m_{n_i}^2+m_{a}^2m_{n_j}^2 +2t_{\beta}^{-2} m_{n_i}^2m_{n_j}^2\right) C_1 \right]\crn
+&
D^*_{ij}m_{n_i}m_{n_j}\left[ B^{(12)}_0 + 2\left(t_{\beta}^{2}+1\right) m_{b}^2C_2+\left[ \left(t_{\beta}^{2}+1\right)m_b^2 +m_{\varphi}^2 +(1+t_{\beta}^{-2})m_{n_j}^2\right] C_0\right.\crn
&\left. \left.+ \left[ m_{a}^2+m_b^2 +t_{\beta}^{-2}(m_{n_i}^2+m_{n_j}^2)\right] C_1 \right]\right\}, \label{eq_d6L}\\
\Delta^{(6)}_{R}=& \frac{g^3m_b}{64\pi^2 m_W^3}\times\frac{c_{\alpha}}{s_{\beta}}\sum_{i,j=1}^9U^{\nu}_{ai}U^{\nu*}_{bj}\left\{D_{ij}\left[m_{n_i}^2 B^{(12)}_0 +\left(t_{\beta}^{2}+1\right) m_{a}^2\left( m_{n_i}^2+m_{n_j}^2 \right) C_1\right.\right. \crn
+& \left. \left[\left(t_{\beta}^{2}+1\right) m_a^2 m_{n_j}^2+m_{\varphi}^2m_{n_i}^2 +\left(t_{\beta}^{-2}+1\right) m_{n_i}^2m_{n_j}^2\right] C_0\right.\crn%
&\left. + \left(m_a^2 m_{n_j}^2+m_{b}^2m_{n_i}^2 +2t_{\beta}^{-2} m_{n_i}^2m_{n_j}^2\right) C_2 \right]\crn
+&
D^*_{ij}m_{n_i}m_{n_j}\left[ B^{(12)}_0 + 2\left(t_{\beta}^{2}+1\right) m_{a}^2C_1+\left[ \left(t_{\beta}^{2}+1\right)m_a^2 +m_{\varphi}^2 +(t_{\beta}^{-2} +1)m_{n_i}^2\right] C_0\right.\crn
&\left. \left.+ \left(m_{a}^2+m_b^2 +t_{\beta}^{-2} (m_{n_i}^2+ m_{n_j}^2)\right) C_2 \right]\right\},\label{eq_d6R}
\end{align}
where $B^{(12)}_{0}=B_{0}(m_h^2;m^2_{n_i},m^2_{n_j})$,   and $C_{0,1,2}=C_{0,1,2}(p_1^2,m_h^2,p_2^2;m^2_{\varphi},m_{n_i}^2,m_{n_j}^2)$,
\begin{align}
\Delta^{(7+8)}_{L} =& \frac{g^3m_a}{64\pi^2 m_W^3}\times \frac{s_{\alpha}}{c_{\beta}}\times \frac{m_b^2}{m_b^2-m_a^2}\sum_{i=1}^9U^{\nu}_{ai}U^{\nu*}_{bi}\crn
\times& \left[\left(2m_W^2 +m_{n_i}^2 \right) \left(B^{(2)}_1-B^{(1)}_1\right) +m_b^2B^{(2)}_1-m_a^2B^{(1)}_1 +2m_{n_i}^2 \left( B^{(2)}_0 -B^{(1)}_0\right)\right], \crn
%
\Delta^{(7+8)}_{R} =&\frac{m_a}{m_b} \Delta^{(7+8)}_{L},\label{eq_d78R}
\end{align}
where $\Delta^{(7+8)}_{L,R}\equiv \Delta^{(7)}_{L,R} +\Delta^{(8)}_{L,R}$, and $B^{(i)}_{0,1}\equiv B_{0,1}(p_i^2;m^2_{n_i},m^2_{W})$,
\begin{align}
%
\Delta^{(9 +10)}_{L} =&\frac{g^3m_a}{64\pi^2 m_W^3}\times \frac{s_{\alpha}}{c_{\beta}}\times \frac{1}{m_b^2-m_a^2} \crn
\times& \sum_{i=1}^9U^{\nu}_{ai}U^{\nu*}_{bi} \left[ m_{n_i}^2\left(2 m^2_{b}B^{(1)}_0 -(m_a^2 +m_b^2)B^{(2)}_0\right)  + \left(t_{\beta}^{-2}\,m_{n_i}^2 +t_{\beta}^{2}\,m_a^2\right) m_b^2\left( B^{(2)}_1 -B^{(1)}_1\right) \right], \crn
%
\Delta^{(9 +10)}_{R} =&\frac{g^3m_b}{64\pi^2 m_W^3}\times \frac{s_{\alpha}}{c_{\beta}}\times \frac{1}{m_b^2-m_a^2} \crn
\times&\sum_{i=1}^9U^{\nu}_{ai}U^{\nu*}_{bi} \left[m_{n_i}^2 \left((m_a^2+m_b^2) B^{(1)}_0 -2 m_a^2 B^{(2)}_0 \right)  +\left(t_{\beta}^{-2}\,m_{n_i}^2+ t_{\beta}^{2}\,m_b^2 \right)m_a^2 \left(B^{(2)}_1- B^{(1)}_1\right)\right],\label{eq_d910R}
\end{align}
where $\Delta^{(9+10)}_{L,R}\equiv \Delta^{(9)}_{L,R} +\Delta^{(10)}_{L,R}$, and $B^{(i)}_{0,1}\equiv B_{0,1}(p_i^2;m^2_{n_i},m^2_{\varphi})$.  The analytic forms of $\Delta^{(i)}_{L,R}$ given here were also cross-checked using the FORM package~\cite{Vermaseren:2000nd, Kuipers:2012rf}.

Divergent cancellation in total $\Delta_{L,R}$ is  proved as follows. Note that divergent part appear only in the $B$-functions. Using notations of divergent part in~\cite{Hue:2015fbb}, $\Delta_{\epsilon}=(1/\epsilon)-\gamma_E+\ln(4\pi)$, we have $\mathrm{div}[B^{(1)}_0]=\mathrm{div}[B^{(2)}_0]=\mathrm{div}[B^{(12)}_0]=\Delta_{\epsilon}$, and $\mathrm{div}[B^{(1)}_1]=\mathrm{div}[B^{(2)}_1]=-\Delta_{\epsilon}/2$.  For $\Delta^{(5),(6)}_{L,R}$, divergences relating with $D^*_{ij}$ vanish because they contain one of the factors $ \sum_{j}U^{\nu}_{bj}m_{n_j}U^{\nu}_{cj}=M^{\nu*}_{bc}=0$ and  $\sum_{i}U^{\nu*}_{ai}m_{n_i}U^{\nu*}_{ci} =M^{\nu}_{ac}=0$ with all $a,b,c=1,2,3$ \cite{Thao:2017qtn}. Ignoring the overall factor $\frac{g^2 m_a}{64 \pi^2m_W^3} $  the divergent part of $\Delta^{(1)}_L$ is
\begin{equation}
 \mathrm{div}\left[\Delta^{(1)}_L\right]=\sin(\beta-\alpha) \sum_{i=1}^9U^{\nu}_{ai}U^{\nu*}_{bi}\Delta_{\epsilon}\left(\dfrac{3}{2}m_{n_i}^2+ \frac{1}{2} m_b^2\right)= \frac{3}{2}\sin(\beta-\alpha)\Delta_{\epsilon} \sum_{i=1}^9U^{\nu}_{ai}U^{\nu*}_{bi}m_{n_i}^2,\label{dvid1L}
\end{equation}
where the term containing $m^2_b$ is canceled because of the Glashow-Ilipolouos-Maiani (GIM) mechanism, $\sum_{i=1}^9U^{\nu}_{ai}U^{\nu*}_{bi}=\delta_{ab}=0$ with $a\ne b$.   In the similar calculation, we have
\begin{align}
\mathrm{div}\left[\Delta^{(2)}_L\right] &=\frac{\Delta_{\epsilon}}{2}t^{-1}_{\beta}\cos(\beta-\alpha) \sum_{i=1}^9U^{\nu}_{ai}U^{\nu*}_{bi} m_{n_i}^2,\crn
\mathrm{div}\left[\Delta^{(3)}_L\right] &=-\Delta_{\epsilon}t_{\beta}\cos(\beta-\alpha) \sum_{i=1}^9U^{\nu}_{ai}U^{\nu*}_{bi} m_{n_i}^2,\crn
\mathrm{div}\left[\Delta^{(4)}_L\right] &=\mathrm{div}\left[\Delta^{(7+8)}_L\right]=0,\crn
\mathrm{div}\left[\Delta^{(5)}_L\right] &=-\frac{3\Delta_{\epsilon}}{2}\times \dfrac{c_{\alpha}}{s_{\beta}} \sum_{i=1}^9U^{\nu}_{ai}U^{\nu*}_{bi}m_{n_i}^2,\crn
\mathrm{div}\left[\Delta^{(6)}_L\right] &=\Delta_{\epsilon}\times \dfrac{c_{\alpha}}{s_{\beta}} \sum_{i=1}^9U^{\nu}_{ai}U^{\nu*}_{bi}m_{n_i}^2,\crn
\mathrm{div}\left[\Delta^{(9+10)}_L\right] &=\Delta_{\epsilon}\times \dfrac{s_{\alpha}}{c_{\beta}} \sum_{i=1}^9U^{\nu}_{ai}U^{*\nu}_{bi}m_{n_i}^2,
\label{divdil}
\end{align}
where we have used some mediate calculations for shortening the formulas of  div$\left[\Delta^{(5,6)}_{L}\right]$, for example,
\begin{align*}
\sum_{i,j=1}^9U^{\nu}_{ai}U^{\nu*}_{bj}\left(\sum_{c=1}^3U^{\nu}_{cj}U^{\nu*}_{ci}\right) m^2_{n_j} &= \sum_{j=1}^9U^{\nu*}_{bj}m_{n_j}^2\sum_{c=1}^3U^{\nu}_{cj}\sum_{i=1}^9U^{\nu}_{ai}U^{\nu*}_{ci}= \sum_{j=1}^9U^{\nu}_{aj}U^{\nu*}_{bj}m_{n_j}^2, \crn
\sum_{i,j=1}^9U^{\nu}_{ai}U^{\nu*}_{bj}\left(\sum_{c=1}^3U^{\nu*}_{cj}U^{\nu}_{ci}\right) m_{n_i}m_{n_j} &=\sum_{c=1}^3\sum_{i,j=1}^9\left(U^{\nu}_{ai}m_{n_i}U^{\nu}_{ci} \right)   \left(U^{\nu*}_{cj}m_{n_j} U^{\nu*}_{bj}\right)= 0.
\end{align*}
The above results are obtained exactly because of the unitary property of the neutrino mixing matrix $U^{\nu}$. For $\mathrm{div}\left[\Delta^{(2,3)}_L\right]$, we have
\bea t^{-1}_{\beta}\cos(\beta-\alpha)&=& \frac{\left(1-s^2_{\beta}\right)c_{\alpha}+c_{\beta}s_{\beta}s_{\alpha}}{s_{\beta}} =\dfrac{c_{\alpha}}{s_{\beta}}- \sin(\beta-\alpha),\crn
t_{\beta}\cos(\beta-\alpha)&=& \frac{\left(1-c^2_{\beta}\right)s_{\alpha}+c_{\beta}s_{\beta}c_{\alpha}}{c_{\beta}} =\dfrac{s_{\alpha}}{c_{\beta}}+ \sin(\beta-\alpha).
\eea
Now it is easy to derive that div$[\Delta_{L,R}]=0$ for $a\neq b$.  For diagonal case of $m_D$ given by the first line of~\eqref{Majoranamass1} in this work,  every formula in ~\eqref{divdil} is automatically equal to zero because $\sum_{i=1}^9U^{\nu}_{ai}U^{\nu*}_{bi}m_{n_i}^2=\left(m_D^{\dagger}m_D\right)_{ab}=0$ for  $a\neq b$, $a,b=1,2,3$.

\end{document}